\documentclass[12pt]{article}
\usepackage{a4wide}
\usepackage{latexsym}
\usepackage{amsmath}
\usepackage{amsfonts}
\usepackage{amscd}
\usepackage{cite}
\usepackage{placeins}

\usepackage{pslatex}
\usepackage{graphicx}
\usepackage[latin1,utf8]{inputenc}
\usepackage[T1]{fontenc}

\allowdisplaybreaks

\newcommand{\bq}{\begin{eqnarray}}
\newcommand{\eq}{\end{eqnarray}}
\newcommand{\eps}{\varepsilon}

\newcommand{\curveone}{(a)}
\newcommand{\curvetwo}{(b)}
\newcommand{\curvethree}{(c)}

\newcommand{\iterintmodular}[2]{F\left(#1;#2\right)}
\newcommand{\iterint}[1]{I_\gamma\left(#1;\lambda\right)}

\begin{document}

\thispagestyle{empty}

\begin{flushright}
  MITP/18-045
\end{flushright}

\vspace{1.5cm}

\begin{center}
  {\Large\bf Analytic results for the planar double box integral relevant to top-pair production with a closed top loop \\
  }
  \vspace{1cm}
  {\large Luise Adams, Ekta Chaubey and Stefan Weinzierl \\
  \vspace{1cm}
      {\small \em PRISMA Cluster of Excellence, Institut f{\"u}r Physik, }\\
      {\small \em Johannes Gutenberg-Universit{\"a}t Mainz,}\\
      {\small \em D - 55099 Mainz, Germany}\\
  } 
\end{center}

\vspace{2cm}

\begin{abstract}\noindent
  {
In this article we give the details on the analytic calculation of the master integrals 
for the planar double box integral relevant to top-pair production with a closed top loop.
We show that these integrals can be computed systematically to all order in the dimensional regularisation parameter $\varepsilon$.
This is done by transforming the system of differential equations into a form linear in $\varepsilon$,
where the $\varepsilon^0$-part is a strictly lower triangular matrix.
Explicit results in terms of iterated integrals are presented for the terms relevant to NNLO calculations.
   }
\end{abstract}

\vspace*{\fill}

\newpage

\section{Introduction}
\label{sect:intro}

Precision particle physics relies on our ability to compute the relevant quantum corrections.
We are now entering an era of precision physics, where two-loop corrections to processes with massive particles 
are required.
It is well known that starting from two loops not all Feynman integrals may be expressed in terms of multiple
polylogarithms.
The simplest Feynman integral which cannot be expressed in terms of multiple polylogarithms is given by
the two-loop equal-mass sunrise integral \cite{Broadhurst:1993mw,Berends:1993ee,Bauberger:1994nk,Bauberger:1994by,Bauberger:1994hx,Caffo:1998du,Laporta:2004rb,Kniehl:2005bc,Groote:2005ay,Groote:2012pa,Bailey:2008ib,MullerStach:2011ru,Adams:2013nia,Bloch:2013tra,Adams:2014vja,Adams:2015gva,Adams:2015ydq,Remiddi:2013joa,Bloch:2016izu,Groote:2018rpb}.
This integral is related to an elliptic curve and can be expressed to all orders in the 
dimensional regularisation parameter $\eps$ in terms of iterated integrals of modular forms \cite{Adams:2017ejb}.
Integrals, which do not evaluate to multiple polylogarithms are now an active field of 
research in particle physics \cite{Bloch:2014qca,Remiddi:2016gno,Adams:2016xah,Adams:2017ejb,Bogner:2017vim,Adams:2018yfj,Sogaard:2014jla,Bonciani:2016qxi,vonManteuffel:2017hms,Primo:2017ipr,Ablinger:2017bjx,Bourjaily:2017bsb,Hidding:2017jkk,Passarino:2017EPJC,Remiddi:2017har,Broedel:2017kkb,Broedel:2017siw,Broedel:2018iwv,Lee:2017qql,Lee:2018ojn}
and string theory \cite{Broedel:2014vla,Broedel:2015hia,Broedel:2017jdo,DHoker:2015wxz,Hohenegger:2017kqy,Broedel:2018izr}.
The equal mass sunrise integral is an integral which depends on a single scale $p^2/m^2$.

In realistic scattering processes we face in general Feynman integrals, which depend on multiple scales.
A prominent example is given by the planar double box integral for $t \bar{t}$-production with a closed top loop.
This integral enters the next-to-next-to-leading order (NNLO) contribution for the process $pp \rightarrow t \bar{t}$.
Until quite recently, it has not been known analytically.
The existing NNLO calculation for the process $pp \rightarrow t \bar{t}$
treats this integral numerically \cite{Czakon:2013goa,Baernreuther:2013caa,Czakon:2008zk,Czakon:2007wk,Czakon:2005rk}.
This integral is clearly a cornerstone and should be understood for further progress on the analytical side.
In a recent letter we reported how this integral can be treated analytically \cite{Adams:2018bsn}.
With this longer article we would like to give all the technical details.

Our starting point is the method of 
differential equations \cite{Kotikov:1990kg,Kotikov:1991pm,Remiddi:1997ny,Gehrmann:1999as,Argeri:2007up,MullerStach:2012mp,Henn:2013pwa,Henn:2014qga,Ablinger:2015tua,Adams:2017tga,Bosma:2017hrk} for the master integrals.
In the modern incarnation of this method one tries to find a basis of master integrals $\vec{J}$, such that the system
of differential equations is in $\eps$-form \cite{Henn:2013pwa}:
\bq
\label{eps_form}
 d \vec{J}
 & = &
 \eps A \vec{J},
\eq
where $A$ does not depend on $\eps$.
If such a form is achieved, a solution in terms of iterated integrals is immediate, supplemented by appropriate boundary conditions.
This strategy has successfully been applied to many Feynman integrals evaluating to multiple polylogarithms.
The difficulty of Feynman integral calculations is therefore reduced to finding the right basis of master integrals.
In the case of multiple polylogarithms the transformation from a Laporta basis $\vec{I}$ to the basis $\vec{J}$ is algebraic
in the kinematic variables. If the transformation is rational in the kinematic variables, several algorithms exist to find such a 
transformation \cite{Gehrmann:2014bfa,Argeri:2014qva,Lee:2014ioa,Prausa:2017ltv,Gituliar:2017vzm,Meyer:2016slj,Meyer:2017joq,Lee:2017oca,Adams:2017tga,Becchetti:2017abb}. Less is known in the genuine algebraic case (i.e. involving roots) \cite{Lee:2017oca}.
In ref.~\cite{Adams:2018yfj} we showed that an $\eps$-form can even be achieved for the equal-mass sunrise / kite system,
essential steps leading to an $\eps$-form have been discussed in \cite{Adams:2015ydq,Remiddi:2016gno}.
This is made possible by enlarging the set of transformations from the Laporta basis $I$ to the basis $J$
from algebraic functions in the kinematic variables towards algebraic functions in the kinematic variables, the periods of the elliptic curve
and their derivatives.

We may slightly relax the form of the differential equation and consider
\bq
\label{linear_form}
 d \vec{J}
 & = &
 \left( A^{(0)} + \eps A^{(1)} \right) \vec{J},
\eq
where $A^{(0)}$ is strictly lower-triangular and $A^{(0)}$ and $A^{(1)}$ are independent of $\eps$.
This does not spoil the property, that the system of differential equations is easily solved in terms of iterated integrals.
Since $A^{(0)}$ is strictly lower triangular, one can easily transform the system to an $\eps$-form 
by introducing primitives for the terms occurring in the $\eps^0$-part.
In this paper we give for the system of the double box integral a transformation matrix, which transforms the system from a
pre-canonical form to a form linear in $\eps$, as in eq.~(\ref{linear_form}).
The transformation is rational in the kinematic variables, the periods of the elliptic curves and their derivatives.
A subsequent transformation, which brings the system into $\eps$-form is possible, however this would introduce additional
transcendental functions.

We choose the basis of master integrals $\vec{J}$ such that
$A^{(0)}$ vanishes if either $t=m^2$ or $s=\infty$. In the former case 
the solution reduces to multiple polylogarithms,
in the latter case 
the solution reduces to iterated integrals of modular forms already encountered in the sunrise / kite system.

The attentive reader might have noticed that we put ``elliptic curves'' into the plural.
The system of differential equations for the double box integral is not governed by a single elliptic curve.
We find that there are three elliptic curves involved, originating from different sub-topologies.
We show how the elliptic curves can be extracted from the maximal cuts.
This fact has consequences: Elliptic multiple polylogarithms are defined as iterated integrals
on a punctured elliptic curve \cite{Beilinson:1994,Levin:1997,Levin:2007,Enriquez:2010,Brown:2011,Wildeshaus,Bloch:2013tra,Bloch:2014qca,Adams:2014vja,Adams:2015gva,Adams:2015ydq,Adams:2016xah,Remiddi:2017har,Broedel:2017kkb,Broedel:2017siw,Broedel:2018iwv}. 
Inherent to this definition is the notion of a single elliptic curve.
Since in our problem there are three distinct elliptic curves involved, we do not expect our results to be expressible in terms
of elliptic multiple polylogarithms (which by definition are tied to one specific elliptic curve).
We express our results as iterated integrals of the integration kernels appearing in eq.~(\ref{linear_form}).
Let us however also stress the other side of the medal:
Our analysis also shows that nothing worse than elliptic curves appears in the calculation.

This paper is organised as follows:
In section~\ref{sect:notation} we introduce our notation and review a few basic facts about Feynman integrals.
In section~\ref{sect:iterated_integrals} we discuss iterated integrals.
There are two special cases of iterated integrals, which we briefly review: multiple polylogarithms and iterated integrals of modular forms.
In section~\ref{sect:polylogs} we discuss in detail the kinematic variables related to the subset of Feynman integrals, which
evaluate to multiple polylogarithms.
Section~\ref{sect:elliptic_curves} is devoted to elliptic curves.
After a discussion of the generic quartic case, we show how to extract the elliptic curves from the maximal cuts.
We discuss the three occurring elliptic curves in detail.
In section~\ref{sect:masters} we define the basis of master integrals $\vec{J}$.
The system of differential equations for this set of master integrals is given in section~\ref{sect:differential_equation}.
In section~\ref{sect:boundary} we show how the boundary values are obtained.
The results for the master integrals up to order $\eps^4$ are given in section~\ref{sect:results}.
Finally, our conclusions are given in section~\ref{sect:conclusions}.
The article is supplemented by an appendix.
In appendix~\ref{sect:master_topologies} we show Feynman graphs for all master topologies.
In appendix~\ref{sect:extra_relation} we give an extra relation, which reduces the number of master integrals in the sector $93$
from five to four.
In appendix~\ref{sect:modular_forms} we collected useful information on the modular forms occurring in the $s\rightarrow \infty$ limit.
Appendix~\ref{sect:boundary_constants} lists the full set of bondary constants.
Appendix~\ref{sect:supplement} describes the supplementary electronic file attached to this article, which gives the definition
of the master integrals, the differential equation and the results in electronic form.


\section{Notation, definitions and review of basics facts}
\label{sect:notation}

\begin{figure}
\begin{center}
\includegraphics[scale=1.0]{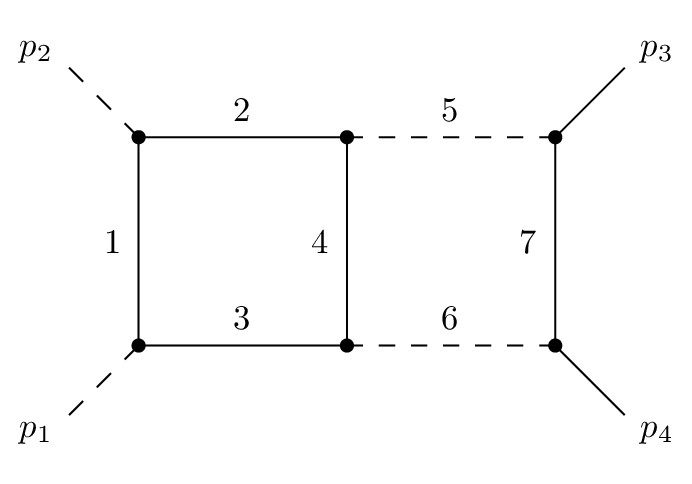}
\end{center}
\caption{
The planar double box.
Solid lines correspond to massive propagators of mass $m$, dashed lines correspond to massless propagators.
All external momenta are out-going and on-shell: $p_1^2=p_2^2=0$ and $p_3^2=p_4^2=m^2$.
}
\label{fig_double_box_graph}
\end{figure}

\begin{figure}
\begin{center}
\includegraphics[scale=1.0]{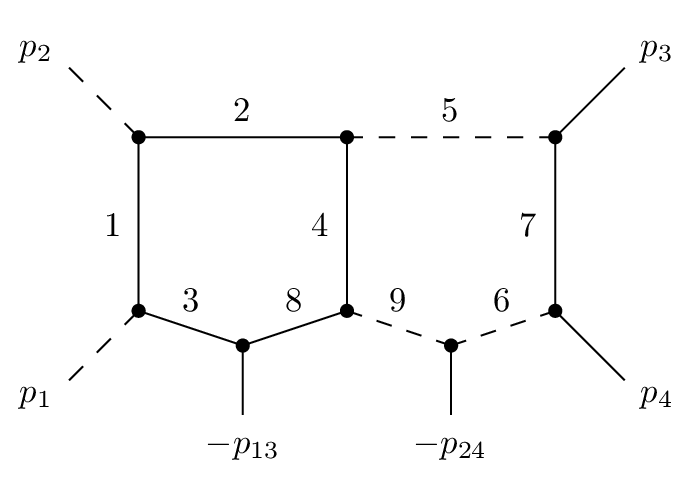}
\end{center}
\caption{
The auxiliary topology with nine propagators.
}
\label{fig_auxiliary_graph}
\end{figure}

We consider the planar double box integral shown in fig.~(\ref{fig_double_box_graph}).
This integral is relevant to the NNLO corrections for $t\bar{t}$-production at the LHC.
In fig.~(\ref{fig_double_box_graph}) the solid lines correspond to propagators with a mass $m$,
while dashed lines correspond to massless propagators.
All external momenta are out-going and on-shell.
Thus we have
\bq
 p_1 + p_2 + p_3 + p_4 = 0,
 & & 
 p_1^2=p_2^2=0, \;\;\;\;\;\; p_3^2=p_4^2=m^2.
\eq
We further set
\bq
 s = \left(p_1+p_2\right)^2,
 & &
 t = \left(p_2+p_3\right)^2.
\eq
Since there are two independent loop momenta and three independent external momenta we have 
nine independent scalar products involving the loop momenta.
We therefore consider an auxiliary topology with nine propagators, shown in fig.~(\ref{fig_auxiliary_graph}).
In $D$-dimensional Minkowski space the integral family for this auxiliary topology
is given by 
\bq
\label{def_integral}
 I_{\nu_1 \nu_2 \nu_3 \nu_4 \nu_5 \nu_6 \nu_7 \nu_8 \nu_9}\left( D, s, t, m^2, \mu^2 \right)
 & = &
 e^{2 \gamma_E \eps}
 \left(\mu^2\right)^{\nu-D}
 \int \frac{d^Dk_1}{i \pi^{\frac{D}{2}}} \frac{d^Dk_2}{i \pi^{\frac{D}{2}}}
 \prod\limits_{j=1}^9 \frac{1}{ P_j^{\nu_j} },
\eq
where $\gamma_E$ denotes the Euler-Mascheroni constant, 
$\mu$ is an arbitrary scale introduced to render the Feynman integral dimensionless, 
the quantity $\nu$ is given by
\bq
 \nu & = &
 \sum\limits_{j=1}^9 \nu_j
\eq
and
\begin{align}
 P_1 & = -\left(k_1+p_2\right)^2 + m^2,
 &
 P_2 & = -k_1^2 + m^2,
 &
 P_3 & = -\left(k_1+p_1+p_2\right)^2 + m^2,
 \nonumber \\
 P_4 & = -\left(k_1+k_2\right)^2 + m^2,
 &
 P_5 & = -k_2^2,
 &
 P_6 & = -\left(k_2+p_3+p_4\right)^2,
 \nonumber \\
 P_7 & = -\left(k_2+p_3\right)^2 + m^2,
 &
 P_8 & = -\left(k_1+p_2-p_3\right)^2 + m^2,
 &
 P_9 & = -\left(k_2-p_2+p_3\right)^2.
\end{align}
The original double box integral corresponds to $\nu_8=\nu_9=0$.
It will be convenient to introduce a short-hand notation:
If $\nu_8=\nu_9=0$, we may suppress these indices.
Furthermore we will not always write explicitly the dependency on the variables $s$, $t$, $m^2$
and $\mu^2$.
Thus
\bq
 I_{\nu_1 \nu_2 \nu_3 \nu_4 \nu_5 \nu_6 \nu_7}\left( D \right)
 & = &
 I_{\nu_1 \nu_2 \nu_3 \nu_4 \nu_5 \nu_6 \nu_7 0 0}\left( D, s, t, m^2, \mu^2 \right).
\eq 
We are interested in the Laurent expansion of these integrals in $\eps$, where $\eps=(4-D)/2$
denotes the dimensional regularisation parameter.
Thus we write
\bq
\label{Laurent_expansion}
 I_{\nu_1 \nu_2 \nu_3 \nu_4 \nu_5 \nu_6 \nu_7}\left( 4-2\eps \right)
 & = &
 \sum\limits_{j=j_{\mathrm{min}}}^\infty
 \eps^j \;  I_{\nu_1 \nu_2 \nu_3 \nu_4 \nu_5 \nu_6 \nu_7}^{(j)}.
\eq
A sector (or topology) is defined by the set of propagators with positive exponents.
We define a sector id (or topology id) by
\bq
\label{def_sector_id}
 \mathrm{id}
 & = & \sum\limits_{j=1}^9 2^{j-1} \Theta(\nu_j).
\eq
Most parts of our paper are valid for arbitrary values of $s$ and $t$.
In detail, there are no restrictions on $s$ and $t$ in section~\ref{sect:notation} to section~\ref{sect:boundary}.
In particular, the system of differential equations is valid for all values of $s$ and $t$.
The results in sections~\ref{sect:s_t_independent}-\ref{sect:s_t_dependent} are given in terms of iterated integrals.
These are valid for all values of $s$ and $t$, if a proper analytic continuation around branch cuts 
according to Feynman's $i\eps$-prescription is understood.
In a neighbourhood of the boundary point (which we take as $s=\infty$ and $t=m^2$) no analytic continuation is needed.
For the analytic continuation we have to choose the integration path such that it avoids the singularities of the 
integration kernels according to Feynman's $i\eps$-prescription. At the same time we have to ensure that the
integration kernels are continuous along the integration path.
The integration kernels will involve the periods of the elliptic curves. We have to ensure that the periods vary continuously
along the integration path.
We express the periods in a neighbourhood of the boundary point in terms of complete elliptic integrals. 
The complete elliptic integral $K(k)$, when viewed as a function of $k^2$ has a branch cut along $[1,\infty[$.
It may happen that the image of the integration path in $k^2$-space crosses this cut.
If this happens, we have to compensate for the discontinuity of $K(k)$ by taking the monodromy around $k^2=1$ into account.
This has been discussed in \cite{Bogner:2017vim}.
In section~\ref{sect:numerical_checks} we perform a numerical check. We expand all integrands in power series and integrate term by term.
This is limited to the region of convergence of the power series expansions.

\subsection{Chains and cycles}

It will be useful to group the internal propagators $P_j$ into chains \cite{Kinoshita:1962ur}.
Two propagators belong to the same chain, if their momenta differ only by a linear combination 
of the external momenta.
Obviously, each internal line can only belong to one chain. 
In fig.~(\ref{fig_auxiliary_graph}) we have three chains:
\bq
 C^{(1)} \; = \; \left\{ P_8, P_3, P_1, P_2 \right\},
 \;\;\;\;\;\;\;\;\;
 C^{(2)} \; = \; \left\{ P_9, P_6, P_7, P_5 \right\},
 \;\;\;\;\;\;\;\;\;
 C^{(3)} \; = \; \left\{ P_4 \right\}.
\eq
We define a cycle to be a closed circuit in the diagram. 
We can denote a cycle by specifying the chains which belong to the cycle.
In the two-loop diagram of fig.~(\ref{fig_auxiliary_graph}) there are three different cycles, given by
\bq
 C^{(13)}, \;\;\; C^{(23)}, \;\;\; C^{(12)}.
\eq
Here we used the notation that $C^{(ij)}$ denotes the cycle consisting of the chains $C^{(i)}$ and $C^{(j)}$.
A cycle corresponds to a one-loop sub-graph.
The discussion carries over to sub-topologies of eq.~(\ref{def_integral}) 
by deleting the appropriate propagators from the chains $C^{(1)}$, $C^{(2)}$ and $C^{(3)}$.
If a chain is empty, the two-loop integral factorises into two one-loop integrals.
For non-empty chains $C^{(1)}$, $C^{(2)}$ and $C^{(3)}$ we are interested 
in a cycle with a minimum number of propagators.
A cycle with a minimum number of propagators simplifies the calculation of the maximal cut of a
Feynman integral within the loop-by-loop approach.
The choice of such a cycle may not be unique, for example for the sunrise topology all three cycles have two propagators.
For the auxiliary topology shown in fig.~(\ref{fig_auxiliary_graph}) and all
non-trivial sub-topologies (i.e. sub-topologies which are not products of one-loop integrals) it is always
possible
to choose either
\bq
 C_1 \;\; = \;\; C^{(13)}
 & \mbox{or} &
 C_2 \;\; = \;\; C^{(23)}
\eq
as a cycle with a minimal number of propagators.
This is due to the fact that the chain $C^{(3)}$ contains already the minimal number of propagators for
a non-trivial chain.

\subsection{The Feynman parameter representation}

It is useful to discuss the Feynman integral representation for the double box integral.
The Feynman parameter integral for $\nu_8=\nu_9=0$ reads
\bq
 I_{\nu_1 \nu_2 \nu_3 \nu_4 \nu_5 \nu_6 \nu_7}\left( D \right)
 & = &
 e^{2 \gamma_E \eps}
 \frac{\Gamma(\nu-D)}{\prod\limits_{j=1}^{7}\Gamma(\nu_j)}
 \int\limits_{\sigma} 
 \left( \prod\limits_{j=1}^{7} x_j^{\nu_j-1} \right)\,\frac{{\mathcal U}^{\nu-\frac{3}{2}D}}
 {{\mathcal F}^{\nu-D}}
 \omega,
\eq
where the integration is over
\bq
 \sigma & = & \left\{ \left[ x_1 : ... : x_7 \right] \in {\mathbb R} {\mathbb P}^6 | x_i \ge 0 \right\}.
\eq
The differential form $\omega$ is given by
\bq
 \omega & = & 
 \sum\limits_{j=1}^7 (-1)^{j-1}
  \; x_j \; dx_1 \wedge ... \wedge \widehat{dx_j} \wedge ... \wedge dx_7,
\eq
where the hat indicates that the corresponding term is omitted.
The graph polynomials are given by
\bq
\label{def_U_and_F}
{\mathcal U} & = & \left( x_1+x_2+x_3 \right) \left( x_5+x_6+x_7 \right) + x_4 \left( x_1+x_2+x_3+x_5+x_6+x_7 \right),
 \nonumber \\
{\mathcal F} & = & \left[ x_2 x_3 \left( x_4+x_5+x_6+x_7 \right)
                        + x_5 x_6 \left( x_1+x_2+x_3+x_4 \right)
                        + x_2 x_4 x_6 + x_3 x_4 x_5 \right] \left( \frac{-s}{\mu^2} \right)
 \nonumber \\
 & &
      + x_1 x_4 x_7 \left( \frac{-t}{\mu^2} \right)
      + x_7 \left[ \left( x_2 + x_3 \right) x_4 + \left(x_5+x_6\right)\left(x_1+x_2+x_3+x_4\right) \right] \left( \frac{-m^2}{\mu^2} \right)
 \nonumber \\
 & &
 + \left( x_1 + x_2 + x_3 + x_4 + x_7 \right) {\mathcal U} \frac{m^2}{\mu^2}.
\eq
The graph polynomial ${\mathcal U}$ reads
in expanded form
\bq 
 {\mathcal U} & = &
  x_1 x_5 + x_1 x_6 + x_1 x_7
+ x_2 x_5 + x_2 x_6 + x_2 x_7
+ x_3 x_5 + x_3 x_6 + x_3 x_7
 \nonumber \\
 & &
+ x_1 x_4 + x_2 x_4 + x_3 x_4 + x_4 x_5 + x_4 x_6 + x_4 x_7.
\eq
Let us further define the derivatives of the graph polynomial ${\mathcal F}$
with respect to $s$ and $t$ by
\bq
 {\mathcal F}_s' & = &
 -\mu^2 \frac{d}{ds} {\mathcal F}
 \; = \;
 x_2 x_3 \left( x_4+x_5+x_6+x_7 \right)
                        + x_5 x_6 \left( x_1+x_2+x_3+x_4 \right)
                        + x_2 x_4 x_6 + x_3 x_4 x_5,
 \nonumber \\
 {\mathcal F}_t' & = &
 -\mu^2 \frac{d}{dt} {\mathcal F}
 \; = \;
 x_1 x_4 x_7.
\eq
The graph polynomials for sub-topologies are obtained from ${\mathcal U}$ and ${\mathcal F}$ by setting the Feynman parameters
to zero which correspond to propagators not present in the sub-topology.

\subsection{Dimensional shift relations and differential equations}

Let us introduce an operator ${\bf i}^+$, which raises the power of the propagator $i$ by one, e.g.
\bq
 {\bf 1}^+ I_{\nu_1 \nu_2 \nu_3 \nu_4 \nu_5 \nu_6 \nu_7}(D)
 & = &
 I_{(\nu_1+1) \nu_2 \nu_3 \nu_4 \nu_5 \nu_6 \nu_7}(D).
\eq
In addition we define two operators ${\bf D}^\pm$, which shift the dimension of space-time by two through
\bq
 {\bf D}^\pm I_{\nu_1 \nu_2 \nu_3 \nu_4 \nu_5 \nu_6 \nu_7}\left( D \right)
 & = &
 I_{\nu_1 \nu_2 \nu_3 \nu_4 \nu_5 \nu_6 \nu_7}\left( D\pm 2 \right).
\eq
The dimensional shift relations for $\nu_1,...,\nu_7 \ge 0$ read \cite{Tarasov:1996br,Tarasov:1997kx}
\bq
 {\bf D}^- I_{\nu_1 \nu_2 \nu_3 \nu_4 \nu_5 \nu_6 \nu_7}\left(D\right)
 & = &
 {\mathcal U}\left( \nu_1 {\bf 1}^+, \nu_2 {\bf 2}^+, \nu_3 {\bf 3}^+, \nu_4 {\bf 4}^+, \nu_5 {\bf 5}^+ , \nu_6 {\bf 6}^+, \nu_7 {\bf 7}^+ \right)
 I_{\nu_1 \nu_2 \nu_3 \nu_4 \nu_5 \nu_6 \nu_7}\left(D\right).
 \nonumber \\
\eq
The dimensional shift relations 
for integrals with irreducible numerators (i.e. $\nu_8<0$ or $\nu_9<0$) can be obtained as follows:
One first converts to a basis of master integrals with $\nu_8=\nu_9=0$ (and raised propagators), 
applies the dimensional shift relations
to the latter and converts back to the original basis.

The differential equations for $\nu_1,...,\nu_7 \ge 0$ can be obtained from
\bq
\label{generic_dgl}
 \mu^2 \frac{d}{ds} I_{\nu_1 \nu_2 \nu_3 \nu_4 \nu_5 \nu_6 \nu_7}\left(D\right)
 & = &
 {\bf D}^+
 {\mathcal F}_s'\left( \nu_1 {\bf 1}^+, ..., \nu_7 {\bf 7}^+ \right)
 I_{\nu_1 \nu_2 \nu_3 \nu_4 \nu_5 \nu_6 \nu_7}\left(D\right),
 \nonumber \\
 \mu^2 \frac{d}{dt} I_{\nu_1 \nu_2 \nu_3 \nu_4 \nu_5 \nu_6 \nu_7}\left(D\right)
 & = &
 {\bf D}^+
 {\mathcal F}_t'\left( \nu_1 {\bf 1}^+, ..., \nu_7 {\bf 7}^+ \right)
 I_{\nu_1 \nu_2 \nu_3 \nu_4 \nu_5 \nu_6 \nu_7}\left(D\right).
\eq
The right-hand side is given by integrals in $(D+2)$ dimensions with three propagators raised by
an additional unit.
Reducing these integrals to the basis in $D$ dimensions gives the differential equation.
Let us mention that eq.~(\ref{generic_dgl}) allows us to write the differential equations
in a compact way. However, from a computational point of view it is not the most advantageous representation,
since it requires reduction of integrals with large $\nu$.

We use the programs {\tt Reduze} \cite{vonManteuffel:2012np},
{\tt Kira} \cite{Maierhoefer:2017hyi},
{\tt Fire} \cite{Smirnov:2014hma}
and {\tt LiteRed} \cite{Lee:2012cn,Lee:2013mka}
to reduce the integrals to master integrals.
These programs are based on integration-by-parts identities \cite{Tkachov:1981wb,Chetyrkin:1981qh}
and implement the Laporta algorithm \cite{Laporta:2001dd}.
There are 44 master integrals. We may choose a basis of master integrals
for which the auxiliary propagators $P_8$ and $P_9$ are not present, at
the expense of having higher powers of the propagators for the
propagators $P_1$ - $P_7$.
It is therefore possible to label these master integrals by
$I_{\nu_1 \nu_2 \nu_3 \nu_4 \nu_5 \nu_6 \nu_7 }$.
In some sectors we have more than one master integral.
The list of master integrals is shown in table~\ref{table_master_integrals}.
\begin{table}
\begin{center}
\begin{tabular}{|c|r|r|l|l|c|}
\hline
 number of   & block & sector & master integrals & master integrals & kinematic \\
 propagators &       &        & basis $\vec{I}$  & basis $\vec{J}$  & dependence \\
\hline
\hline
 $2$ & $1$ & $9$ & $I_{1001000}$ & $J_{1}$ & $-$ \\
\hline
\hline
 $3$ & $2$ & $14$ & $I_{0111000}$ & $J_{2}$ & $s$ \\
     & $3$ & $28$ & $I_{0011100}$, $I_{0021100}$ & $J_{3}, J_{4}$ & $s$ \\
     & $4$ & $49$ & $I_{1000110}$ & $J_{5}$ & $s$ \\
     & $5$ & $73$ & $I_{1001001}$, $I_{2001001}$ & $J_{6}, J_{7}$ & $t$ \\
     & $6$ & $74$ & $I_{0101001}$ & $J_{8}$ & $-$ \\
\hline
\hline
 $4$ & $7$ & $15$ & $I_{1111000}$ & $J_{9}$ & $s$ \\
     & $8$ & $29$ & $I_{1011100}$ & $J_{10}$ & $s$ \\
     & $9$ & $54$ & $I_{0110110}$ & $J_{11}$ & $s$ \\
     & $10$ & $57$ & $I_{1001110}, I_{2001110}$ & $J_{12}, J_{13}$ & $s$ \\
     & $11$ & $75$ & $I_{1101001}$ & $J_{14}$ & $t$ \\
     & $12$ & $78$ & $I_{0111001}, I_{0211001}$ & $J_{15}, J_{16}$ & $s$ \\
     & $13$ & $89$ & $I_{1001101}$ & $J_{17}$ & $t$ \\
     & $14$ & $92$ & $I_{0011101}, I_{0021101}$ & $J_{18}, J_{19}$ & $s$ \\
     & $15$ & $113$ & $I_{1000111}$ & $J_{20}$ & $s$ \\
\hline
\hline
 $5$ & $16$ & $55$ & $I_{1110110}$ & $J_{21}$ & $s$ \\
     & $17$ & $59$ & $I_{1101110}$ & $J_{22}$ & $s$ \\
     & $18$ & $62$ & $I_{0111110}$ & $J_{23}$ & $s$ \\
     & $19$ & $79$ & $I_{1111001}, I_{2111001}, I_{1211001}$ & $J_{24}, J_{25}, J_{26}$ & $s, t$ \\
     & $20$ & $93$ & $I_{1011101}, I_{2011101}, I_{1021101},$ & $J_{27}, J_{28}, J_{29},$ & $s, t$ \\
     &     &      & $I_{1012101}$ & $J_{30}$ & \\
     & $21$ & $118$ & $I_{0110111}$ & $J_{32}$ & $s$ \\
     & $22$ & $121$ & $I_{1001111}, I_{2001111}, I_{1001112}$ & $J_{33}, J_{34}, J_{35}$ & $s, t$ \\
\hline
\hline
 $6$ & $23$ & $63$ & $I_{1111110}$ & $J_{36}$ & $s$ \\
     & $24$ & $119$ & $I_{1110111}$ & $J_{37}$ & $s$ \\
     & $25$ & $123$ & $I_{1101111}, I_{1101211}$ & $J_{38}, J_{39}$ & $s, t$ \\
     & $26$ & $126$ & $I_{0111111}$ & $J_{40}$ & $s$ \\
\hline
\hline
 $7$ & $27$ & $127$ & $I_{1111111}, I_{2111111}, I_{1211111},$ & $J_{41}, J_{42}, J_{43},$ & $s, t$ \\
     &      &      & $I_{1111112}, I_{3111111}$ & $J_{44}, J_{45}$ & \\
\hline
\end{tabular}
\end{center}
\caption{Overview of the set of master integrals.
The first column denotes the number of propagators, the second column labels consecutively the sectors or topologies,
the third column gives the sector id (defined in eq.~(\ref{def_sector_id})),
the fourth column lists the master integrals in the basis $\vec{I}$,
the fifth column the corresponding ones in the basis $\vec{J}$.
The last column denotes the kinematic dependence. 
}
\label{table_master_integrals}
\end{table}
The master topologies are shown in appendix~\ref{sect:master_topologies}.

We may change the basis of master integrals.
Let us denote by $\vec{I}$ the vector of the pre-canonical master integrals.
The differential equation for $\vec{I}$ reads for $\mu=m$
\bq
 d\vec{I} \;\; = \;\; A \vec{I},
 & &
 A \;\; = \;\; A_s \; \frac{ds}{m^2} + A_t \; \frac{dt}{m^2}.
\eq
The matrix-valued one-form $A$ satisfies the integrability condition
\bq
\label{integrability}
 dA - A \wedge A & = & 0.
\eq
Under a change of basis \cite{Henn:2013pwa}
\bq
 \vec{J} & = & U \vec{I},
\eq
one obtains
\bq
 d \vec{J}
 & = &
 A' \vec{J},
\eq
where the matrix $A'$ is related to $A$ by
\bq
 A' & = & U A U^{-1} - U d U^{-1}.
\eq
A comment is in order: The statement that there are $44$ master integrals (and not $45$) is already a non-trivial statement.
We first run 
{\tt Reduze},
{\tt Kira} and
{\tt Fire} (the latter in combination with 
{\tt LiteRed} \cite{Lee:2012cn,Lee:2013mka} and without).
Taking trivial symmetry relations into account, all programs give with standard settings 45 master integrals.
However, the reductions disagree for the three most complicated topologies
and at first sight the results of two of the three programs seem to violate the integrability condition eq.~(\ref{integrability}).
All these symptoms are resolved once an additional relations is taken into account.
The relation is given in appendix~\ref{sect:extra_relation}.
This relation reduces the number of master integrals in sector $93$ from $5$ to $4$ (and in turn the total number of master integrals
from $45$ to $44$).
Imposing this relation, the results from {\tt Reduze}, {\tt Kira} and
{\tt Fire} agree and the integrability condition is satisfied.
We have found this relation by comparing the output of the three programs above.
All differences are proportional
to a single equation.
In addition, we verified numerically the first few terms in the $\eps$-expansion of this relation.
This extra relation comes from a higher sector (i.e. sector $123$).
We would like to mention that {\tt Reduze} is able to find the relation and can be forced to use this relation with the command
\verb|distribute_external|\footnote{We thank L. Tancredi for pointing this out.}.
We also would like to mention that the new version $1.1$ of {\tt Kira} 
gives $44$ master integrals\footnote{We thank P. Maierhoefer and J. Usovitsch.}.
Let us also mention that {\tt MINT} \cite{Lee:2013hzt} and {\tt AZURITE} \cite{Georgoudis:2016wff} are programs, which can be used
to count the number of master integrals.
{\tt MINT} analyses critical points, {\tt AZURITE} is based on syzygy relations.
Applied to our problem, {\tt MINT} reports $4$ master integrals for sector $93$, {\tt AZURITE} gives $5$.


\section{Iterated integrals}
\label{sect:iterated_integrals}

Let us first review Chen's definition of iterated integrals \cite{Chen}:
Let $M$ be a $n$-dimensional manifold and
\bq
 \gamma & : & \left[0,1\right] \rightarrow M
\eq
a path with start point ${x}_i=\gamma(0)$ and end point ${x}_f=\gamma(1)$.
Suppose further that $\omega_1$, ..., $\omega_k$ are differential $1$-forms on $M$.
Let us write
\bq
 f_j\left(\lambda\right) d\lambda & = & \gamma^\ast \omega_j
\eq
for the pull-backs to the interval $[0,1]$.
For $\lambda \in [0,1]$ the $k$-fold iterated integral 
of $\omega_1$, ..., $\omega_k$ along the path $\gamma$ is defined
by
\bq
 I_{\gamma}\left(\omega_1,...,\omega_k;\lambda\right)
 & = &
 \int\limits_0^{\lambda} d\lambda_1 f_1\left(\lambda_1\right)
 \int\limits_0^{\lambda_1} d\lambda_2 f_2\left(\lambda_2\right)
 ...
 \int\limits_0^{\lambda_{k-1}} d\lambda_k f_k\left(\lambda_k\right).
\eq
We define the $0$-fold iterated integral to be
\bq
 I_{\gamma}\left(;\lambda\right)
 & = &
 1.
\eq
We have
\bq
 \frac{d}{d\lambda}
 I_{\gamma}\left(\omega_1,\omega_2,...,\omega_k;\lambda\right)
 & = &
 f_1\left(\lambda\right) \;
 I_{\gamma}\left(\omega_2,...,\omega_k;\lambda\right).
\eq
Let us now specialise to our case of interest: Without loss of generality we may set $\mu=m$ in
eq.~(\ref{def_integral}).
Then the Feynman integrals $I_{\nu_1 \nu_2 \nu_3 \nu_4 \nu_5 \nu_6 \nu_7}^{(j)}$ appearing
in the Laurent expansion of eq.~(\ref{Laurent_expansion})
depend only on two dimensionless ratios, which may be taken as
\bq
\label{s_t_coordinates}
 \frac{s}{m^2},
 & &
 \frac{t}{m^2}.
\eq
In other words, we may view the integrals $I_{\nu_1 \nu_2 \nu_3 \nu_4 \nu_5 \nu_6 \nu_7}^{(j)}$ as functions
on $M={\mathbb P}^2({\mathbb C})$, where
\bq
 \left[ s : t : m^2 \right]
\eq
denote the homogeneous coordinates.
We will express $I_{\nu_1 \nu_2 \nu_3 \nu_4 \nu_5 \nu_6 \nu_7}^{(j)}$ 
as iterated integrals on ${\mathbb P}^2({\mathbb C})$.

Note that we are free to choose any convenient coordinates on $M$.
One possibility is given by eq.~(\ref{s_t_coordinates}).
We will refer to this choice as $(s,t)$-coordinates.

A second possibility is given by the set $(x,y)$, where $x$ and $y$ are related to $s$ and $t$ by
\bq
 \frac{s}{m^2} \; = \; - \frac{\left(1-x\right)^2}{x},
 & &
 \frac{t}{m^2} \; = \; y.
\eq
We will refer to this choice as $(x,y)$-coordinates.

The $(s,t)$-coordinates and the $(x,y)$-coordinates will be our main coordinate systems,
although not the only ones.
The $(s,t)$-coordinates are closest to physics, however in these coordinates we encounter square roots
already in very simple sub-topologies.
The $(x,y)$-coordinates rationalise the most prominent square root $\sqrt{-s(4m^2-s)}$.
However, this is not the only occurring square root.
For example, we also encounter the square root $\sqrt{-s(-4m^2-s)}$.
In order to simultaneously rationalise both square roots we use the coordinates $(\tilde{x},y)$, where
$\tilde{x}$ is defined by
\bq
 \frac{s}{m^2} & = & - \frac{\left(1+\tilde{x}^2\right)^2}{\tilde{x}\left(1-\tilde{x}^2\right)}.
\eq
However, there is a price to pay: Rationalising square roots will 
increase the degree of the polynomials in intermediate stages of the calculation.
For this reason we work bottom-up and treat each sub-topology in a coordinate system adapted to
this sub-topology.

On top of this there are several elliptic topologies. Here we use the modular parameter $\tau$
of the associated elliptic curve as one variable.

\subsection{Multiple polylogarithms}

Multiple polylogarithms are a special case of iterated integrals.
For $z_k \neq 0$ they are defined by \cite{Goncharov_no_note,Goncharov:2001,Borwein,Moch:2001zr}
\bq
 \label{Gfuncdef}
 G(z_1,...,z_k;y)
 & = &
 \int\limits_0^y \frac{dy_1}{y_1-z_1}
 \int\limits_0^{y_1} \frac{dy_2}{y_2-z_2} ...
 \int\limits_0^{y_{k-1}} \frac{dy_k}{y_k-z_k}.
\eq
The number $k$ is referred to as the depth of the integral representation
or the weight of the multiple polylogarithm.
Let us introduce the short-hand notation
\bq
\label{Gshorthand}
 G_{m_1,...,m_k}(z_1,...,z_k;y)
 & = &
 G(\underbrace{0,...,0}_{m_1-1},z_1,...,z_{k-1},\underbrace{0...,0}_{m_k-1},z_k;y),
\eq
where all $z_j$ for $j=1,...,k$ are assumed to be non-zero.
This allows us to relate the integral representation of the multiple polylogarithms 
to the sum representation of the multiple polylogarithms.
The sum representation is defined by
\bq 
\label{def_multiple_polylogs_sum}
 \mathrm{Li}_{m_1,...,m_k}(x_1,...,x_k)
  & = & \sum\limits_{n_1>n_2>\ldots>n_k>0}^\infty
     \frac{x_1^{n_1}}{{n_1}^{m_1}}\ldots \frac{x_k^{n_k}}{{n_k}^{m_k}}.
\eq
The number $k$ is referred to as the depth of the sum representation of the multiple polylogarithm,
the weight is now given by $m_1+m_2+...m_k$.
The relations between the two representations are given by
\bq
\label{Gintrepdef}
 \mathrm{Li}_{m_1,...,m_k}(x_1,...,x_k)
 & = & 
 (-1)^k 
 G_{m_1,...,m_k}\left( \frac{1}{x_1}, \frac{1}{x_1 x_2}, ..., \frac{1}{x_1...x_k};1 \right),
 \nonumber \\
 G_{m_1,...,m_k}(z_1,...,z_k;y) 
 & = & 
 (-1)^k \; \mathrm{Li}_{m_1,...,m_k}\left(\frac{y}{z_1}, \frac{z_1}{z_2}, ..., \frac{z_{k-1}}{z_k}\right).
\eq
Note that in the integral representation one variable is redundant due 
to the following scaling relation (recall $z_k \neq 0$):
\bq
\label{G_scaling_relation}
 G(z_1,...,z_k;y) 
 & = & 
 G(x z_1, ..., x z_k; x y).
\eq
If one further sets $g(z;y) = 1/(y-z)$, then one has
\bq
 \frac{d}{dy} G(z_1,...,z_k;y) 
 & = & 
 g(z_1;y) G(z_2,...,z_k;y)
\eq
and
\bq
\label{Grecursive}
 G(z_1,z_2,...,z_k;y) 
 & = & 
 \int\limits_0^y dy_1 \; g(z_1;y_1) G(z_2,...,z_k;y_1).
\eq
One can slightly enlarge the set of multiple polylogarithms 
and define $G(0,...,0;y)$ with $k$ zeros for $z_1$ to $z_k$ to be
\bq
\label{trailingzeros}
 G(0,...,0;y) 
 & = & 
 \frac{1}{k!} \left( \ln y \right)^k.
\eq
This permits us to allow trailing zeros in the sequence
$(z_1,...,z_k)$ by defining the function $G$ with trailing zeros via eq.~(\ref{Grecursive}) 
and eq.~(\ref{trailingzeros}).
Please note that the scaling relation eq.~(\ref{G_scaling_relation})
does not hold for multiple polylogarithms with trailing zeros.
Using the shuffle product it is possible to remove trailing zeros and to express any
multiple polylogarithms as a linear combination of terms which involve
multiple polylogarithms without trailing zeros and powers of $\ln y$.

It will be convenient to introduce the following notation:
For differential one-forms
\bq
 \omega_j 
 & = &
 \sum\limits_{r=1}^{r_j} c_{j,r} \; \frac{dy}{y-z_{j,r}}
\eq
we define $G(\omega_1,...,\omega_k;y)$ recursively through
\bq
\label{def_G_omega}
 G\left(\omega_1,\omega_2,...,\omega_k;y\right) 
 & = &
 \sum\limits_{r=1}^{r_1} 
 c_{1,r}
 \int\limits_0^y
 dy_1 \;
 g\left(z_{1,r},y_1\right)
 \;
 G\left(\omega_2,...,\omega_k;y_1\right).
\eq
Methods for the numerical evaluation of multiple polylogarithms can be found in \cite{Vollinga:2004sn}. 

\subsection{Iterated integrals of modular forms}

A second special case of iterated integrals are iterated integrals of modular forms.
Let $f_1(\tau)$, $f_2(\tau)$, ..., $f_k(\tau)$ be modular forms of a congruence subgroup.

The (full) modular group $\text{SL}_2(\mathbb{Z})$ is the group of $(2 \times 2)$-matrices over the integers with unit determinant:
\bq
\text{SL}_2(\mathbb{Z}) & = &
\left\{ 
\left( \begin{array}{cc}
a & b \\ 
c & d
\end{array}  \right)  
 \bigg\vert \
a,b,c,d \in \mathbb{Z},\ ad-bc=1
\right\}.
\eq
The standard congruence subgroups of the modular group $\text{SL}_2(\mathbb{Z})$ are defined by
\bq
\Gamma_0(N) &=& \left\{ \left( \begin{array}{cc}
a & b \\ 
c & d
\end{array}  \right) \in \text{SL}_2(\mathbb{Z}): c \equiv 0\ \text{mod}\ N \right\}, 
 \nonumber \\
\Gamma_1(N) &=& \left\{ \left( \begin{array}{cc}
a & b \\ 
c & d
\end{array}  \right) \in \text{SL}_2(\mathbb{Z}): a,d \equiv 1\ \text{mod}\ N, \; c \equiv 0\ \text{mod}\ N  \right\},
 \nonumber \\
\Gamma(N) &=& \left\{ \left( \begin{array}{cc}
a & b \\ 
c & d
\end{array}  \right) \in \text{SL}_2(\mathbb{Z}): a,d \equiv 1\ \text{mod}\ N, \; b,c \equiv 0\ \text{mod}\ N \right\}.
\eq
The most prominent example in our application will be the congruence subgroup $\Gamma_1(6)$.
Let us further assume that $f_k(\tau)$ vanishes at the cusp $\tau=i\infty$.
We define the $k$-fold iterated integral by
\bq
\label{iter_int_modular_forms}
 \iterintmodular{f_1,f_2,...,f_k}{q}
 & = &
 \left(2 \pi i \right)^k
 \int\limits_{i \infty}^{\tau} d\tau_1
 \;
 f_1\left(\tau_1\right)
 \int\limits_{i \infty}^{\tau_1} d\tau_2
 \;
 f_2\left(\tau_2\right)
 ...
 \int\limits_{i \infty}^{\tau_{k-1}} d\tau_k
 \;
 f_k\left(\tau_k\right),
 \;\;\;\;\;\;
 q \; = \; e^{2\pi i \tau}.
 \;\;\;\;\;\;
\eq
The case where $f_k(\tau)$ does not vanishes at the cusp $\tau=i\infty$ is discussed in \cite{Adams:2017ejb,Brown:2014aa}
and is similar to trailing zeros in the case of multiple polylogarithms.
This is easily seen by changing the variable from $\tau$ to $q$:
\bq
 2 \pi i \int\limits_{i \infty}^\tau d\tau_1 \; f\left(\tau_1\right)
 & = &
 \int\limits_0^q \frac{dq_1}{q_1} \; f\left(\tau_1\left(q_1\right)\right),
 \;\;\;\;\;\;
 \tau_1\left(q_1\right)
 \; = \;
 \frac{1}{2\pi i} \ln q_1
\eq
Modular forms have a Fourier expansion around the cusp $\tau=i\infty$:
\bq
 f_j\left(\tau\right)
 & = &
 \sum\limits_{n=0}^\infty a_{j,n} q^n.
\eq
$f_j(\tau)$ vanishes at $\tau=i\infty$ if $a_{j,0}=0$.
Using the Fourier expansion and integrating term-by-term one obtains the
$q$-series of the iterated integral of modular forms corresponding to eq.~(\ref{iter_int_modular_forms}):
\bq
 \iterintmodular{f_1,f_2,...,f_k}{q}
 & = &
 \sum\limits_{n_1=0}^\infty ... \sum\limits_{n_k=0}^\infty
 \;
 \frac{a_{1,n_1}}{n_1+...+n_k} \; ... \; \frac{a_{k-1,n_{k-1}}}{n_{k-1}+n_k} \; \frac{a_{k,n_k}}{n_k}
 \;
 q^{n_1+...+n_k}.
\eq


\section{The kinematic variables for the multiple polylogarithms}
\label{sect:polylogs}

A large fraction of the integrals under consideration
will only depend on the kinematic variable $s$, but not on $t$.
These integrals are expressible in terms of multiple polylogarithms.
Furthermore one finds that all integrals under consideration
are expressible in terms of multiple polylogarithms
in the special kinematic configuration $t=m^2$.
Let us now introduce several variables related to the multiple polylogarithms.
They all replace the kinematic variable $s$ and rationalise one or several square roots.
The difference lies in the square roots they rationalise.

Let us start with the simplest case relevant to integrals with a singular point at $s=4m^2$.
The variable $x$ replaces $s$ and is defined by \cite{Fleischer:1998nb,Kotikov:2007vr,Bonciani:2010ms,Henn:2013woa}
\bq
 \frac{s}{m^2} \;\; = \;\; - \frac{\left(1-x\right)^2}{x},
 & &
 x \;\; = \;\;
 \frac{1}{2} \left( \frac{-s}{m^2} + 2 - \sqrt{ \frac{-s}{m^2} } \sqrt{4 -\frac{s}{m^2}} \right).
 \;\;\;\;
\eq
The interval $s \in ]-\infty,0]$ is mapped to $x \in [0,1]$, with the point $s=-\infty$ being mapped to $x=0$ and the point
$s=0$ being mapped to $x=1$.
This change of variables rationalises the square root $\sqrt{-s(4m^2-s)}$.
In more detail we have
\bq
\label{diff_forms_x}
 \frac{ds}{s}
 \; = \;
 \frac{2 dx}{x-1} - \frac{dx}{x},
 \;\;\;\;\;\;
 \frac{ds}{s-4m^2}
 \; = \;
 \frac{2 dx}{x+1} - \frac{dx}{x},
 \;\;\;\;\;\;
 \frac{ds}{\sqrt{-s\left(4m^2-s\right)}}
 \; = \;
 \frac{dx}{x}.
\eq
We will encounter sub-topologies, which have a singular point at $s=-4m^2$. 
The simplest example is given by sector $57$.
For these integrals we change from the variable $s$ to a variable $x'$ defined through
\bq
 \frac{s}{m^2} \; = \; -\frac{\left(1+x'\right)^2}{x'},
 & &
 x' \; = \;
 \frac{1}{2} \left( \frac{-s}{m^2} - 2 - \sqrt{-\frac{s}{m^2}} \sqrt{ -4 -\frac{s}{m^2}} \right).
 \;\;\;\;
\eq
The interval $s \in ]-\infty,-4m^2]$ is mapped to $x' \in [0,1]$.
The point $s=-\infty$ is mapped to $x'=0$, the point $s=0$ is mapped to $x'=-1$.
This change of variables rationalises the square root $\sqrt{-s(-4m^2-s)}$, but not the square root $\sqrt{-s(4m^2-s)}$.
We have
\bq
\label{diff_forms_xprime}
 \frac{ds}{s}
 \; = \;
 \frac{2 dx'}{x'+1} - \frac{dx'}{x'},
 \;\;\;\;\;\;
 \frac{ds}{s+4m^2}
 \; = \;
 \frac{2dx'}{x'-1} - \frac{dx'}{x'},
 \;\;\;\;\;\;
 \frac{ds}{\sqrt{-s\left(-4m^2-s\right)}}
 \; = \;
 \frac{dx'}{x'}.
\eq
We further have
\bq
 x' & = & -1 + \frac{\left(1-x\right)^2}{2x} - \frac{1-x}{2 x} \sqrt{x^2-6x+1},
 \nonumber \\
 x & = & 1 + \frac{\left(1+x'\right)^2}{2x'} - \frac{1+x'}{2 x'} \sqrt{x'^2+6x'+1}.
\eq
In order to rationalise simultaneously the two square roots $\sqrt{-s(4m^2-s)}$ and $\sqrt{-s(-4m^2-s)}$
we introduce a variable $\tilde{x}$ through
\bq
 x \; = \;\tilde{x} \frac{\left(1-\tilde{x}\right)}{\left(1+\tilde{x}\right)},
 & &
 \tilde{x}
 \; = \; 
 \frac{1}{2} \left( 1-x-\sqrt{x^2-6x+1}\right).
\eq
Expressing $x'$ in terms of $\tilde{x}$ yields
\bq
 x' & = & \tilde{x} \frac{\left(1+\tilde{x}\right)}{\left(1-\tilde{x}\right)}.
\eq
We have 
\begin{alignat}{3}
\label{def_omega}
 &
 \omega_0
 & \; = \; &
 \frac{ds}{s} 
 & \; = \; & 
 \frac{2 \left(2\tilde{x}\right)d\tilde{x}}{\tilde{x}^2+1} - \frac{d\tilde{x}}{\tilde{x}-1} - \frac{d\tilde{x}}{\tilde{x}+1} - \frac{d\tilde{x}}{\tilde{x}},
 \nonumber \\
 &
 \omega_4
 & \; = \; &
 \frac{ds}{s-4m^2}
 & \; = \; & 
 \frac{2 \left( 2\tilde{x}-2 \right)d\tilde{x}}{\tilde{x}^2-2\tilde{x}-1} - \frac{d\tilde{x}}{\tilde{x}-1} - \frac{d\tilde{x}}{\tilde{x}+1} - \frac{d\tilde{x}}{\tilde{x}},
 \nonumber \\
 &
 \omega_{-4}
 & \; = \; &
 \frac{ds}{s+4m^2}
 & \; = \; &
 \frac{2 \left( 2\tilde{x}+2 \right)d\tilde{x}}{\tilde{x}^2+2\tilde{x}-1} - \frac{d\tilde{x}}{\tilde{x}-1} - \frac{d\tilde{x}}{\tilde{x}+1} - \frac{d\tilde{x}}{\tilde{x}},
 \nonumber \\
 &
 \omega_{0,4}
 & \; = \; &
 \frac{ds}{\sqrt{-s\left(4m^2-s\right)}}
 & \; = \; &
 \frac{d\tilde{x}}{\tilde{x}-1} - \frac{d\tilde{x}}{\tilde{x}+1} + \frac{d\tilde{x}}{\tilde{x}},
 \nonumber \\
 &
 \omega_{-4,0}
 & \; = \; &
 \frac{ds}{\sqrt{-s\left(-4m^2-s\right)}}
 & \; = \; &
 -\frac{d\tilde{x}}{\tilde{x}-1} + \frac{d\tilde{x}}{\tilde{x}+1} + \frac{d\tilde{x}}{\tilde{x}}.
\end{alignat}
In eq.~(\ref{def_omega}) we defined for later convenience the differential forms $\omega_0$, $\omega_4$, $\omega_{-4}$,
$\omega_{0,4}$ and $\omega_{-4,0}$.
Note that
\bq
\label{omega_to_multiple_polylogs}
 \frac{2\tilde{x} d\tilde{x}}{\tilde{x}^2+1}
 & = &
 \frac{d\tilde{x}}{\tilde{x}-i}
 +
 \frac{d\tilde{x}}{\tilde{x}+i},
 \nonumber \\
 \frac{\left( 2\tilde{x}-2 \right)d\tilde{x}}{\tilde{x}^2-2\tilde{x}-1}
 & = & 
 \frac{d\tilde{x}}{\tilde{x}-\left(1+\sqrt{2}\right)}
 +
 \frac{d\tilde{x}}{\tilde{x}-\left(1-\sqrt{2}\right)},
 \nonumber \\
 \frac{\left( 2\tilde{x}+2 \right)d\tilde{x}}{\tilde{x}^2+2\tilde{x}-1}
 & = &
 \frac{d\tilde{x}}{\tilde{x}-\left(-1+\sqrt{2}\right)}
 +
 \frac{d\tilde{x}}{\tilde{x}-\left(-1-\sqrt{2}\right)}.
\eq
From eq.~(\ref{diff_forms_x}), eq.~(\ref{diff_forms_xprime}) and eq.~(\ref{def_omega})
we may read off the alphabets ${\mathcal A}$, ${\mathcal A}'$ and $\tilde{{\mathcal A}}$ for the variables $x$, $x'$ and $\tilde{x}$,
respectively.
We have
\bq
 {\mathcal A} & = & \left\{ -1, 0, 1 \right\},
 \nonumber \\
 {\mathcal A}' & = & \left\{ -1, 0, 1 \right\},
 \nonumber \\
 \tilde{{\mathcal A}} & = & \left\{ -1, 0, 1, i, -i, 1+\sqrt{2}, 1-\sqrt{2}, -1+\sqrt{2}, -1-\sqrt{2} \right\}.
\eq
Thus, iterated integrals in the variable $x$ involving the differential forms of eq.~(\ref{diff_forms_x})
may be expressed in terms of the smaller class of harmonic polylogarithms \cite{Vermaseren:1998uu,Remiddi:1999ew}.
The same holds true for iterated integrals in the variable $x'$ involving the differential forms of eq.~(\ref{diff_forms_xprime}).
On the other hand, iterated integrals in the variable $\tilde{x}$ involving the differential forms of eq.~(\ref{def_omega})
have a larger alphabet and are expressed in terms of multiple polylogarithms \cite{Goncharov_no_note,Goncharov:2001,Borwein,Moch:2001zr}.

In practice, the results for the more complicated integrals are most compactly expressed by introducing the notation of eq.~(\ref{def_G_omega}).
All  sub-topologies, which depend only on $s$, can be expressed as iterated integrals with integration kernels given by the five differential one-forms
\bq
 \left\{ \omega_0, \omega_4, \omega_{-4}, \omega_{0,4}, \omega_{-4,0} \right\}.
\eq
In addition, for $t=m^2$ (or equivalently $y=1$)
all master integrals can be expressed as iterated integrals with these integration kernels.
From eq.~(\ref{def_omega}) and eq.~(\ref{omega_to_multiple_polylogs}) it is clear that
all iterated integrals in these integration kernels 
are expressible in terms of multiple polylogarithms.


\section{Elliptic curves}
\label{sect:elliptic_curves}

In this section we discuss elliptic curves.
We start with a review of the general quartic case in section~\ref{sect:general_quartic}.
The relevant elliptic curves are extracted from the maximal cuts.
This is done in section~\ref{sect:maximal_cuts}.
We find three different elliptic curves, which we label $E^{\curveone}$, $E^{\curvetwo}$ and $E^{\curvethree}$.
These curves are discussed individually in sections~\ref{sect:curve_73} - \ref{sect:curve_121}.


\subsection{The general quartic case}
\label{sect:general_quartic}

Let us consider the elliptic curve
\bq
 E
 & : &
 w^2 - \left(z-z_1\right) \left(z-z_2\right) \left(z-z_3\right) \left(z-z_4\right)
 \; = \; 0,
\eq
where the roots $z_j$ may depend on variables $x=(x_1,...,x_n)$:
\bq
 z_j & = & z_j\left(x\right),
 \;\;\;\;\;\;
 j \in \{1,2,3,4\}.
\eq
We use the notation
\bq
 d f\left(x\right)
 & = &
 \sum\limits_{i=1}^n \left( \frac{\partial f}{\partial x_i} \right) dx_i.
\eq
We set
\bq
 Z_1 \; = \; \left(z_2-z_1\right)\left(z_4-z_3\right),
 \;\;\;\;\;\;
 Z_2 \; = \; \left(z_3-z_2\right)\left(z_4-z_1\right),
 \;\;\;\;\;\;
 Z_3 \; = \; \left(z_3-z_1\right)\left(z_4-z_2\right).
\eq 
Note that we have
\bq
 Z_1 + Z_2 & = & Z_3.
\eq
We define the modulus and the complementary modulus of the elliptic curve $E$ by
\bq
 k^2 
 \; = \; 
 \frac{Z_1}{Z_3},
 & &
 \bar{k}^2 
 \; = \;
 1 - k^2 
 \; = \;
 \frac{Z_2}{Z_3}.
\eq
Note that there are six possibilities of defining $k^2$.
Our standard choice for the periods and quasi-periods is
\bq
\label{def_generic_periods}
 \psi_1 
 \; = \; 
 \frac{4 K\left(k\right)}{Z_3^{\frac{1}{2}}},
 & &
 \psi_2
 \; = \; 
 \frac{4 i K\left(\bar{k}\right)}{Z_3^{\frac{1}{2}}},
 \nonumber \\
 \phi_1 
 \; = \; 
 \frac{4 \left[ K\left(k\right) - E\left(k\right) \right]}{Z_3^{\frac{1}{2}}},
 & &
 \phi_2
 \; = \; 
 \frac{4 i E\left(\bar{k}\right)}{Z_3^{\frac{1}{2}}}.
\eq
These periods satisfy the first-order system of differential equations
\bq
 d 
 \left( \begin{array}{c} \psi_i \\ \phi_i \end{array} \right)
 & = &
 \left( \begin{array}{cc}
 - \frac{1}{2} d \ln Z_2 & \frac{1}{2} d \ln \frac{Z_2}{Z_1} \\
 - \frac{1}{2} d \ln \frac{Z_2}{Z_3} & \frac{1}{2} d \ln \frac{Z_2}{Z_3^2} \\
 \end{array}
 \right)
 \left( \begin{array}{c} \psi_i \\ \phi_i \end{array} \right),
 \;\;\;\;\;\; i \in \{1,2\},
\eq
and the Legendre relation
\bq
 \psi_1 \phi_2 - \psi_2 \phi_1
 & = &
 \frac{8 \pi i}{Z_3}.
\eq
The parameter $\tau$ and the nome squared $q$ are defined by
\bq
 \tau \; = \; \frac{\psi_2}{\psi_1},
 & &
 q \; = \; e^{2 i \pi \tau}.
\eq
We have
\bq
 2 \pi i \; d\tau
 \; = \; 
 d\ln q
 & = &
 \frac{2 \pi i}{\psi_1^2} \; \frac{4 \pi i}{Z_3} d\ln\frac{Z_2}{Z_1}.
\eq
Let us now consider a path
$\gamma : [0,1] \rightarrow {\mathbb C}^n$ such that $x_i=x_i(\lambda)$, where the variable $\lambda$ parametrises the path.
A specific example is the path 
$\gamma_\alpha : [0,1] \rightarrow {\mathbb C}^n$, indexed by $\alpha=[\alpha_1:...:\alpha_n] \in {\mathbb C} {\mathbb P}^{n-1}$ and given explicitly by
\bq
\label{def_path}
 x_i\left(\lambda\right) & = & x_i(0) + \alpha_i \lambda,
 \;\;\;\;\;\; 1 \le i \le n.
\eq
For a path $\gamma$ we may view the periods $\psi_1$ and $\psi_2$ as functions of the variable $\lambda$.
We then have
\bq
 \left[ \frac{d^2}{d\lambda^2} + p_{1,\gamma} \frac{d}{d\lambda} + p_{0,\gamma} \right] \psi_i & = & 0,
 \;\;\;\;\;\;\;\;\;\;\;\;\;\;\;
 i \in \{1,2\},
\eq
where
\bq
\label{def_p1_p0}
 p_{1,\gamma} & = & 
 \frac{d}{d\lambda} \ln Z_3 
 - \frac{d}{d\lambda} \ln\left( \frac{d}{d\lambda} \ln \frac{Z_2}{Z_1} \right),
 \\
 p_{0,\gamma} & = &
 \frac{1}{2}
 \left( \frac{d}{d\lambda} \ln Z_1 \right)
 \left( \frac{d}{d\lambda} \ln Z_2 \right)
 - \frac{1}{2} \frac{ \left(\frac{d}{d\lambda} Z_1\right)\left(\frac{d^2}{d\lambda^2} Z_2\right)
                     -\left(\frac{d^2}{d\lambda^2} Z_1\right)\left(\frac{d}{d\lambda} Z_2\right) }
                    { Z_1\left(\frac{d}{d\lambda} Z_2\right) - Z_2\left(\frac{d}{d\lambda} Z_1\right)}
 \nonumber \\
 & &
 +
 \frac{1}{4Z_3} 
 \left[ 
   \frac{1}{Z_1} \left( \frac{d}{d\lambda} Z_1 \right)^2
   +
   \frac{1}{Z_2} \left( \frac{d}{d\lambda} Z_2 \right)^2
 \right].
 \nonumber
\eq
This defines the Picard-Fuchs operator along the path $\gamma$:
\bq
 L_\gamma & = & \frac{d^2}{d\lambda^2} + p_{1,\gamma} \frac{d}{d\lambda} + p_{0,\gamma}.
\eq
The Wronskian is defined by
\bq
\label{def_Wronskian}
 W_\gamma & = & 
 \psi_1 \frac{d}{d\lambda} \psi_2 - \psi_2 \frac{d}{d\lambda} \psi_1
 \; = \;
 \frac{4 \pi i}{Z_3} 
   \frac{d}{d\lambda} \ln \frac{Z_2}{Z_1}.
\eq
We have
\bq
 \frac{d}{d\lambda} W_\gamma & = & - p_{1,\gamma} W_\gamma,
 \nonumber \\
 2\pi i d\tau & = & \frac{2\pi i \; W_\gamma}{\psi_1^2} d\lambda.
\eq
Let us now specify to the case, where the base space is given by the two variables $(x,y)$.
Eq.~(\ref{def_p1_p0}) and eq.~(\ref{def_Wronskian}) allows us to obtain the Picard-Fuchs operator and the Wronskian
from the roots $z_1$, $z_2$, $z_3$ and $z_4$
for the variation of the elliptic curve along the paths 
\bq
 \gamma_\alpha & : & [0,1] \rightarrow {\mathbb C}^2, 
 \nonumber \\
 & & x\left(\lambda\right) \; = \; x + \alpha_1 \lambda, 
 \;\;\;\;\;\;\;\;\;
 y\left(\lambda\right) \; = \; y + \alpha_2 \lambda. 
\eq
We define the Wronskians $W_x$ and $W_y$ at the point $(x,y)$ as the derivatives in the directions
$x$ and $y$, respectively.
Thus
\bq
 W_x \; = \; 
 \psi_1 \frac{d}{dx} \psi_2 - \psi_2 \frac{d}{dx} \psi_1,
 & &
 W_y \; = \; 
 \psi_1 \frac{d}{dy} \psi_2 - \psi_2 \frac{d}{dy} \psi_1.
\eq
We have
\bq
 W_{\gamma_\alpha} & = & \alpha_1 W_x + \alpha_2 W_y,
 \nonumber \\
 2 \pi i d\tau
 & = &
 \frac{2\pi i W_{\gamma_\alpha}}{\psi_1^2} d\lambda
 \;\; = \;\; 
 \frac{2\pi i}{\psi_1^2} \left( W_x dx + W_y dy \right).
\eq
We may use the Picard-Fuchs operator to eliminate second derivatives:
\bq
 \frac{d^2}{dx^2} \psi_1
 & = &
 - p_{1,x} \frac{d}{dx} \psi_1 - p_{0,x} \psi_1,
 \\
 \frac{d^2}{dy^2} \psi_1
 & = &
 - p_{1,y} \frac{d}{dy} \psi_1 - p_{0,y} \psi_1,
 \nonumber \\
 2 \frac{d^2}{dx dy} \psi_1
 & = &
 - \left( p_{1,x+y} - p_{1,x} \right)\frac{d}{dx} \psi_1 
 - \left( p_{1,x+y} - p_{1,y} \right) \frac{d}{dy} \psi_1 
 - \left( p_{0,x+y} - p_{0,x} - p_{0,y} \right) \psi_1,
 \nonumber 
\eq
where the subscript $x+y$ refers to the path with $(\alpha_1,\alpha_2)=(1,1)$.
This leaves us with
\bq
 \psi_1, 
 \;\;\;\;\;\;\;\;\;
 \frac{d}{dx} \psi_1,
 \;\;\;\;\;\;\;\;\;
 \frac{d}{dy} \psi_1.
\eq
There is a further relation, since we may exchange any derivative of $\psi_1$ in favour of $\phi_1$:
\bq
 \frac{\frac{1}{2} \left(\frac{d}{dx} \ln Z_2\right) \psi_1 + \frac{d}{dx} \psi_1}{\frac{1}{2} \frac{d}{dx} \ln \frac{Z_2}{Z_1}}
 \;\; = \;\;
 \phi_1
 \;\; = \;\;
 \frac{\frac{1}{2} \left(\frac{d}{dy} \ln Z_2\right) \psi_1 + \frac{d}{dy} \psi_1}{\frac{1}{2} \frac{d}{dy} \ln \frac{Z_2}{Z_1}}.
\eq
This yields
\bq
\label{eq_first_derivatives}
\lefteqn{
 \frac{1}{2} \left( \frac{d}{dy} \ln \frac{Z_2}{Z_1} \right) \frac{d}{dx} \psi_1
 -
 \frac{1}{2} \left( \frac{d}{dx} \ln \frac{Z_2}{Z_1} \right) \frac{d}{dy} \psi_1
 = } & &
 \nonumber \\
 & &
 \frac{1}{4}
 \left[
  \left(\frac{d}{dx} \ln Z_2 \right) \left(\frac{d}{dy} \ln Z_1 \right)
  -
  \left(\frac{d}{dy} \ln Z_2 \right) \left(\frac{d}{dx} \ln Z_1 \right)
 \right]
 \psi_1.
\eq
Using eq.~(\ref{eq_first_derivatives}) we may eliminate one derivative, say $\frac{d}{dx} \psi_1$.
This leaves us with
\bq
 \psi_1, 
 \;\;\;\;\;\;\;\;\;
 \frac{d}{dy} \psi_1,
\eq
as expected, since the first cohomology group of an elliptic curve is two dimensional.


\subsection{Maximal cuts}
\label{sect:maximal_cuts}

In order to identify the elliptic curves associated to a Feynman integrals 
we use maximal cuts in the 
Baikov representation \cite{Baikov:1996iu,Lee:2009dh,Kosower:2011ty,CaronHuot:2012ab,Frellesvig:2017aai,Bosma:2017ens,Harley:2017qut}.
For the maximal cut of an integral we use the loop-by-loop approach \cite{Frellesvig:2017aai}.
Let us first assume that all propagators occur to the power one, although we allow irreducible numerators.
We first consider a one-loop sub-graph with a minimal number of propagators.
This is equivalent to the statement that the dimension of the sub-space spanned by the external momenta
for this sub-graph is minimal.
Let us assume that this sub-graph contains $(e+1)$ propagators $P_1$, ..., $P_{e+1}$, therefore
the sub-space spanned by the external momenta for this sub-graph has dimension $e$.
We change the integration variables for this sub-graph according to
\bq
\label{baikov_measure}
 \frac{d^Dk}{i \pi^{\frac{D}{2}}} & = &
 u \;
 \frac{2^{-e} \pi^{-\frac{e}{2}}}{\Gamma\left(\frac{D-e}{2}\right)}
 G\left(p_1,...,p_e\right)^{\frac{1+e-D}{2}} G\left(k,p_1,...,p_e\right)^{\frac{D-e-2}{2}}
 \prod\limits_{j=1}^{e+1} dP_j,
\eq
where the momenta $p_1$, ..., $p_e$
denote the linearly independent external momenta for this sub-graph,
the Gram determinant (in Minkowski space) is defined by
\bq
 G\left(p_1,...,p_e\right) & = & \det\left(- p_i \cdot p_j\right)_{1\le i,j \le e},
\eq
and $u$ denotes an (irrelevant) phase ($|u|=1$).
The maximal cuts are solutions of the homogeneous differential equations \cite{Primo:2016ebd}.
Any such solution remains a solution upon multiplication with a non-zero constant.
Therefore the phase $u$ is not relevant.

We then repeat this procedure for the second loop, replacing $p_1$, ..., $p_e$
by the set of independent external momenta for the full graph.
For an integral of the form
\bq
 I
 & = &
 e^{2 \gamma_E \eps}
 \left(\mu^2\right)^{n-D}
 \int \frac{d^Dk_1}{i \pi^{\frac{D}{2}}} \frac{d^Dk_2}{i \pi^{\frac{D}{2}}}
 \; N\left(k_1,k_2\right) \;
 \prod\limits_{j=1}^n  \frac{1}{P_j },
\eq
where 
$N(k_1,k_2)$ is a polynomial in $k_1$ and $k_2$,
a maximal cut is given by
\bq
 \mathrm{MaxCut}_{\mathcal C} \; I
 & = &
 e^{2 \gamma_E \eps}
 \left(\mu^2\right)^{n-D}
 \int\limits_{\mathcal C} \frac{d^Dk_1}{i \pi^{\frac{D}{2}}} \frac{d^Dk_2}{i \pi^{\frac{D}{2}}}
 \; N\left(k_1,k_2\right) \;
 \prod\limits_{j=1}^n  \delta\left(P_j\right),
\eq
where the integration measure is re-written according to eq.~(\ref{baikov_measure})
and the integration is over a (yet to be) specified contour in the variables $P_j$ not eliminated 
by the delta distributions.
The exact definition of the integration contour is not relevant for the extraction of the elliptic curve from the
maximal cut.
We aim for a one-dimensional integral representation for the maximal cut with a constant in the numerator and a square
root of a quartic polynomial in the denominator. This defines the elliptic curve.
The possible choices for the integration contour are then given by an integration between any pair of roots of the quartic
polynomial.
This integration gives a period of the elliptic curve. The result for any choice of integration contour may be expressed 
as a linear combination of two independent periods.
In practice we label/order the roots and define the periods by eq.~(\ref{def_generic_periods}).

For integrals with $\nu_j > 1$ we may compute a maximal cut by first converting to a basis with $\nu_j=1$
and possibly irreducible numerators and then computing the maximal cut in this basis.
Alternatively, we may interpret the delta distribution $\delta(P_j)$ as a contour integration along a small circle around $P_j=0$.
We may therefore compute the maximal cut from residues.

Let us look at a few examples.
We start with the equal mass sunrise integral (sector $73$) in two space-time dimensions.
Starting with the sub-loop $C_1$ first we obtain
\bq
\label{maxcut_73_12}
\lefteqn{
 \mathrm{MaxCut}_{\mathcal C} \; I_{1001001}\left(2-2\eps\right)
 = } & &
 \\
 & &
 \frac{u \mu^2}{\pi^2}
 \int\limits_{\mathcal C} 
 \frac{dP'}{\left(P' -t + 2 m^2 \right)^{\frac{1}{2}} \left(P' - t + 6 m^2 \right)^{\frac{1}{2}} \left(P'^2 + 6 m^2 P' - 4 m^2 t + 9 m^4\right)^{\frac{1}{2}}}
 +
 {\mathcal O}\left(\eps\right).
 \nonumber
\eq
For the sunrise integral we could equally well start with the sub-loop $C_2$. Doing so we find
\bq
\label{maxcut_73_21}
\lefteqn{
 \mathrm{MaxCut}_{\mathcal C} \; I_{1001001}\left(2-2\eps\right)
 = } & &
 \\
 & &
 \frac{u \mu^2}{\pi^2}
 \int\limits_{\mathcal C} 
 \frac{dP}{\left(P -t \right)^{\frac{1}{2}} \left(P - t + 4 m^2 \right)^{\frac{1}{2}} \left(P^2 + 2 m^2 P - 4 m^2 t + m^4\right)^{\frac{1}{2}}}
 +
 {\mathcal O}\left(\eps\right).
 \nonumber
\eq
The two representations are related by $P'=P-2m^2$.

Let us now look at the maximal cut of the double box integral (sector $127$), this time in four space-time dimensions.
We have
\bq
\label{maxcut_127_21}
\lefteqn{
 \mathrm{MaxCut}_{\mathcal C} \; I_{1111111}\left(4-2\eps\right)
 = } & &
 \\
 & &
 \frac{u \mu^6}{4 \pi^4 s^2}
 \int\limits_{\mathcal C} 
 \frac{dP}{\left(P -t \right)^{\frac{1}{2}} \left(P - t + 4 m^2 \right)^{\frac{1}{2}} \left(P^2 + 2 m^2 P - 4 m^2 t + m^4 - \frac{4m^2\left(m^2-t\right)^2}{s} \right)^{\frac{1}{2}}}
 +
 {\mathcal O}\left(\eps\right).
 \nonumber
\eq
One recognises in eq.~(\ref{maxcut_73_12}), eq.~(\ref{maxcut_73_21}) and eq.~(\ref{maxcut_127_21})
the typical period integrals of an elliptic curve.
We note that the integrand of eq.~(\ref{maxcut_127_21}) differs from the one of eq.~(\ref{maxcut_73_21}).
The difference is given by the additional term
\bq
 - \frac{4m^2\left(m^2-t\right)^2}{s}.
\eq
This term vanishes in the limit $s\rightarrow \infty$.

Let us now look at the maximal cut in the sector $79$. We find with $P=P_8$
\bq
\label{maxcut_79_21}
\lefteqn{
 \mathrm{MaxCut}_{\mathcal C} \; I_{1112001}\left(4-2\eps\right)
 = 
 } & &
 \\
 & &
 \frac{u \mu^4}{4 \pi^3 s}
 \int\limits_{\mathcal C} 
 \frac{dP}{\left(P -t \right)^{\frac{1}{2}} \left(P - t + 4 m^2 \right)^{\frac{1}{2}} \left(P^2 + 2 m^2 P - 4 m^2 t + m^4 - \frac{4m^2\left(m^2-t\right)^2}{s} \right)^{\frac{1}{2}}}
 +
 {\mathcal O}\left(\eps\right).
 \nonumber
\eq
Up to the prefactor, this is the same maximal cut integral as in eq.~(\ref{maxcut_127_21}).
Therefore the sectors $79$ and $127$ are associated to the same elliptic curve.

Our next example is the maximal cut in the sector $121$.
Here we find with $P=P_9+2m^2$
\bq
\label{maxcut_121_12}
\lefteqn{
 \mathrm{MaxCut}_{\mathcal C} \; I_{2001111}\left(4-2\eps\right)
 = 
 \frac{u \mu^4}{4 \pi^3 \left(-s\right)^{\frac{1}{2}} \left(4m^2-s\right)^{\frac{1}{2}} }
 } & &
 \\
 & &
 \times
 \int\limits_{\mathcal C} 
 \frac{dP}{\left(P -t \right)^{\frac{1}{2}} \left(P - t + 4 m^2 \right)^{\frac{1}{2}} \left(P^2 + 2 m^2 \frac{\left(s+4t\right)}{\left(s-4m^2\right)} P
           + m^2 \left(m^2-4t\right) \frac{s}{s-4m^2} - \frac{4 m^2 t^2}{s-4m^2} \right)^{\frac{1}{2}}}
 +
 {\mathcal O}\left(\eps\right).
 \nonumber
\eq
This corresponds to an elliptic curve different from the one found in sectors $79$ and $127$.
In the limit $s\rightarrow \infty$ the maximal cut integral reduces again up to a prefactor to the one of eq.~(\ref{maxcut_73_21}).

The most complicated example is the maximal cut in sector $93$.
For this sector we find first within the loop-by-loop approach a two-fold integral representation in $P_2$ and $P_8$ for the maximal cut.
The integrand has a single pole at $P_2=0$. 
Choosing as a contour for the $P_2$-integration a small circle around this pole leads (with $P=P_8$) to 
\bq
\label{maxcut_93_21}
\lefteqn{
 \frac{1}{\eps} \; \mathrm{MaxCut}_{\mathcal C} \; I_{1012101}\left(4-2\eps\right)
 = 
 } & &
 \\
 & &
 \frac{u \mu^4}{\pi^2 s}
 \int\limits_{\mathcal C} 
 \frac{dP}{\left(P -t \right)^{\frac{1}{2}} \left(P - t + 4 m^2 \right)^{\frac{1}{2}} \left(P^2 + 2 m^2 P - 4 m^2 t + m^4 - \frac{4m^2\left(m^2-t\right)^2}{s} \right)^{\frac{1}{2}}}
 +
 {\mathcal O}\left(\eps\right).
 \nonumber
\eq
We recognise again the elliptic curve of sector $79$ and $127$.

Our last example is the maximal cut in the sector $123$.
Here we find with $P=P_9+2m^2$
\bq
\label{maxcut_123_12}
\lefteqn{
 \mathrm{MaxCut}_{\mathcal C} \; I_{1101111}\left(4-2\eps\right)
 = 
 \frac{u \mu^4}{4 \pi^3 \left(-s\right)^{\frac{1}{2}} \left(4m^2-s\right)^{\frac{1}{2}}}
 } & &
 \\
 & &
 \times
 \int\limits_{\mathcal C} 
 \frac{dP}{\left(P -t \right) \left(P^2 + 2 m^2 \frac{\left(s+4t\right)}{\left(s-4m^2\right)} P
           + m^2 \left(m^2-4t\right) \frac{s}{s-4m^2} - \frac{4 m^2 t^2}{s-4m^2} \right)^{\frac{1}{2}}}
 +
 {\mathcal O}\left(\eps\right).
 \nonumber
\eq
The denominator may be viewed as a square root of a quartic polynomial, where two roots coincide.
This does not involve an elliptic curve and corresponds to genus zero.


\subsection{The elliptic curve associated to sector $73$}
\label{sect:curve_73}

From eq.~(\ref{maxcut_73_21}) we may read off the elliptic curve for the sunrise integral:
\bq
\label{E_73_maxcut}
 E^{\curveone}
 & : &
 w^2 - \left(z - \frac{t}{\mu^2} \right) 
       \left(z - \frac{t - 4 m^2}{\mu^2} \right) 
       \left(z^2 + \frac{2 m^2}{\mu^2} z + \frac{m^4 - 4 m^2 t}{\mu^4} \right)
 \; = \; 0.
\eq
The roots of the quartic polynomial are
\bq
 z^{\curveone}_1 \; = \; \frac{t-4m^2}{\mu^2},
 \;\;\;
 z^{\curveone}_2 \; = \; \frac{-m^2-2m\sqrt{t}}{\mu^2},
 \;\;\;
 z^{\curveone}_3 \; = \; \frac{-m^2+2m\sqrt{t}}{\mu^2},
 \;\;\;
 z^{\curveone}_4 \; = \; \frac{t}{\mu^2}.
\eq
This curve has the $j$-invariant
\bq
 j\left(E^{\curveone}\right)
 & = &
 \frac{\left(3m^2+t\right)^3 \left(3 m^6 + 75 m^4 t - 15 m^2 t^2 + t^3\right)^3}
      {m^6 t \left(m^2-t\right)^6 \left(9m^2-t\right)^2}.
\eq
Two elliptic curves over ${\mathbb C}$ are isomorphic, if and only if they have the same $j$-invariant.
Let us now consider a path $\gamma_\beta$ in $(s,t)$-space parametrised by 
\bq
\label{def_path_beta}
 s \; = \; s_0 + \beta_1 \lambda \mu^2,
 & &
 t \; = \; t_0 + \beta_2 \lambda \mu^2.
\eq
For the Wronskian and the Picard-Fuchs operator $\frac{d}{d\lambda^2} + p^{\curveone}_{1,\gamma_\beta} \frac{d}{d\lambda} + p^{\curveone}_{0,\gamma_\beta}$ we find
\bq
 W^{\curveone}_{\gamma_\beta} & = & 
 2 \pi i \mu^6 \frac{3 \beta_2}{t \left(t-m^2\right) \left(t-9m^2\right)},
 \nonumber \\
 p^{\curveone}_{1,\gamma_\beta} & = & 
 - \mu^2 \left( \beta_1 \frac{d}{ds} + \beta_2 \frac{d}{dt} \right) \ln W^{\curveone}_{\gamma_\beta},
 \nonumber \\
 p^{\curveone}_{0,\gamma_\beta} & = &
 \mu^{10}
 \frac{2\pi i}{W^{\curveone}_{\gamma_\beta}} 
 \frac{3 \beta_2^3 \left(t-3m^2\right)}{t^2 \left(t-m^2\right)^2 \left(t-9m^2\right)^2}.
\eq
Eq.~(\ref{eq_first_derivatives}) reduces to the trivial equation
\bq
 0 & = & 0.
\eq
We have
\bq
\label{modular_lambda_lambdabar_73}
 16 \frac{\eta\left(\frac{\tau^{\curveone}}{2}\right)^{24} \eta\left(2\tau^{\curveone}\right)^{24}}{\eta\left(\tau^{\curveone}\right)^{48}}
 \;\; = \;\;
 \left( k^{\curveone} \bar{k}^{\curveone} \right)^2
 \;\; = \;\;
 16 \frac{m^3 \sqrt{t} \left(m-\sqrt{t}\right)^3 \left(3m+\sqrt{t}\right)}{\left(m+\sqrt{t}\right)^6 \left(3m-\sqrt{t}\right)^2}.
\eq
Dedekind's eta function is defined by
\bq
 \eta (\tau) \; = \; e^{\frac{i \pi \tau}{12}} \prod\limits_{n=1}^{\infty} (1-e^{2\pi i n \tau}) 
 \; = \;
 q^{\frac{1}{24}} \prod\limits_{n=1}^{\infty} (1-q^n),
 \qquad q=e^{2\pi i \tau}.
\eq
For a path $\gamma_\alpha$ in $(x,y)$-space
\bq
 x \; = \; \alpha_1 \lambda,
 & &
 y \; = \; 1 + \alpha_2 \lambda
\eq
we may use eq.~(\ref{modular_lambda_lambdabar_73}) to express $\lambda$ as a power series in $q^{\curveone}$ and vice versa.
The point $(x,y)=(0,1)$ corresponds to $\tau^{\curveone}=i\infty$.

For $y=1$ we have
\bq
 \left. \psi_1^{\curveone} \right|_{y=1}
 \; = \;
 \frac{\pi}{2},
 & &
 \left. \frac{d}{dy} \psi_1^{\curveone} \right|_{y=1}
 \; = \;
 - \frac{\pi}{8}.
\eq


\subsection{The elliptic curve associated to sector $79$, sector $93$ and sector $127$}
\label{sect:curve_127}

From eq.~(\ref{maxcut_127_21}) we obtain the elliptic curve associated to the double box integral:
\bq
 E^{\curvetwo}
 & : &
 w^2 - \left(z - \frac{t}{\mu^2} \right) 
       \left(z - \frac{t - 4 m^2}{\mu^2} \right) 
       \left(z^2 + \frac{2 m^2}{\mu^2} z + \frac{m^4 - 4 m^2 t}{\mu^4} - \frac{4m^2\left(m^2-t\right)^2}{\mu^4 s} \right)
 \; = \; 0.
 \nonumber \\
\eq
The roots of the quartic polynomial are now
\bq
 & &
 z^{\curvetwo}_1 \; = \; \frac{t-4m^2}{\mu^2},
 \;\;\;
 z^{\curvetwo}_2 \; = \; \frac{-m^2-2m\sqrt{t + \frac{\left(m^2-t\right)^2}{s}}}{\mu^2},
 \;\;\;
 z^{\curvetwo}_3 \; = \; \frac{-m^2+2m\sqrt{t + \frac{\left(m^2-t\right)^2}{s}}}{\mu^2},
 \nonumber \\
 & &
 z^{\curvetwo}_4 \; = \; \frac{t}{\mu^2}.
\eq
The $j$-invariant is given by
\bq
\lefteqn{
 j\left(E^{\curvetwo}\right)
 = } & &
 \\
 & &
 \!\!\!\!\!\!\!\!\!
 \frac{\left\{ s \left(3m^2+t\right) \left[ s \left(3m^6 + 75 m^4 t - 15 m^2 t + t^3\right) + 8 m^2 \left(m^2-t\right)^2 \left(9m^2-t\right) \right]  + 16 m^4 \left(m^2-t\right)^4 
       \right\}^3}
      {s m^6 \left(s-4m^2\right)^2 \left[s t + \left(m^2-t\right)^2 \right] \left(m^2-t\right)^6 \left[ s \left( 9m^2 - t \right) - 4 m^2 \left( m^2-t \right) \right]^2}.
 \nonumber
\eq
For the Wronskian and the Picard-Fuchs operator $\frac{d}{d\lambda^2} + p^{\curvetwo}_{1,\gamma_\beta} \frac{d}{d\lambda} + p^{\curvetwo}_{0,\gamma_\beta}$ 
we find for the path $\gamma_\beta$ defined in eq.~(\ref{def_path_beta})
\bq
\label{def_Wronskian_127}
 W^{\curvetwo}_{\gamma_\beta} & = & 
 2 \pi i \mu^6 
 \frac{ \beta_1 \left(m^2-t\right) \left[ s\left(t+3m^2\right)-4m^2\left(m^2-t\right)\right] 
      + \beta_2 s \left(s-4m^2\right) \left(2m^2-3s-2t\right)}
      {\left(s-4m^2\right) \left(t-m^2\right) \left[s t +\left(m^2-t\right)^2\right] \left[s \left(9 m^2-t\right) -4 m^2 \left(m^2-t\right)\right]},
 \nonumber \\
 p^{\curvetwo}_{1,\gamma_\beta} & = & 
 - \mu^2 \left( \beta_1 \frac{d}{ds} + \beta_2 \frac{d}{dt} \right) \ln W^{\curvetwo}_{\gamma_\beta},
 \\
 p^{\curvetwo}_{0,\gamma_\beta} & = &
 \mu^{10}
 \frac{2\pi i}{W^{\curvetwo}_{\gamma_\beta}} 
 \frac{N^{\curvetwo}}{s^2 \left(s-4m^2\right)^2 \left(t-m^2\right)^2 \left[s t +\left(m^2-t\right)^2\right]^2 \left[s \left(9 m^2-t\right) -4 m^2 \left(m^2-t\right)\right]^2},
 \nonumber 
\eq
with
\bq
 N^{\curvetwo} 
 & = &
 2 \beta_1^3 
 \left(m^2-t\right)^4 m^2 \left(8 m^8-10 m^6 s-16 m^6 t+9 m^4 s^2+4 m^4 s t+8 m^4 t^2+8 m^2 s^2 t
 \right. \nonumber \\
 & & \left.
 +6 m^2 s t^2-s^2 t^2\right) 
 \nonumber \\
 & &
 + 
 \beta_1^2 \beta_2 s \left(m^2-t\right)^2 
   \left(96 m^{12}-248 m^{10} s-288 m^{10} t+276 m^8 s^2+504 m^8 s t+288 m^8 t^2
   \right. \nonumber \\
   & & \left. 
   -63 m^6 s^3-360 m^6 s^2 t-264 m^6 s t^2-96 m^6 t^3+119 m^4 s^3 t+84 m^4 s^2 t^2+8 m^4 s t^3
   \right. \nonumber \\
   & & \left. 
   -18 m^2 s^4 t-25 m^2 s^3 t^2+2 s^4 t^2+s^3 t^3\right) 
 \nonumber \\
 & &
 +
 2 \beta_1 \beta_2^2 s^2 \left(4 m^2-s\right) \left(m^2-t\right)^2 
    \left(24 m^8-78 m^6 s-48 m^6 t+88 m^4 s^2+84 m^4 s t
   \right. \nonumber \\
   & & \left. 
          +24 m^4 t^2
          -18 m^2 s^3-24 m^2 s^2 t-6 m^2 s t^2+s^3 t\right) 
 \nonumber \\
 & &
 +
 \beta_2^3 s^3 \left(4 m^2-s\right)^2
   \left(8 m^8-30 m^6 s-24 m^6 t+36 m^4 s^2+62 m^4 s t+24 m^4 t^2-9 m^2 s^3
   \right. \nonumber \\
   & & \left. 
   -42 m^2 s^2 t-34 m^2 s t^2-8 m^2 t^3+3 s^3 t+6 s^2 t^2+2 s t^3\right).
\eq
Eq.~(\ref{eq_first_derivatives}) yields
\bq
\label{eliminate_derivative_s_t_127}
 \psi^{\curvetwo}_1
 & = &
 \frac{\left(s-4m^2\right)\left(3s+2t-2m^2\right)}{t-m^2}
 \frac{d}{ds} \psi^{\curvetwo}_1 
 -
 \frac{s \left(t+3m^2\right)-4m^2\left(m^2-t\right)}{s}
 \frac{d}{dt} \psi^{\curvetwo}_1.
\eq
In $(x,y)$-space this translates to
\bq
\label{eliminate_derivative_x_y_127}
 \psi^{\curvetwo}_1
 & = &
 - \frac{\left(x+1\right)\left(3x^2-2xy-4x+3\right)}{\left(x-1\right)\left(y-1\right)}
 \frac{d}{dx} \psi^{\curvetwo}_1 
 -
 \frac{x^2y + 3 x^2 - 6 xy - 2x + y + 3}{\left(x-1\right)^2}
 \frac{d}{dy} \psi^{\curvetwo}_1.
 \;\;\;\;\;\;
\eq
We have with
\bq
 \chi^{\curvetwo} & = &
 \sqrt{t + \frac{\left(m^2-t\right)^2}{s}}
\eq
the relation
\bq
\label{modular_lambda_lambdabar_127}
 16 \frac{\eta\left(\frac{\tau^{\curvetwo}}{2}\right)^{24} \eta\left(2\tau^{\curvetwo}\right)^{24}}{\eta\left(\tau^{\curvetwo}\right)^{48}}
 & = &
 \left( k^{\curvetwo} \bar{k}^{\curvetwo} \right)^2
 \\
 & = &
 16 \frac{m^3 \chi^{\curvetwo} \left(m^2+t-2m\chi^{\curvetwo}\right) \left(3m^2-t-2m\chi^{\curvetwo}\right)}{\left(m^2+t+2m\chi^{\curvetwo}\right)^2 \left(3m^2-t+2m\chi^{\curvetwo}\right)^2}.
 \nonumber
\eq
For a path $\gamma_\alpha$ in $(x,y)$-space
\bq
 x \; = \; \alpha_1 \lambda,
 & &
 y \; = \; 1 + \alpha_2 \lambda
\eq
we may use eq.~(\ref{modular_lambda_lambdabar_127}) to express $\lambda$ as a power series in $q^{\curvetwo}$ and vice versa.
The point $(x,y)=(0,1)$ corresponds to $\tau^{\curvetwo}=i\infty$.

For $y=1$ we have
\bq
 \left. \psi_1^{\curvetwo} \right|_{y=1}
 \; = \;
 \frac{\pi}{2},
 & &
 \left. \frac{d}{dy} \psi_1^{\curvetwo} \right|_{y=1}
 \; = \;
 - \frac{\pi}{8}.
\eq


\subsection{The elliptic curve associated to sector $121$}
\label{sect:curve_121}

From eq.~(\ref{maxcut_121_12}) we obtain the elliptic curve associated to sector $121$:
\bq
 E^{\curvethree}
 & : &
 w^2 - \left(z - \frac{t}{\mu^2} \right) 
       \left(z - \frac{t - 4 m^2}{\mu^2} \right) 
       \left(z^2 + \frac{2 m^2 \left(s+4t\right)}{\mu^2 \left(s-4m^2\right)} z + \frac{s m^2 \left(m^2-4t\right) - 4 m^2 t^2}{\mu^4 \left(s-4m^2\right)} \right)
 \; = \; 0.
 \nonumber \\
\eq
The roots of the quartic polynomial are now
\bq
 z^{\curvethree}_1 & = & \frac{t-4m^2}{\mu^2},
 \nonumber \\
 z^{\curvethree}_2 & = & \frac{1}{\mu^2} \left( - m^2 \frac{\left(s+4t\right)}{\left(s-4m^2\right)} - \frac{2}{4m^2-s} \sqrt{s m^2 \left( st + \left(m^2-t\right)^2 \right) } \right),
 \nonumber \\
 z^{\curvethree}_3 & = & \frac{1}{\mu^2} \left( - m^2 \frac{\left(s+4t\right)}{\left(s-4m^2\right)} + \frac{2}{4m^2-s} \sqrt{s m^2 \left( st + \left(m^2-t\right)^2 \right) } \right),
 \nonumber \\
 z^{\curvethree}_4 & = & \frac{t}{\mu^2}.
\eq
The $j$-invariant is given by
\bq
 j\left(E^{\curvethree}\right)
 & = &
 \frac{\left\{ s \left(3m^2+t\right) \left(3m^6 + 75 m^4 t - 15 m^2 t + t^3\right) + 192 m^6 \left(m^2-t\right)^2 
       \right\}^3}
      {m^6 \left[s t + \left(m^2-t\right)^2 \right] \left(m^2-t\right)^4 \left[ s \left(m^2-t\right) \left( 9m^2 - t \right) - 64 m^6 \right]^2}.
 \nonumber
\eq
For the Wronskian and the Picard-Fuchs operator $\frac{d}{d\lambda^2} + p^{\curvethree}_{1,\gamma_\beta} \frac{d}{d\lambda} + p^{\curvethree}_{0,\gamma_\beta}$ 
we find for the path $\gamma_\beta$ defined in eq.~(\ref{def_path_beta})
\bq
\label{def_Wronskian_121}
\lefteqn{
 W^{\curvethree}_{\gamma_\beta} =  
 2 \pi i \mu^6 
 } & &
 \nonumber \\
 & &
 \frac{ \left(s-4m^2\right) 
        \left\{\beta_1 \left(t-m^2\right)\left(t^2-6m^2t-3m^4\right) 
               + \beta_2 s \left( 3 s \left(t-m^2\right) + 2 \left(t-m^2\right)^2 + 16 m^4 \right)
        \right\}}
      {s \left(t-m^2\right) \left[s t +\left(m^2-t\right)^2\right] \left[s \left(m^2-t\right) \left(9 m^2-t\right) - 64 m^6\right]},
 \nonumber \\
\lefteqn{
 p^{\curvethree}_{1,\gamma_\beta} =  
 - \mu^2 \left( \beta_1 \frac{d}{ds} + \beta_2 \frac{d}{dt} \right) \ln W^{\curvethree}_{\gamma_\beta},
 } & &
 \\
\lefteqn{
 p^{\curvethree}_{0,\gamma_\beta} = 
 \mu^{10}
 \frac{2\pi i}{W^{\curvethree}_{\gamma_\beta}} 
 \frac{N^{\curvethree}}{s^3 \left(s-4m^2\right) \left(t-m^2\right)^2 \left[s t +\left(m^2-t\right)^2\right]^2 \left[s \left(m^2-t\right) \left(9 m^2-t\right) - 64 m^6\right]^2},
} & & \nonumber 
\eq
with
\bq
\lefteqn{
 N^{\curvethree} 
 = } & & \nonumber \\
 & &
-2 \beta_1^3 m^2 \left(m^2-t\right)^2 \left(3 m^4+6 m^2 t-t^2\right) 
 \left(32 m^{12}-62 m^{10} s-64 m^{10} t+3 m^8 s^2-24 m^8 s t
 \right. \nonumber \\
 & & \left.
 +32 m^8 t^2
 -26 m^6 s^2 t-20 m^6 s t^2-36 m^4 s^2 t^2-24 m^4 s t^3-6 m^2 s^2 t^3+2 m^2 s t^4+s^2 t^4\right) 
 \nonumber \\
 & &
 - \beta_1^2 \beta_2 s \left(m^2-t\right) 
   \left(9600 m^{18}-8232 m^{16} s-17920 m^{16} t+1656 m^{14} s^2+18736 m^{14} s t
 \right. \nonumber \\
 & & \left.
   +7424 m^{14} t^2
   -81 m^{12} s^3-8724 m^{12} s^2 t-7512 m^{12} s t^2+512 m^{12} t^3+1356 m^{10} s^3 t
 \right. \nonumber \\
 & & \left.
   +2972 m^{10} s^2 t^2-416 m^{10} s t^3+384 m^{10} t^4-54 m^8 s^4 t-1045 m^8 s^3 t^2+632 m^8 s^2 t^3+1896 m^8 s t^4
 \right. \nonumber \\
 & & \left.
   +96 m^6 s^4 t^2-256 m^6 s^3 t^3-720 m^6 s^2 t^4-400 m^6 s t^5-28 m^4 s^4 t^3+37 m^4 s^3 t^4+92 m^4 s^2 t^5
 \right. \nonumber \\
 & & \left.
   +24 m^4 s t^6-16 m^2 s^4 t^4-12 m^2 s^3 t^5-4 m^2 s^2 t^6+2 s^4 t^5+s^3 t^6\right) 
 \nonumber \\
 & &
 -2 \beta_1 \beta_2^2 s^2 \left(4 m^2-s\right) 
  \left(544 m^{16}-518 m^{14} s-768 m^{14} t+84 m^{12} s^2+1252 m^{12} s t-192 m^{12} t^2
 \right. \nonumber \\
 & & \left.
  -420 m^{10} s^2 t-58 m^{10} s t^2+512 m^{10} t^3+39 m^8 s^3 t+416 m^8 s^2 t^2+536 m^8 s t^3-96 m^8 t^4
 \right. \nonumber \\
 & & \left.
  -76 m^6 s^3 t^2-176 m^6 s^2 t^3-250 m^6 s t^4+34 m^4 s^3 t^3+108 m^4 s^2 t^4+68 m^4 s t^5+4 m^2 s^3 t^4
 \right. \nonumber \\
 & & \left.
  -12 m^2 s^2 t^5-6 m^2 s t^6-s^3 t^5\right) 
 \nonumber \\
 & &
 + \beta_2^3 s^3 \left(4 m^2-s\right)^2 
   \left(736 m^{12}-542 m^{10} s-672 m^{10} t+120 m^8 s^2+562 m^8 s t-96 m^8 t^2-9 m^6 s^3
 \right. \nonumber \\
 & & \left.
    -206 m^6 s^2 t-340 m^6 s t^2+32 m^6 t^3+21 m^4 s^3 t+122 m^4 s^2 t^2+76 m^4 s t^3-15 m^2 s^3 t^2
 \right. \nonumber \\
 & & \left.
    -42 m^2 s^2 t^3-14 m^2 s t^4+3 s^3 t^3+6 s^2 t^4+2 s t^5\right).
\eq
Eq.~(\ref{eq_first_derivatives}) yields
\bq
\label{eliminate_derivative_s_t_121}
 \psi^{\curvethree}_1
 & = &
 \frac{s \left(s-4m^2\right)\left[ 3s \left(t-m^2\right) + 2 \left(t-m^2\right)^2 +16m^4\right]}{\left(t+3m^2\right)\left[s\left(t-m^2\right)+8m^4\right]}
 \frac{d}{ds} \psi^{\curvethree}_1 
 \nonumber \\
 & &
 -
 \frac{\left(s-4m^2\right)\left(t-m^2\right) \left(t^2-6m^2t-3m^4\right)}{\left(t+3m^2\right)\left[s\left(t-m^2\right)+8m^4\right]}
 \frac{d}{dt} \psi^{\curvethree}_1.
\eq
In $(x,y)$-space this translates to
\bq
\label{eliminate_derivative_x_y_121}
 \psi^{\curvethree}_1
 & = &
 - \frac{\left(x+1\right)\left(x-1\right)\left(3x^2y -2 xy^2 -3x^2 -2 xy -12x+3y-3\right)}{\left(y+3\right)\left(x^2y-x^2-2xy-6x+y-1\right)}
 \frac{d}{dx} \psi^{\curvethree}_1 
 \nonumber \\
 & &
 -
 \frac{\left(x+1\right)^2\left(y-1\right)\left(y^2-6y-3\right)}{\left(y+3\right)\left(x^2y-x^2-2xy-6x+y-1\right)}
 \frac{d}{dy} \psi^{\curvethree}_1.
\eq
We have with
\bq
 \chi^{\curvethree} & = & \sqrt{s m^2 \left( st + \left(m^2-t\right)^2 \right) }
\eq
the relation
\bq
\label{modular_lambda_lambdabar_121}
\lefteqn{
 16 \frac{\eta\left(\frac{\tau^{\curvethree}}{2}\right)^{24} \eta\left(2\tau^{\curvethree}\right)^{24}}{\eta\left(\tau^{\curvethree}\right)^{48}}
 \; = \;
 \left( k^{\curvethree} \bar{k}^{\curvethree} \right)^2
} & &
 \\
 & = &
 16 \frac{m^2 \left(4m^2-s\right) \chi^{\curvethree} \left( s m^2 + s t + 2 \chi^{\curvethree} \right) \left(3 s m^2 - s t - 16 m^4 + 2 \chi^{\curvethree} \right)}
{\left(s m^2 + s t - 2 \chi^{\curvethree} \right)^2 \left(3 s m^2 - s t - 16 m^4 - 2 \chi^{\curvethree} \right)^2}.
 \nonumber
\eq
For a path $\gamma_\alpha$ in $(x,y)$-space
\bq
 x \; = \; \alpha_1 \lambda,
 & &
 y \; = \; 1 + \alpha_2 \lambda
\eq
we may use eq.~(\ref{modular_lambda_lambdabar_121}) to express $\lambda$ as a power series in $q^{\curvethree}$ and vice versa.
The point $(x,y)=(0,1)$ corresponds to $\tau^{\curvethree}=i\infty$.

For $y=1$ we have
\bq
 \left. \psi_1^{\curvethree} \right|_{y=1}
 \; = \;
 \frac{\pi}{2} \frac{\left(1+x\right)}{\left(1-x\right)},
 & &
 \left. \frac{d}{dy} \psi_1^{\curvethree} \right|_{y=1}
 \; = \;
 - \frac{\pi}{8} \frac{\left(1+x\right)}{\left(1-x\right)}.
\eq

\subsection{Modular forms}

The integrals which only depend on $t$, but not on $s$, are all related to the elliptic curve $E^{\curveone}$.
Associated to the curve $E^{\curveone}$ are modular forms of $\Gamma_1(6)$.
The differential one-forms relevant to the integrals dependent on $t$ but not on $s$ are of the form
\bq
 f \; \left(2\pi i\right) d\tau_6^{\curveone},
\eq
where 
\bq
 \tau_6^{\curveone} & = &
 \frac{1}{6} \frac{\psi_2^{\curveone}}{\psi_1^{\curveone}},
\eq
which we substitute for $y$ (or $t$).
Furthermore, $f$ is a modular form of $\Gamma_1(6)$ from the set
\bq
\label{set_modular_forms_1}
 \left\{ 1, f_2, f_3, f_4, g_{2,1} \right\}.
\eq
The modular weights are given by $0$, $2$, $3$, $4$ and $2$, respectively.
The non-trivial modular forms are given by
\bq
\label{def_modular_forms_1}
 f_2 
 & = &
 - \frac{1}{4} \left( 3 y^2 - 10 y - 9 \right) \left( \frac{\psi_1^{\curveone}}{\pi} \right)^2,
 \nonumber \\
 f_3
 & = &
 -\frac{3}{2} y \left(y-1\right) \left(y-9\right) \left( \frac{\psi_1^{\curveone}}{\pi} \right)^3,
 \nonumber \\
 f_4
 & = &
 \frac{1}{16} \left(y+3\right)^4 \left( \frac{\psi_1^{\curveone}}{\pi} \right)^4,
 \nonumber \\
 g_{2,1} 
 & = &
 - \frac{1}{2} y \left( y - 9 \right) \left( \frac{\psi_1^{\curveone}}{\pi} \right)^2.
\eq
In appendix~\ref{sect:modular_forms} we collected useful information on the occurring modular forms of $\Gamma_1(6)$ 
and give alternative representations.

\subsection{The high-energy limit}

In the high-energy limit $s\rightarrow \infty$ (or equivalently $x=0$)
the elliptic curves $E^{\curvetwo}$ and $E^{\curvethree}$ degenerate to the elliptic curve $E^{\curveone}$.
In this limit we therefore have only one elliptic curve $E^{\curveone}$.
In this limit we may express all master integrals in terms of iterated integrals of modular forms.
The resulting set of modular forms is slightly larger than eq.~(\ref{set_modular_forms_1}).
In order to present this set, we first set
\bq
 g_{n,r} & = & - \frac{1}{2} \frac{y\left(y-1\right)\left(y-9\right)}{y-r} \left( \frac{\psi^{\curveone}_1}{\pi} \right)^n,
 \nonumber \\
 h_{n,s} & = & - \frac{1}{2} y\left(y-1\right)^{1+s}\left(y-9\right) \left( \frac{\psi^{\curveone}_1}{\pi} \right)^n.
\eq
Relevant to the high-energy limit is the set
\bq
\label{modular_forms}
 \left\{ 1, g_{2,0}, g_{2,1}, g_{2,9}, g_{3,1}, h_{3,0}, g_{4,0}, g_{4,1}, g_{4,9}, h_{4,0}, h_{4,1} \right\}.
\eq
These are again modular forms of $\Gamma_1(6)$ in the variable $\tau_6^{\curveone}$.
Additional details are given in appendix~\ref{sect:modular_forms}.
All integration kernels reduce in the high-energy limit to ${\mathbb Q}$-linear combinations of elements of this set 
(times $\left(2\pi i\right) d\tau_6^{\curveone}$).


\section{Master integrals}
\label{sect:masters}

In this section we define the 44 master integrals in the basis $\vec{J}$.
They are labelled $J_1 - J_{30}$ and $J_{32} - J_{45}$.
An integral $J_{31}$ is missing due to the extra relation given in eq.~(\ref{extra_relation}).
In the following we set $\mu=m$ and $D=4-2\eps$.

The basis $\vec{J}$ is constructed as follows:
The master integrals, which only depend on $s$ do not pose any problems and 
can in principal be constructed with the algorithms of \cite{Lee:2014ioa,Prausa:2017ltv,Gituliar:2017vzm,Meyer:2016slj,Meyer:2017joq,Lee:2017oca}.
The master integrals, which only depend on $t$ are similar to the kite / sunrise system
and can be constructed along the lines of \cite{Adams:2017ejb,Adams:2018yfj}.
This leaves the master integrals, which depend on $s$ and $t$.
Here we first analyse the diagonal blocks.
Combining the information from the maximal cuts with the technique based on the factorisation properties 
of Picard-Fuchs operators \cite{Adams:2017tga} we first obtain an intermediate basis with the desired properties up to sub-topologies.
We then use a slightly modified version of the algorithm of Meyer \cite{Meyer:2016slj,Meyer:2017joq} for the non-diagonal blocks
and fix the sub-topologies.
\begin{alignat}{2}
 \mbox{Sector 9:} \;\;\;\; &
 J_{1}
 & = \;\; & 
 \eps^2 \; {\bf D}^- I_{1001000},
 \nonumber \\
 \mbox{Sector 14:} \;\;\;\; &
 J_{2}
 & = \;\; & 
 \eps^2 \frac{\left(1-x\right)\left(1+x\right)}{2x} \; {\bf D}^- I_{0111000},
 \nonumber \\
 \mbox{Sector 28:} \;\;\;\; &
 J_{3}
 & = \;\; & 
 \eps^2 \frac{\left(1-x^2\right)}{x} \left[ I_{0021200} + \frac{1}{2} I_{0022100} \right],
 \nonumber \\
 &
 J_{4}
 & = \;\; & 
 \eps^2 \frac{\left(1-x\right)^2}{x} I_{0022100},
 \nonumber \\
 \mbox{Sector 49:} \;\;\;\; &
 J_{5}
 & = \;\; & 
 - \frac{\left(1-x\right)^2}{2x} \eps^2 \; {\bf D}^- I_{1000110},
 \nonumber \\
 \mbox{Sector 73:} \;\;\;\; &
 J_{6}
 & = \;\; & 
 \eps^2 \frac{\pi}{\psi^{\curveone}_1} \; {\bf D}^- I_{1001001},
 \nonumber \\
 &
 J_{7}
 & = \;\; & 
 \frac{6}{\eps} \frac{\left(\psi^{\curveone}_1\right)^2}{2 \pi i W^{\curveone}_y} \frac{d}{dy} J_6 
 - \frac{1}{4} \left(3y^2-10y-9\right) \left( \frac{\psi^{\curveone}_1}{\pi} \right)^2 J_6,
 \nonumber \\
 \mbox{Sector 74:} \;\;\;\; &
 J_{8}
 & = \;\; & 
 4 \eps^2 \; {\bf D}^- I_{0101001},
 \nonumber \\
 \mbox{Sector 15:} \;\;\;\; &
 J_{9}
 & = \;\; & 
 - \eps^3 \left(1-\eps\right) \frac{\left(1-x\right)^2}{x} I_{1111000},
 \nonumber \\
 \mbox{Sector 29:} \;\;\;\; &
 J_{10}
 & = \;\; & 
 - \eps^3 \frac{\left(1-x\right)^2}{x} I_{1012100},
 \nonumber \\
 \mbox{Sector 54:} \;\;\;\; &
 J_{11}
 & = \;\; & 
 - \eps^2 \frac{\left(1+x\right)\left(1-x\right)^3}{4 x^2} \; {\bf D}^- I_{0110110},
 \nonumber \\
 \mbox{Sector 57:} \;\;\;\; &
 J_{12}
 & = \;\; & 
 \eps^2 \frac{\left(1-x'^2\right)}{x'} \left[ -\frac{\left(1+x'\right)^2}{x'} I_{2001210} - \frac{\eps}{2\left(1+2\eps\right)} I_{2002000} \right],
 \nonumber \\
 &
 J_{13}
 & = \;\; & 
 -\eps^3 \frac{\left(1+x'\right)^2}{x'} I_{2001110},
 \nonumber \\
 \mbox{Sector 75:} \;\;\;\; &
 J_{14}
 & = \;\; &
 \eps^3 \left(1-y\right) I_{1102001}, 
 \nonumber \\
 \mbox{Sector 78:} \;\;\;\; &
 J_{15}
 & = \;\; & 
 \eps^2 \frac{\left(1-x^2\right)^2}{x^2} I_{0212001}
 - \frac{3 \eps^2 \left(1-x\right)^2}{2 x} I_{0202001},
 \nonumber \\
 &
 J_{16}
 & = \;\; & 
 \eps^3 \frac{\left(1-x^2\right)}{x} I_{0112001},
 \nonumber \\
 \mbox{Sector 89:} \;\;\;\; &
 J_{17}
 & = \;\; & 
 \eps^3 \left(1-y\right) I_{2001101}, 
 \nonumber \\
 \mbox{Sector 92:} \;\;\;\; &
 J_{18}
 & = \;\; & 
 \eps^2 \frac{\left(1+x\right)^2}{x} I_{0021102}
 -
 \eps^2 \frac{\left(1+x\right)}{x} \left( I_{0021200} + \frac{1}{2} I_{0022100} \right),
 \nonumber \\
 &
 J_{19}
 & = \;\; & 
 \eps^3 \frac{\left(1-x^2\right)}{x} I_{0021101},
 \nonumber \\
 \mbox{Sector 113:} \;\;\;\; &
 J_{20}
 & = \;\; & 
 \eps^3 \left(1-\eps\right) \frac{\left(1-x^2\right)}{x} I_{1000111},
 \nonumber \\
 \mbox{Sector 55:} \;\;\;\; &
 J_{21}
 & = \;\; & 
 \eps^3 \left( 1-2\eps \right) \frac{\left(1-x\right)^2}{x} I_{1110110},
 \nonumber \\
 \mbox{Sector 59:} \;\;\;\; &
 J_{22}
 & = \;\; & 
 \eps^4 \frac{\left(1-x\right)^2}{x} I_{1101110},
 \nonumber \\
 \mbox{Sector 62:} \;\;\;\; &
 J_{23}
 & = \;\; & 
 \eps^3 \left(1-2\eps\right) \frac{\left(1-x\right)^2}{x} I_{0111110},
 \nonumber \\
 \mbox{Sector 79:} \;\;\;\; &
 J_{24}
 & = \;\; & 
 \eps^3 \frac{\left(1-x\right)^2}{x} \frac{\pi}{\psi^{\curvetwo}_1} I_{1112001},
 \nonumber \\
 &
 J_{25}
 & = \;\; & 
 \eps^3 \left(1-2\eps\right) \frac{\left(1-x\right)^2}{x} I_{1111001}
 - \frac{1}{3} \left(y-9\right) \frac{\psi^{\curvetwo}_1}{\pi} J_{24},
 \nonumber \\
 &
 J_{26}
 & = \;\; & 
 \frac{6}{\eps} 
 \frac{\left(\psi^{\curvetwo}_1\right)^2}{2 \pi i W^{\curvetwo}_y} \frac{d}{dy} J_{24}
 - \frac{1}{4} \left(3y^2-10y-9\right)\left( \frac{\psi^{\curvetwo}_1}{\pi} \right)^2 J_{24}
 \nonumber \\
 & & &
 - \frac{1}{24} \left(y^2-30y-27\right) \frac{\psi^{\curvetwo}_1}{\pi} \frac{\psi^{\curveone}_1}{\pi} J_{6},
 \nonumber \\
 \mbox{Sector 93:} \;\;\;\; &
 J_{27}
 & = \;\; & 
 \eps^3 \frac{\left(1-x\right)^2}{x} \frac{\pi}{\psi^{\curvetwo}_1} I_{1012101},
 \nonumber \\
 &
 J_{28}
 & = \;\; & 
 \eps^3 \left[ 1-y + \frac{(1-x)^2}{x} \right] \left( I_{1021101} + I_{2011101} \right)
 - \frac{1}{6} \left(y-3\right) \frac{\psi^{\curvetwo}_1}{\pi} J_{27},
 \nonumber \\
 &
 J_{29}
 & = \;\; & 
 \eps^4 \left[ 1-y + \frac{(1-x)^2}{x} \right] I_{1011101},
 \nonumber \\
 &
 J_{30}
 & = \;\; & 
 \frac{6}{\eps} 
 \frac{\left(\psi^{\curvetwo}_1\right)^2}{2 \pi i W^{\curvetwo}_y} \frac{d}{dy} J_{27}
 - \frac{1}{4} \left( 3y^2-10y-9\right) \left( \frac{\psi^{\curvetwo}_1}{\pi} \right)^2 J_{27}
 \nonumber \\
 & & &
 - \frac{1}{12} \left(y^2-30y-27\right) \frac{\psi^{\curvetwo}_1}{\pi} \frac{\psi^{\curveone}_1}{\pi} J_{6},
 \nonumber \\
 \mbox{Sector 118:} \;\;\;\; &
 J_{32}
 & = \;\; & 
 \eps^3 \left(1-2\eps\right) \frac{\left(1-x\right)^2}{x} I_{0110111}
 + 2 \eps^3 \left(1-\eps\right) \frac{\left(1-x\right)}{x} I_{1000111},
 \nonumber \\
 \mbox{Sector 121:} \;\;\;\; &
 J_{33}
 & = \;\; & 
 \eps^3 \frac{\left(1-x^2\right)}{x} \frac{\pi}{\psi^{\curvethree}_1} I_{2001111},
 \nonumber \\
 &
 J_{34}
 & = \;\; & 
 \eps^3 \left(1-2\eps\right) \frac{\left(1-x^2\right)}{x} I_{1001111}
 - \frac{1}{3} \left(y-9\right) \frac{\left(1+2x\right)}{\left(1+x\right)} \frac{\psi^{\curvethree}_1}{\pi} J_{33},
 \nonumber \\
 &
 J_{35}
 & = \;\; & 
 \frac{6}{\eps} 
 \frac{\left(\psi^{\curvethree}_1\right)^2}{2 \pi i W^{\curvethree}_y} \frac{d}{dy} J_{33}
 - \frac{1}{4} \left( 3y^2 -10y-9\right) \frac{\left(1-x\right)^2}{\left(1+x\right)^2} \left( \frac{\psi^{\curvethree}_1}{\pi} \right)^2 J_{33}
 \nonumber \\
 & & &
 + \frac{1}{8} \left(y^2-2y+9\right) \frac{\left(1-x\right)}{\left(1+x\right)} \frac{\psi^{\curvethree}_1}{\pi} \frac{\psi^{\curveone}_1}{\pi} J_{6},
 \nonumber \\
 \mbox{Sector 63:} \;\;\;\; &
 J_{36}
 & = \;\; & 
 \eps^4 \frac{\left(x+1\right) \left(x-1\right)^3}{x^2} I_{1111110},
 \nonumber \\
 \mbox{Sector 119:} \;\;\;\; &
 J_{37}
 & = \;\; & 
 \eps^4 \frac{\left(x+1\right)\left(x-1\right)^3}{x^2} I_{1110111},
 \nonumber \\
 \mbox{Sector 123:} \;\;\;\; &
 J_{38}
 & = \;\; & 
 2 \eps^4 \frac{\left(x^2-1\right)}{x} \left[ I_{11011110(-1)} - \left(y-2\right) I_{1101111} \right]
 - \frac{4}{x-1} J_{22},
 \nonumber \\
 &
 J_{39}
 & = \;\; & 
 \eps^4 \left(1-y\right) \frac{\left(1-x\right)^2}{x} I_{1101111},
 \nonumber \\
 \mbox{Sector 126:} \;\;\;\; &
 J_{40}
 & = \;\; & 
 \eps^4 \frac{\left(x+1\right)\left(x-1\right)^3}{x^2} I_{0111111},
 \nonumber \\
 \mbox{Sector 127:} \;\;\;\; &
 J_{41}
 & = \;\; & 
 \eps^4 \frac{\left(1-x\right)^4}{x^2} \frac{\pi}{\psi^{\curvetwo}_1} I_{1111111},
 \nonumber \\
 &
 J_{42}
 & = \;\; & 
 8 \eps^4 \frac{\left(1-x\right)^2}{x} I_{1111111(-1)(-1)}
 - 8 \eps^4 \frac{\left(y-2\right)\left(1-x\right)^2}{x} I_{1111111(-1)0}
 \nonumber \\
 & & &
 - 8 \eps^4 \frac{y \left(1-x\right)^2}{x} I_{11111110(-1)}
 - \frac{8 x}{\left(1-x\right)^2} \frac{\psi^{\curvetwo}_1}{\pi} J_{41}
 \nonumber \\
 & & &
 -4\,{\frac { \left( x-1 \right)  \left( {x}^{2}-2\,xy+1 \right) }{ \left( x+1 \right) ^{3}}} J_{40}
 -4\,{\frac {{x}^{2}-2\,xy+1}{ \left( x-1 \right)  \left( x+1 \right) }} J_{37}
 \nonumber \\
 & & &
 -4\,{\frac {{x}^{2}-2\,xy-4\,x+1}{ \left( x-1 \right)  \left( x+1 \right) }} J_{36}
 - \frac{4}{3} \left(y+3\right) \frac{\psi^{\curvetwo}_1}{\pi} J_{27}
 + \frac{8}{3} \left(y+3\right) \frac{\psi^{\curvetwo}_1}{\pi} J_{24}
 \nonumber \\
 & & &
 - \left( 4 + \frac{32 \eps}{\left(1-2\eps\right)} {\frac {\left({x}^{4}-y{x}^{3}-xy+1\right)}{ \left( x-1 \right) ^{2} \left( x+1 \right) ^{2}}} \right) J_{23}
 -16\,{\frac { \left( y-1 \right) x}{ \left( x-1 \right) ^{2}}} J_{22}
 \nonumber \\
 & & &
 -8\,{\frac { \left( x-1 \right)  \left( {x}^{2}-xy+x+1 \right) }{ \left( x+1 \right) ^{3}}} J_{19}
 + 4\,{\frac { \left( x-1 \right) ^{2}}{ \left( y-1 \right) x}} J_{17}
 \nonumber \\
 & & &
 + 16\,{\frac { \left( x-1 \right)  \left( {x}^{2}-xy+x+1 \right) }{ \left( x+1 \right) ^{3}}} J_{16}
 -\frac{4}{3}\,{\frac {\left(6\,{x}^{2}-5\,xy-7\,x+6\right)}{ \left( y-1 \right) x}} J_{14},
 \nonumber \\
 &
 J_{43}
 & = \;\; & 
 \frac{6}{\eps} 
 \frac{\left(\psi^{\curvetwo}_1\right)^2}{2 \pi i W^{\curvetwo}_y} \frac{d}{dy} J_{41}
 - \frac{1}{4} \left( 3y^2-10y-9\right) \left( \frac{\psi^{\curvetwo}_1}{\pi} \right)^2 J_{41}
 + 4 y \frac{\left(1-x\right)}{\left(1+x\right)} \frac{\psi^{\curvetwo}_1}{\pi} \frac{\psi^{\curvethree}_1}{\pi} J_{33}
 \nonumber \\
 & & &
 + \frac{2}{3} y \left(y-9\right) \left( \frac{\psi^{\curvetwo}_1}{\pi} \right)^2 J_{27}
 + \frac{2}{3} y \left(y-3\right) \left( \frac{\psi^{\curvetwo}_1}{\pi} \right)^2 J_{24},
 \nonumber \\
 &
 J_{44}
 & = \;\; & 
 \eps^4 \frac{\left(1-x\right)^4}{x^2} I_{1111111(-1)0}
 - \frac{1}{3} \left( 2 y -3 \right) \frac{\psi^{\curvetwo}_1}{\pi} J_{41}
 - \eps^4 \frac{\left(1-x\right)^4}{x^2} I_{0111111}
 \nonumber \\
 & & &
 + \eps^4 \frac{\left(1-x\right)^2 \left(x^2-2xy+1\right)}{x^2} I_{1101111},
 \nonumber \\
 &
 J_{45}
 & = \;\; & 
 \eps^4 \frac{\left(x-1\right)^2\left(x+1\right)^2}{x^2} I_{11111110(-1)}
 - \frac{\left(2\,{x}^{2}y-9\,{x}^{2}+8\,xy-6\,x+2\,y-9\right)}{3 \left(x-1\right)^2} \frac{\psi^{\curvetwo}_1}{\pi} J_{41}
 \nonumber \\
 & & &
 - \frac{2}{x-1} J_{36}
 + \frac{1}{2} \left( \frac{1}{x} + x - \frac{2}{3} y \right) \frac{\psi^{\curvetwo}_1}{\pi} J_{27}
 - \left( \frac{1}{x} + x - \frac{2}{3} y \right) \frac{\psi^{\curvetwo}_1}{\pi} J_{24}.
\end{alignat}


\section{The system of differential equations}
\label{sect:differential_equation}

In the basis $\vec{J}$ the system of differential equations is linear in $\eps$, i.e. of the form
\bq
 d \vec{J}
 & = &
 \left( A^{(0)} + \eps A^{(1)} \right) \vec{J},
\eq
The matrices $A^{(0)}$ and $A^{(1)}$ are independent of $\eps$.
Furthermore, $A^{(0)}$ is strictly lower-triangular,
i.e.
\bq
 A^{(0)}_{ij} & = & 0 \;\;\;\;\;\; \mbox{for} \;\;\;j \ge i.
\eq
As usual, $A^{(1)}$ is block-triangular.

The system of differential equations simplifies for $t=m^2$ (corresponding to $y=1$)
as well as for $s=\infty$ (corresponding to $x=0$).
In both limits the matrix $A^{(0)}$ vanishes.
In the former case ($t=m^2$) the integration kernels are linear combinations of the one-forms given in eq.~(\ref{def_omega})
and the solution for the master integrals can be expressed in terms of multiple polylogarithms.
In the latter case $(s=\infty$) the integration kernels are of the form
\bq
 f \; \left(2\pi i\right) d\tau_6^{\curveone},
\eq
where $f$ is a modular form of the congruence subgroup $\Gamma_1(6)$ from the set given in eq.~(\ref{modular_forms}).
In this case the solution for the master integrals can be expressed in terms of iterated integrals of modular forms.
For all modular forms from the set given in eq.~(\ref{modular_forms}) the modular weight 
can be inferred from the scaling behaviour under a rescaling of the periods.
One has
\bq
 \mbox{modular weight}
 & = &
 \mbox{scaling power} + 2,
\eq
where the additional $2$ is due to the Jacobian obtained by replacing $dy$ by $d\tau_6^{\curveone}$.

We may view the entries of
\bq
 A & = & A^{(0)} + \eps A^{(1)}
\eq
as differential one-forms rational in 
\bq
 \eps, \tilde{x}, y, \psi_1^{\curveone}, \psi_1^{\curvetwo}, \psi_1^{\curvethree},
 \partial_y \psi_1^{\curveone}, \partial_y \psi_1^{\curvetwo}, \partial_y \psi_1^{\curvethree}.
\eq
We observe that each entry of $A$ is homogeneous under a simultaneous rescaling of all periods
and their derivatives
\bq
 \psi_1^{(r)} \rightarrow \lambda \; \psi_1^{(r)},
 & &
 \partial_y \psi_1^{(r)} \rightarrow \lambda \; \partial_y \psi_1^{(r)},
 \;\;\;\;\;\;\;\;\;\;\;\;
 r \in \{ a,b,c \}.
\eq
This allows us to group the entries of $A$ according to the scaling behaviour
under a simultaneous rescaling of all periods and their derivatives.
We define a $m$-weight as
\bq
 \mbox{$m$-weight}
 & = &
 \mbox{scaling power} + 2.
\eq
This is an ad hoc definition, which we find useful for bookkeeping.
No further properties are implied.
In the limit $x=0$ the $m$-weight agrees with the modular weight.

Let us note that the requirements that 
\begin{description}
\item{-} $A$ is linear in $\eps$,
\item{-} $A^{(0)}$ is strictly lower-triangular,
\item{-} $A^{(0)}$ vanishes for $x=0$ or $y=1$,
\item{-} $A^{(1)}$ reduces to integration kernels for multiple polylogarithms for $y=1$,
\item{-} $A^{(1)}$ reduces to modular forms for $x=0$, 
\end{description}
do not fix uniquely the set of master integrals $\vec{J}$ and the matrix $A$.
There is still a ``gauge freedom'' of transformations left, leaving these conditions intact.


\subsection{Integration kernels}
\label{sect:integration_kernels}

Let us now discuss the integration kernels appearing in the matrix $A$.
The entries of the matrix $A$ can be written as linear combinations (with rational coefficients) of fewer 
basic building blocks, such that no further linear relations with rational coefficients exist among these building blocks.
We call these building blocks ${\mathbb Q}$-independent integration kernels.
Let us explain this concept with an example. Let us restrict to the subset of Feynman integrals which only depend on $s$.
For this subset of Feynman integrals, all entries of $A$ are linear combinations of 
\bq
 \frac{d\tilde{x}}{\tilde{x}-c},
\eq
with
\bq
 c & \in &
 \tilde{{\mathcal A}} \;\; = \;\; \left\{ -1, 0, 1, i, -i, 1+\sqrt{2}, 1-\sqrt{2}, -1+\sqrt{2}, -1-\sqrt{2} \right\}.
\eq
The alphabet $\tilde{{\mathcal A}}$ has nine letters.
However, in the matrix $A$ only specific linear combinations of these one-forms appear.
All entries of $A$ can be expressed as ${\mathbb Q}$-linear combinations of
\bq
\label{example_Q_independent_set}
 \left\{ \omega_0, \omega_4, \omega_{-4}, \omega_{0,4}, \omega_{-4,0} \right\},
\eq
where the $\omega$'s have been defined in eq.~(\ref{def_omega}).
Thus the set of ${\mathbb Q}$-independent integration kernels contains for this example
only five elements, given by eq.~(\ref{example_Q_independent_set}).
It is clear that the results in terms of iterated integrals are shorter, if we work with a 
${\mathbb Q}$-independent set of integration kernels.

Let us now return to the general case.
For our choice of basis $\vec{J}$ we find 107 ${\mathbb Q}$-independent integration kernels.
It is a matter of personal taste, if one considers this number to be large or small.
On the one hand, it can be considered a large number as it limits our possibilities to present explicit results
on paper: For iterated integrals of depth $4$ we face a priori $107^4$ combinations.
On the other hand, it can be considered a small number, given the fact that we are dealing with a matrix of size $44 \times 44$.
The matrix $A$ can be considered to be sparse.

Let us mention explicitly that the number $107$ is the number for our choice of master integrals $\vec{J}$.
Other choices of master integrals (respecting the criteria given at the beginning of this section) may lead to
a different number of ${\mathbb Q}$-independent integration kernels.

We group the integration kernels according to their $m$-weight.
We have integration kernels with $m$-weight $0$, $1$, $2$, $3$ and $4$.
The complexity of the expressions increases with the $m$-weight.
The $\eps^0$-part $A^{(0)}$ contains only integration kernels of $m$-weight $3$ and $4$.

Our naming scheme is as follows: We keep the notation for the integration kernels, which already occurred
in the special cases $t=m^2$ or $s=\infty$.
This concerns
\bq
 \left\{ \omega_0, \omega_4, \omega_{-4}, \omega_{0,4}, \omega_{-4,0}, f_2, f_3, f_4, g_{2,1} \right\}.
\eq
Integration kernels appearing in the $\eps^0$-part $A^{(0)}$ are denoted by
\bq
 a^{(r)}_{n,j},
\eq
where $n$ gives the $m$-weight, $(r)$ indicates the periods appearing in the integration kernel and $j$ indexes
different integration kernels with the same $n$ and $(r)$.
We denote integration kernels appearing in the $\eps^1$-part $A^{(1)}$ generically by
\bq
 \eta^{(r)}_{n,j}.
\eq
The superscript $(r)$ and the second subscript $j$ are optional.
The interpretation of the super- and subscripts is as above.
For dlog-forms we use the notation
\bq
 d_{2,j}.
\eq
These are necessarily of $m$-weight $2$.

Let us now discuss for all $m$-weights typical examples, the cases of $m$-weight $0$ and $1$ are discussed completely.
The full list of integration kernels is given in the supplementary electronic file attached to this article.
The full list consists of the integration kernels
\bq
 & &
 \left\{ 
  \omega_0, \omega_4, \omega_{-4}, \omega_{0,4}, \omega_{-4,0}, 
  f_2, f_3, f_4, g_{2,1},
  \eta_0^{(r)}, 
  \eta^{(b)}_{1,1-4}, \eta^{(c)}_{1,1-3},
  d_{2,1-5},
  \eta_{2,1-12},
  \eta^{(\frac{r}{s})}_2,
 \right. \nonumber \\
 & & \left.
  a^{(b)}_{3,1-4}, a^{(c)}_{3,1-3},
  \eta^{(a)}_{3,1-3}, \eta^{(b)}_{3,1-24}, \eta^{(c)}_{3,1-11}, 
  a^{(a,b)}_{4,1}, a^{(a,c)}_{4,1}, a^{(b,b)}_{4,1-5}, a^{(c,c)}_{4,1}, a^{(b,c)}_{4,1},
 \right. \nonumber \\
 & & \left.
  \eta^{(a,b)}_{4,1-3}, \eta^{(a,c)}_{4,1}, \eta^{(b,b)}_{4,1-5}, \eta^{(c,c)}_{4,1}, \eta^{(b,c)}_{4,1}
 \right\}.
\eq
with $r,s \in \{ a,b,c \}$ and $r \neq s$.

\subsubsection{$m$-weight 0}

At $m$-weight $0$ we have three integration kernels. They are given by
\bq
 \eta_0^{\curveone} & = & 2 \pi i \; d\tau_6^{\curveone},
 \nonumber \\
 \eta_0^{\curvetwo} & = & 2 \pi i \; d\tau_6^{\curvetwo},
 \nonumber \\
 \eta_0^{\curvethree} & = & 2 \pi i \; d\tau_6^{\curvethree}.
\eq
These are exactly the integration kernels we expect at $m$-weight $0$.
We expressed them (compactly) in terms of the variables $\tau_6^{\curveone}$, $\tau_6^{\curvetwo}$ or $\tau_6^{\curvethree}$, respectively.
Of course we may re-write them in terms of the variables $x$ and $y$.
For example, for $\eta_0^{\curvetwo}$ one has 
\bq
 \eta_0^{\curvetwo} & = & 
 \frac{2}{3} \,
 {\frac {\left( x-1 \right)  \left( {x}^{2}y+3\,{x}^{2}-6\,xy-2\,x+y+3 \right) }
        { \left( {x}^{2}y-9\,{x}^{2}+2\,xy+14\,x+y-9 \right)  \left( xy-1 \right)  \left( x-y \right)  \left( x+1 \right) }} 
 \left( \frac{\pi}{\psi_1^{\curvetwo}}\right)^2 dx
 \nonumber \\
 & &
 -
 \frac{2}{3} \,
 {\frac { \left( x-1 \right) ^{2} \left( 3\,{x}^{2}-2\,xy-4\,x+3 \right)}
 { \left( y-1 \right)  \left( {x}^{2}y-9\,{x}^{2}+2\,xy+14\,x+y-9 \right)  \left( xy-1 \right)  \left( x-y \right) }}
 \left( \frac{\pi}{\psi_1^{\curvetwo}}\right)^2 dy.
\eq

\subsubsection{$m$-weight 1}

We find $7$ integration kernels of $m$-weight $1$, four of them are associated to the elliptic curve $E^{\curvetwo}$, three of them to the
elliptic curve $E^{\curvethree}$.
There are no integration kernels of $m$-weight $1$ for the elliptic curve $E^{\curveone}$.
The integration kernels of $m$-weight associated to the elliptic curve $E^{\curvetwo}$ are
\bq
 \eta_{1,1}^{\curvetwo}
 & = &
 {\frac { \left( x-1 \right) }{ \left( 3\,{x}^{2}-2\,xy-4\,x+3 \right)  \left( x+1 \right) }}
 \frac{\pi}{\psi_1^{\curvetwo}} dx,
 \nonumber \\
 \eta_{1,2}^{\curvetwo}
 & = &
 {\frac { \left( x-1 \right)  \left( {x}^{2}{y}^{2}-9\,{x}^{2}y+6\,x{y}^{2}-2\,xy+{y}^{2}+12\,x-9\,y \right) }
        { \left( x+1 \right)  \left( {x}^{2}y-9\,{x}^{2}+2\,xy+14\,x+y-9 \right) \left( xy-1 \right)  \left( x-y \right) }}
 \frac{\pi}{\psi_1^{\curvetwo}} dx
 \nonumber \\
 & &
 -
 {\frac { x \left( x-1 \right) ^{2} \left( y-3 \right) }{ \left( {x}^{2}y-9\,{x}^{2}+2\,xy+14\,x+y-9 \right) \left( xy-1 \right)  \left( x-y \right) }}
 \frac{\pi}{\psi_1^{\curvetwo}} dy,
 \nonumber \\
 \eta_{1,3}^{\curvetwo}
 & = &
 {\frac { \left( {x}^{2}y-9\,{x}^{2}-6\,xy+22\,x+y-9 \right) }
        { \left( x+1 \right)  \left( {x}^{2}y-9\,{x}^{2}+2\,xy+14\,x+y-9 \right)  \left( x-1 \right) }}
 \frac{\pi}{\psi_1^{\curvetwo}} dx
 \nonumber \\
 & &
 +
 2\,{\frac {x}{ \left( {x}^{2}y-9\,{x}^{2}+2\,xy+14\,x+y-9\right) }}
 \frac{\pi}{\psi_1^{\curvetwo}} dy,
 \nonumber \\
 \eta_{1,4}^{\curvetwo}
 & = &
 {\frac {x \left( y-1 \right)  \left( -6\,xy+y+{x}^{2}y-2\,x+3+3\,{x}^{2} \right) }
 { \left( x+1 \right)  \left( x-1 \right)  \left( xy-1 \right)  \left( x-y \right)  \left( {x}^{2}y-9\,{x}^{2}+2\,xy+14\,x+y-9 \right) }}
 \frac{\pi}{\psi_1^{\curvetwo}} dx
 \nonumber \\
 & &
 -
 {\frac {x \left( 3\,{x}^{2}-2\,xy-4\,x+3 \right) }
        { \left( {x}^{2}y-9\,{x}^{2}+2\,xy+14\,x+y-9 \right)  \left( xy-1 \right) \left( x-y \right) }}
 \frac{\pi}{\psi_1^{\curvetwo}} dy.
\eq
Associated to the elliptic curve $E^{\curvethree}$ are 
\bq
\lefteqn{
 \eta_{1,1}^{\curvethree}
 = 
 {\frac { \left( x+1 \right)  \left( y+3 \right) }
        { \left( x-1 \right)  \left( 3\,{x}^{2}y-2\,x{y}^{2}-3\,{x}^{2}-2\,xy-12\,x+3\,y-3 \right) }}
 \frac{\pi}{\psi_1^{\curvethree}} dx,
 } & &
 \nonumber \\
\lefteqn{
 \eta_{1,2}^{\curvethree}
 = 
 \left( x+1 \right) 
 \frac{\pi}{\psi_1^{\curvethree}}
 } & & \nonumber \\
 & & 
 \left[
 {\frac { \left( {x}^{2}{y}^{3}+3\,{x}^{2}{y}^{2}-9\,x{y}^{3}-105\,{x}^{2}y+99\,x{y}^{2}+2\,{y}^{3}-27\,{x}^{2}+45\,xy-12\,{y}^{2}+57\,x-54\,y \right) }
        { \left( x-1 \right)  \left( {x}^{2}{y}^{2}-10\,{x}^{2}y-2\,x{y}^{2}+9\,{x}^{2}+20\,xy+{y}^{2}+46\,x-10\,y+9 \right)  \left( xy-1 \right)  \left( x-y \right) }}
 dx
 \right. \nonumber \\
 & & \left.
 +
 {\frac { x \left( 3\,{x}^{2}{y}^{2}-4\,x{y}^{3}-30\,{x}^{2}y+38\,x{y}^{2}-2\,{y}^{3}+27\,{x}^{2}-48\,xy+25\,{y}^{2}+78\,x-84\,y-3 \right) }
        { \left( y-1 \right)  \left( {x}^{2}{y}^{2}-10\,{x}^{2}y-2\,x{y}^{2}+9\,{x}^{2}+20\,xy+{y}^{2}+46\,x-10\,y+9 \right)  \left( x-y \right)  \left( xy-1 \right) }}
 dy
 \right],
 \nonumber \\
\lefteqn{
 \eta_{1,3}^{\curvethree}
 = 
 {\frac { \left( x+1 \right) ^{2} \left( y-3 \right) }
        { \left( x-1 \right)  \left( 3\,{x}^{2}y-2\,x{y}^{2}-3\,{x}^{2}-2\,xy-12\,x+3\,y-3 \right) \sqrt {{x}^{2}-6\,x+1} }}
 \frac{\pi}{\psi_1^{\curvethree}} dx.
}
\eq
Note that in the last expression the square root is rationalised by using the variable $\tilde{x}$.
An alternative form for $\eta_{1,3}^{\curvethree}$ is
\bq
\lefteqn{
 \eta_{1,3}^{\curvethree}
 = 
 - \frac{\pi}{\psi_1^{\curvethree}} d\tilde{x}
 }
 & & \\
 & &
 {\frac { \left( y-3 \right)  \left( {{\tilde{x}}}^{2}-2\,{\tilde{x}}-1 \right) ^{2} }
        { \left( {{\tilde{x}}}^{2}+1 \right)  \left( 3\,{{\tilde{x}}}^{4}y+2\,{{\tilde{x}}}^{3}{y}^{2}-3\,{{\tilde{x}}}^{4}-4\,{{\tilde{x}}}^{3}y+18\,{{\tilde{x}}}^{3}+6\,{{\tilde{x}}}^{2}y-2\,{\tilde{x}}\,{y}^{2}-6\,{{\tilde{x}}}^{2}+4\,{\tilde{x}}\,y-18\,{\tilde{x}}+3\,y-3 \right) }},
 \nonumber
\eq
which is manifestly rational in $\tilde{x}$, $y$ and $\psi_1^{\curvethree}$.

\subsubsection{$m$-weight 2}

The integration kernels of $m$-weight $2$ are numerous and we only list a few typical cases.
The integration kernels for the multiple polylogarithms
\bq
\label{weight_2_omega}
 \omega_0, \; \omega_4, \; \omega_{-4}, \; \omega_{0,4}, \; \omega_{-4,0},
\eq
defined in eq.~(\ref{def_omega}) belong to this class.
Furthermore, the modular forms of modular weight $2$ clearly belong to this class: 
\bq
\label{weight_2_modular}
 f_2 \; \left(2\pi i\right) d\tau_6^{\curveone}
 & = &
 \frac{dy}{y-1} + \frac{dy}{y-9} - \frac{dy}{2y},
 \nonumber \\
 g_{2,1} \; \left(2\pi i\right) d\tau_6^{\curveone}
 & = &
 \frac{dy}{y-1}.
\eq
The differential one-forms in eq.~(\ref{weight_2_omega}) and eq.~(\ref{weight_2_modular})
are all dlog-forms, depending either on $x$ (or alternatively on $\tilde{x}$) or $y$, but not both.
There are further dlog-forms, depending on both variables $x$ and $y$.
These are
\bq
 d_{2,1} & = & d \ln\left(x-y\right) + d \ln\left(xy-1\right),
 \nonumber \\
 d_{2,2} & = & d \ln\left(xy-1\right),
 \nonumber \\
 d_{2,3} & = & d \ln\left({x}^{2}-xy-x+1\right),
 \nonumber \\
 d_{2,4} & = & d \ln\left(3\,{x}^{2}-2\,xy-4\,x+3 \right),
 \nonumber \\
 d_{2,5} & = & d \ln\left({x}^{2}y-9\,{x}^{2}+2\,xy+14\,x+y-9\right).
\eq
There are six differential one-forms involving ratios of periods, one for each ratio.
For example
\bq
\lefteqn{
 \eta_2^{(\frac{b}{a})}
 =
 \frac{1}{2}\,{\frac { \left( x+1 \right) }{ x\, \left( x-1 \right) }}
 \frac{\psi_1^{\curvetwo}}{\psi_1^{\curveone}} dx
 } & &
 \\
 & &
 -\frac{1}{12} \,
 {\frac { \left( 3\,{x}^{2}{y}^{2}+2\,x{y}^{3}-90\,{x}^{2}y+52\,x{y}^{2}-81\,{x}^{2}+138\,xy+3\,{y}^{2}+144\,x-90\,y-81 \right) }
 { y \, \left( y-1 \right)  \left( y-9 \right)  \left( 3\,{x}^{2}-2\,xy-4\,x+3 \right) }}
 \frac{\psi_1^{\curvetwo}}{\psi_1^{\curveone}} dy.
 \nonumber
\eq
In addition there are $12$ differential one-forms of $m$-weight $2$, which do not belong to any class discussed up to now.
An example is given by
\bq
 \eta_{2,1}
 & = &
 {\frac { \left( x-1 \right) }{ \left( x+1 \right) \left( 3\,{x}^{2}-2\,xy-4\,x+3 \right)  }} dx.
\eq
For our choice of master integrals $\vec{J}$ we observe that in the integration kernels of $m$-weight $2$ polynomials in denominator
occur only as a single power, i.e. there are no higher poles in $m$-weight $2$.

\subsubsection{$m$-weight 3}

At $m$-weight $3$ we have first of all the modular form of weight $3$ from the sunrise sector
\bq
 f_3 \; \left(2\pi i\right) d\tau_6^{\curveone}
 & = &
 3 \frac{\psi_1^{\curveone}}{\pi} dy.
\eq
At $m$-weight $3$ we have integration kernels appearing in the $\eps^0$-part $A^{(0)}$, an example is given by
\bq
 a_{3,1}^{\curvetwo}
 & = &
 {\frac { \left( {x}^{2}y-3\,{x}^{2}+4\,xy+y-3 \right)  \left( y-1 \right)}
         { \left( x-1 \right)  \left( 3\,{x}^{2}-2\,xy-4\,x+3 \right) \, \left( x+1 \right) }}
  \frac{\psi_1^{\curvetwo}}{\pi} dx
 \nonumber \\
 & &
 +{\frac { \left( {x}^{2}{y}^{2}-9\,{x}^{2}y+6\,x{y}^{2}-2\,xy+{y}^{2}+12\,x-9\,y \right)  \left( y-1 \right) }
         { \left( x-1 \right) \left( 3\,{x}^{2}-2\,xy-4\,x+3 \right) \, \left( x+1 \right) }}
  \left( \frac{\partial_y \psi_1^{\curvetwo}}{\pi} \right) dx
 \nonumber \\
 & &
 -{\frac {x \left( y-1 \right) }
         {\left( 3\,{x}^{2}-2\,xy-4\,x+3 \right) }}
  \frac{\psi_1^{\curvetwo}}{\pi} dy
 -{\frac { \left( y-3 \right) x \left( y-1 \right) }
         {\left( 3\,{x}^{2}-2\,xy-4\,x+3 \right) }}
  \left( \frac{\partial_y \psi_1^{\curvetwo}}{\pi} \right) dy.
\eq
In addition, there are integration kernels of $m$-weight $3$ in the $\eps^1$-part $A^{(1)}$, an example is given by
\bq
 \eta_{3,1}^{\curvetwo}
 & = &
 4\,{\frac {\left( y-1 \right) }{\left( 3\,{x}^{2}-2\,xy-4\,x+3 \right) }}
  \frac{\psi_1^{\curvetwo}}{\pi} dx
 -
 {\frac {\left( x-1 \right)  \left( x+1 \right) }{\left( 3\,{x}^{2}-2\,xy-4\,x+3 \right) }}
  \frac{\psi_1^{\curvetwo}}{\pi} dy.
\eq

\subsubsection{$m$-weight 4}

At $m$-weight $4$ we have one modular form of weight $4$ from the sunrise sector
\bq
 f_4 \; \left(2\pi i\right) d\tau_6^{\curveone}
 & = &
 - \frac{\left(y+3\right)^4}{8 y \left(y-1\right) \left(y-9\right)} \left( \frac{\psi_1^{\curveone}}{\pi} \right)^2 dy.
\eq
In addition, we encounter integration kernels appearing in the $\eps^0$-part $A^{(0)}$.
An example is given by
\bq
 a_{4,3}^{(b,b)}
 & = &
 -\frac{2}{3}\,
 {\frac { \left( y-3 \right)  \left( y-1 \right)  N_{4,3,1}^{(b,b)} }{ \left( x-1 \right)  \left( x+1 \right)  \left( 3\,{x}^{2}-2\,xy-4\,x+3 \right) ^{2}}} 
 \left( \frac{\psi_1^{\curvetwo}}{\pi} \right)^2 dx 
 \nonumber \\
 & &
 +
 \frac{4}{3}\,
 {\frac { x \left( y-3 \right)  \left( y-1 \right) ^{2} N_{4,3,2}^{(b,b)} }{ \left( x-1 \right)  \left( x+1 \right)  \left( 3\,{x}^{2}-2\,xy-4\,x+3 \right) ^{2}}}
 \left( \frac{\psi_1^{\curvetwo}}{\pi} \right) \left( \frac{\partial_y \psi_1^{\curvetwo}}{\pi} \right) dx 
 \nonumber \\
 & &
 +\frac{2}{3} \,
 {\frac { x \, N_{4,3,3}^{(b,b)} }{ \left( 3\,{x}^{2}-2\,xy-4\,x+3 \right) ^{2}}} 
 \left( \frac{\psi_1^{\curvetwo}}{\pi} \right)^2 dy
 \nonumber \\
 & &
 +
 \frac{4}{3}\,
 {\frac { x \left( y-3 \right)  \left( y-1 \right)  N_{4,3,4}^{(b,b)} }{ \left( 3\,{x}^{2}-2\,xy-4\,x+3 \right) ^{2}}}
 \left( \frac{\psi_1^{\curvetwo}}{\pi} \right) \left( \frac{\partial_y \psi_1^{\curvetwo}}{\pi} \right) dy,
\eq
with
\bq
 N_{4,3,1}^{(b,b)}
 & = &
 3\,{x}^{4}y-2\,{x}^{3}{y}^{2}+9\,{x}^{4}+20\,{x}^{3}y-20\,{x}^{2}{y}^{2}-18\,{x}^{3}+2\,{x}^{2}y-2\,x{y}^{2}-6\,{x}^{2}+20\,xy-18\,x
 \nonumber \\
 & &
 +3\,y+9, 
 \nonumber \\
 N_{4,3,2}^{(b,b)}
 & = &
 {x}^{2}{y}^{2}-24\,{x}^{2}y+18\,x{y}^{2}-9\,{x}^{2}+16\,xy+{y}^{2}+30\,x-24\,y-9,
 \nonumber \\
 N_{4,3,3}^{(b,b)}
 & = &
 9\,{x}^{2}{y}^{2}-8\,x{y}^{3}-6\,{x}^{2}y+10\,x{y}^{2}-27\,{x}^{2}-20\,xy+9\,{y}^{2}+66\,x-6\,y-27,
 \nonumber \\
 N_{4,3,4}^{(b,b)}
 & = &
 3\,{x}^{2}y-4\,x{y}^{2}+9\,{x}^{2}-2\,xy-18\,x+3\,y+9.
\eq
Finally, there are integration kernels of $m$-weight $4$ appearing in the $\eps^1$-part $A^{(1)}$.
An example is given by
\bq
 \eta_{4,3}^{(b,b)}
 & = & 
 \frac{1}{9} \,
 \frac{1}{\left( xy-1 \right)  \left( x-y \right) \left( 3\,{x}^{2}-2\,xy-4\,x+3 \right) ^{2} \left( {x}^{2}y-9\,{x}^{2}+2\,xy+14\,x+y-9 \right) }
 \nonumber \\
 & &
 \left[
 {\frac { \left( y-1 \right) P_{4,3,1}^{(b,b)} }
        { \left( x-1 \right)  \left( x+1 \right) }} dx 
 -
 {\frac { \left( x-1 \right) ^{2}  P_{4,3,2}^{(b,b)} }{ \left( y-1 \right)  }} dy
 \right] \left( \frac{\psi_1^{\curvetwo}}{\pi} \right)^2, 
\eq
with
\bq
 P_{4,3,1}^{(b,b)}
 & = &
 27\,{x}^{8}{y}^{4}-75\,{x}^{7}{y}^{5}+48\,{x}^{6}{y}^{6}-243\,{x}^{8}{y}^{3}+909\,{x}^{7}{y}^{4}-946\,{x}^{6}{y}^{5}+288\,{x}^{5}{y}^{6}+2673\,{x}^{8}{y}^{2}
 \nonumber \\
 & &
 -7182\,{x}^{7}{y}^{3}+5914\,{x}^{6}{y}^{4}-1237\,{x}^{5}{y}^{5}-288\,{x}^{4}{y}^{6}-729\,{x}^{8}y-6966\,{x}^{7}{y}^{2}+17592\,{x}^{6}{y}^{3}
 \nonumber \\
 & &
 -11277\,{x}^{5}{y}^{4}+1828\,{x}^{4}{y}^{5}+288\,{x}^{3}{y}^{6}-3159\,{x}^{7}y+22392\,{x}^{6}{y}^{2}-36146\,{x}^{5}{y}^{3}
 \nonumber \\
 & &
 +15766\,{x}^{4}{y}^{4}-1237\,{x}^{3}{y}^{5}+48\,{x}^{2}{y}^{6}+729\,{x}^{7}+13770\,{x}^{6}y-41898\,{x}^{5}{y}^{2}+43510\,{x}^{4}{y}^{3}
 \nonumber \\
 & &
 -11277\,{x}^{3}{y}^{4}-946\,{x}^{2}{y}^{5}+1134\,{x}^{6}-25929\,{x}^{5}y+52590\,{x}^{4}{y}^{2}-36146\,{x}^{3}{y}^{3}
 \nonumber \\
 & &
 +5914\,{x}^{2}{y}^{4}-75\,x{y}^{5}-9369\,{x}^{5}+30942\,{x}^{4}y-41898\,{x}^{3}{y}^{2}+17592\,{x}^{2}{y}^{3}+909\,x{y}^{4}
 \nonumber \\
 & &
 +15012\,{x}^{4}-25929\,{x}^{3}y+22392\,{x}^{2}{y}^{2}-7182\,x{y}^{3}+27\,{y}^{4}-9369\,{x}^{3}+13770\,{x}^{2}y
 \nonumber \\
 & &
 -6966\,x{y}^{2}-243\,{y}^{3}+1134\,{x}^{2}-3159\,xy+2673\,{y}^{2}+729\,x-729\,y,
 \nonumber \\
 P_{4,3,2}^{(b,b)}
 & = &
 27\,{x}^{6}{y}^{4}-45\,{x}^{5}{y}^{5}-30\,{x}^{4}{y}^{6}+48\,{x}^{3}{y}^{7}+243\,{x}^{6}{y}^{3}-603\,{x}^{5}{y}^{4}+828\,{x}^{4}{y}^{5}-460\,{x}^{3}{y}^{6}
 \nonumber \\
 & &
 +729\,{x}^{6}{y}^{2}-2592\,{x}^{5}{y}^{3}+1899\,{x}^{4}{y}^{4}-126\,{x}^{3}{y}^{5}-30\,{x}^{2}{y}^{6}+729\,{x}^{6}y-4212\,{x}^{5}{y}^{2}
 \nonumber \\
 & &
 +8085\,{x}^{4}{y}^{3}-4686\,{x}^{3}{y}^{4}+828\,{x}^{2}{y}^{5}-2187\,{x}^{5}y+6741\,{x}^{4}{y}^{2}-8864\,{x}^{3}{y}^{3}+1899\,{x}^{2}{y}^{4}
 \nonumber \\
 & &
 -45\,x{y}^{5}-729\,{x}^{5}+7587\,{x}^{4}y-9708\,{x}^{3}{y}^{2}+8085\,{x}^{2}{y}^{3}-603\,x{y}^{4}+810\,{x}^{4}-9954\,{x}^{3}y
 \nonumber \\
 & &
 +6741\,{x}^{2}{y}^{2}-2592\,x{y}^{3}+27\,{y}^{4}-810\,{x}^{3}+7587\,{x}^{2}y-4212\,x{y}^{2}+243\,{y}^{3}+810\,{x}^{2}
 \nonumber \\
 & &
 -2187\,xy+729\,{y}^{2}-729\,x+729\,y.
\eq


\subsection{Singularities}
\label{sect:singularities}

As already mentioned, the integration kernels are rational in
\bq
 \eps, \tilde{x}, y, \psi_1^{\curveone}, \psi_1^{\curvetwo}, \psi_1^{\curvethree},
 \partial_y \psi_1^{\curveone}, \partial_y \psi_1^{\curvetwo}, \partial_y \psi_1^{\curvethree}.
\eq
In the next section we will choose as boundary point the point $(x,y)=(0,1)$ (or equivalently $(s,t)=(\infty,m^2)$).
This motivates the introduction of the variable 
\bq
 \tilde{y}& = & 1-y.
\eq
Our boundary point is then $(\tilde{x},\tilde{y})=(0,0)$.

Of particular interest are the polynomials in $\tilde{x}$ and $\tilde{y}$ appearing in the denominator of the integration kernels.
There aren't too many. Polynomials, which only depend on $\tilde{x}$ are (compare with eq.~(\ref{def_omega}))
\bq
 Q_1 & = & \tilde{x},
 \nonumber \\
 Q_2 & = & \tilde{x}-1,
 \nonumber \\
 Q_3 & = & \tilde{x}+1,
 \nonumber \\
 Q_4 & = & \tilde{x}^2+1,
 \nonumber \\
 Q_5 & = & \tilde{x}^2-2\tilde{x}-1,
 \nonumber \\
 Q_6 & = & \tilde{x}^2+2\tilde{x}-1.
\eq
Polynomials, which only depend on $\tilde{y}$ are
\bq
 Q_7 & = & \tilde{y},
 \nonumber \\
 Q_8 & = & \tilde{y}-1,
 \nonumber \\
 Q_9 & = & \tilde{y}+8.
\eq
Polynomials, which depend on $\tilde{x}$ and $\tilde{y}$ are
\bq
 Q_{10} & = & 
 {{\tilde{x}}}^{2}-{\tilde{x}}\,{\tilde{y}}-{\tilde{y}}+1,
 \nonumber \\
 Q_{11} & = &
 -{{\tilde{x}}}^{2}{\tilde{y}}+{{\tilde{x}}}^{2}+{\tilde{x}}\,{\tilde{y}}+1,
 \nonumber \\
 Q_{12} & = &
 {{\tilde{x}}}^{4}-{{\tilde{x}}}^{3}{\tilde{y}}+2\,{{\tilde{x}}}^{2}+{\tilde{x}}\,{\tilde{y}}+1,
 \nonumber \\
 Q_{13} & = &
 3\,{{\tilde{x}}}^{4}-2\,{{\tilde{x}}}^{3}{\tilde{y}}+6\,{{\tilde{x}}}^{2}+2\,{\tilde{x}}\,{\tilde{y}}+3,
 \nonumber \\
 Q_{14} & = &
 {{\tilde{x}}}^{4}{\tilde{y}}+8\,{{\tilde{x}}}^{4}-4\,{{\tilde{x}}}^{3}{\tilde{y}}+2\,{{\tilde{x}}}^{2}{\tilde{y}}+16\,{{\tilde{x}}}^{2}+4\,{\tilde{x}}\,{\tilde{y}}+{\tilde{y}}+8,
 \nonumber \\
 Q_{15} & = &
 3\,{{\tilde{x}}}^{4}{\tilde{y}}-2\,{{\tilde{x}}}^{3}{{\tilde{y}}}^{2}-16\,{{\tilde{x}}}^{3}+6\,{{\tilde{x}}}^{2}{\tilde{y}}+2\,{\tilde{x}}\,{{\tilde{y}}}^{2}+16\,{\tilde{x}}+3\,{\tilde{y}},
 \nonumber \\
 Q_{16} & = &
 {{\tilde{x}}}^{2}{\tilde{y}}-8\,{\tilde{x}}+{\tilde{y}}+8,
 \nonumber \\
 Q_{17} & = &
 {{\tilde{x}}}^{2}{\tilde{y}}+8\,{{\tilde{x}}}^{2}+8\,{\tilde{x}}+{\tilde{y}}.
\eq
It is helpful to have corresponding expressions in $(s,t)$-space: The polynomials $Q_{10}$ and $Q_{11}$ appear when the expression
$s t + (m^2-t)^2$ is expressed in the variables $\tilde{x}$ and $\tilde{y}$:
\bq
 s t + \left(m^2-t\right)^2
 & = &
 \frac{Q_{10} Q_{11}}{\tilde{x} \left(\tilde{x}-1\right) \left(\tilde{x}+1\right)}.
\eq
The polynomial $Q_{12}$ is related to
\bq
 m^2 -t -s
 & = &
 - \frac{Q_{12}}{\tilde{x} \left(\tilde{x}-1\right) \left(\tilde{x}+1\right)}.
\eq
The polynomials $Q_{13}$ and $Q_{14}$ are related to the elliptic curve $E^{\curvetwo}$.
We have
\bq
 3 s+2 t-2 m^2
 & = &
 \frac{Q_{13}}{\tilde{x} \left(\tilde{x}-1\right) \left(\tilde{x}+1\right)},
 \nonumber \\
 s \left(t-9m^2\right) + 4 m^2 \left(m^2-t\right)
 & = &
 - \frac{Q_{14}}{\tilde{x} \left(\tilde{x}-1\right) \left(\tilde{x}+1\right)},
\eq
The polynomial $Q_{13}$ enters through eq.~(\ref{eliminate_derivative_s_t_127}) or eq.~(\ref{eliminate_derivative_x_y_127}),
the polynomial $Q_{14}$ appears in the Picard-Fuchs operator for $\psi_1^{\curvetwo}$ in eq.~(\ref{def_Wronskian_127}).

The polynomials $Q_{15}$, $Q_{16}$ and $Q_{17}$ are related to the elliptic curve $E^{\curvethree}$.
We have
\bq
 3 s \left(t-m^2\right)+2 \left(t-m^2\right)^2+16 m^4
 & = &
 - \frac{Q_{15}}{\tilde{x} \left(\tilde{x}-1\right) \left(\tilde{x}+1\right)},
 \nonumber \\
 s \left(t-9m^2\right) \left(t-m^2\right) - 64 m^6
 & = &
 \frac{Q_{16}Q_{17}}{\tilde{x} \left(\tilde{x}-1\right) \left(\tilde{x}+1\right)}.
\eq
The polynomial $Q_{15}$ enters through eq.~(\ref{eliminate_derivative_s_t_121}) or eq.~(\ref{eliminate_derivative_x_y_121}),
the polynomials $Q_{16}$ and $Q_{17}$ appears in the Picard-Fuchs operator for $\psi_1^{\curvethree}$ in eq.~(\ref{def_Wronskian_121}).

Note that the polynomials $Q_1$, $Q_7$, $Q_{15}$ and $Q_{17}$ vanish for $(\tilde{x},\tilde{y})=(0,0)$.


\section{Boundary conditions and boundary constants}
\label{sect:boundary}

We integrate the system of differential equations starting from the point $(x,y)=(0,1)$ (corresponding to $s=\infty$ and $t=m^2$).
In order to do so, we need the boundary constants at this point.
The boundary constants may be expressed as a linear combination of transcendental constants.
A basis of these transcendental constants up to weight four is given by \cite{Blumlein:2009}
\bq
 w=1: & & \ln(2),
 \nonumber \\
 w=2: & & \zeta_2, \;\;\; \ln^2(2),
 \nonumber \\
 w=3: & & \zeta_3, \;\;\; \zeta_2 \ln(2), \;\;\; \ln^3(2),
 \nonumber \\
 w=4: & &
 \zeta_4,
 \;\;\;
 \mathrm{Li}_4\left(\frac{1}{2}\right),
 \;\;\;
 \zeta_3 \ln(2),
 \;\;\;
 \zeta_2 \ln^2(2),
 \;\;\;
 \ln^4(2).
\eq
The boundary constants for the master integrals, which are products of one-loop integrals are easily computed.
For example, the master integral $J_1$ is a product of two tadpole integrals.
The tadpole integral is given by
\bq
\label{def_tadpole}
 T_{\nu}\left( D, m^2, \mu^2 \right)
 & = &
 e^{\gamma_E \eps}
 \left(\mu^2\right)^{\nu-\frac{D}{2}}
 \int \frac{d^Dk}{i \pi^{\frac{D}{2}}}
 \frac{1}{\left(-k^2+m^2\right)^{\nu}}
 \;\; = \;\;
 e^{\gamma_E \eps}
 \frac{\Gamma\left(\nu-\frac{D}{2}\right)}{\Gamma\left(\nu\right)}
 \left( \frac{m^2}{\mu^2} \right)^{\frac{D}{2}-\nu}.
 \;\;\;\;
\eq
For $D=2-2\eps$, $\mu=m$  and $\nu=1$ we have
\bq
 T_{1}\left( 2-2\eps \right)
 & = &
 e^{\gamma_E \eps}
 \Gamma\left(\eps\right)
 \; = \; 
  \frac{1}{\eps} \left[ 1 + \frac{1}{2} \zeta_2 \eps^2 - \frac{1}{3} \zeta_3 \eps^3 + \frac{9}{16} \zeta_4 \eps^4
  + {\mathcal O}\left(\eps^5\right) \right].
\eq
In addition we need to calculate explicitly the boundary constants for the integrals, which neither 
depend on $s$ nor on $t$.
There are two such integrals: $J_1$ (which is also a product of tadpoles) and $J_8$ (the sunrise integral
at the pseudo-threshold).
For the latter we proceed as follows:
We start from the Feynman parameter integral.
We have
\bq
 J_8
 & = & 
 6 \eps^2 e^{2\gamma_E \eps} \Gamma\left(1+2\eps\right)
 \int\limits_0^1 dx_2 \int\limits_0^1 dx_4 
 \left[ \frac{1}{x_2-1} - \frac{1}{x_2+1} \right]
 \left[ \frac{1}{x_4+1} - \frac{1}{x_4+x_2} \right]
 \nonumber \\
 & & \times
 \left( x_2+1 \right)^{\eps} 
 \left( x_4+1 \right)^{-2\eps} 
 \left( x_4+x_2 \right)^{-2\eps} 
 \left( x_4+\frac{x_2}{x_2+1} \right)^{\eps}.
\eq
At each order in $\eps$, the $x_4$-integration is easily performed, resulting in multiple polylogarithms $G(z_1,...,z_k;x_2)$,
where the remaining variable $x_2$ appears in the argument list $z_1$, ..., $z_k$.
With the methods of \cite{Vollinga:2004sn,Broedel:2013tta} we convert all 
polylogarithms to a form, where
the parameters $z_1$, ..., $z_k$ do not depend on $x_2$, the simplest example is given by
\bq
 G\left(-x_2;1\right)
 & = &
 G\left(-1;x_2\right)
 -
 G\left(0;x_2\right).
\eq
We may then perform the integration over the variable $x_2$.
The resulting expressions may be simplified with the help of the PSLQ-algorithm \cite{Ferguson:1992}.

For all other master integrals we obtain the boundary constants 
from the behaviour at a specific point, where the master integral vanishes or reduces to simpler integrals.
The specific points which we consider are
$(x,y)=(0,1)$, $(x,y)=(1,1)$ and $(x,y)=(-1,1)$.
For the points $(x,y)=(1,1)$ or $(x,y)=(-1,1)$ we integrate the system along $y=1$ from $x=0$ to $x=\pm1$.
For $y=1$ we only obtain multiple polylogarithms. We evaluate the multiple polylogarithms
to high precision \cite{Vollinga:2004sn}
and use the PSLQ-algorithm to extract the transcendental constants.

The complete list of boundary constants is given in appendix~\ref{sect:boundary_constants}.
 

\section{Results}
\label{sect:results}

For all basis integrals we write
\bq
 J_k
 & = &
 \sum\limits_{j=0}^\infty \eps^j J_k^{(j)}.
\eq
Obviously, the basis of master integrals involves also rather simple sub-topologies, where results for these integrals
are known in the literature \cite{Davydychev:2000na,Davydychev:2003mv,Bytev:2009kb}.

\subsection{Integrals, which do not depend on $s$ nor $t$}
\label{sect:s_t_independent}

The integrals $J_1$ and $J_8$ are independent of $s$ and $t$ (or equivalently independent of $x$ and $y$).
The integrals $J_1$ and $J_8$ are given by
\bq
 J_{1} & = &
 1 + \zeta_2 \eps^2 - \frac{2}{3} \zeta_3 \eps^3 + \frac{7}{4} \zeta_4 \eps^4 
 + {\mathcal O}\left(\eps^5\right),
 \nonumber \\
 J_{8} & = &
 6 \zeta_2 \eps^2 + \eps^3 \left( 21 \zeta_3 - 36 \zeta_2 \ln 2 \right) 
 + \eps^4 \left( 144 \, \mathrm{Li}_4\left(\frac{1}{2}\right) - 78 \zeta_4 + 72 \zeta_2 \ln^2\left(2\right) + 6 \ln^4\left(2\right)
\right)
 \nonumber \\
 & &
 + {\mathcal O}\left(\eps^5\right).
\eq

\subsection{Integrals, which only depend on $s$}
\label{sect:s_dependent}

The integrals $J_2-J_5$, $J_9-J_{13}$, $J_{15}-J_{16}$, $J_{18}-J_{23}$, $J_{32}$, $J_{36}-J_{37}$ and $J_{40}$ depend
only on the variable $s$.
We may group them into three categories.
The first category contains most of these integrals.
More concretely, the integrals in the first category are given by 
$J_2-J_5$, $J_9-J_{11}$, $J_{15}-J_{16}$, $J_{18}-J_{21}$, $J_{23}$, $J_{32}$, $J_{37}$ and $J_{40}$.
These integrals are most naturally expressed in terms of harmonic polylogarithms in the variable $x$.

The integrals $J_{12}$ and $J_{13}$ belong to the second category. These integrals have a singular point at
$s=-4m^2$ but not at $s=4m^2$. These two integrals are most naturally expressed
in terms of harmonic polylogarithms in the variable $x'$.

The integrals $J_{22}$ and $J_{36}$ belong to the third category.
These integrals have a singular point at $s=-4m^2$ and at $s=4m^2$.
These two integrals are expressed in terms of multiple polylogarithms in the variable $\tilde{x}$.

\subsubsection{Integrals, which are expressed in the variable $x$}

The integrals $J_2-J_5$, $J_9-J_{11}$, $J_{15}-J_{16}$, $J_{18}-J_{21}$, $J_{23}$, $J_{32}$, $J_{37}$ and $J_{40}$ 
are most naturally expressed in terms of harmonic polylogarithms in the variable $x$.
\bq
 J_{2}^{(0)} 
 & = &
 0,
 \nonumber \\
 J_{2}^{(1)} 
 & = &
 -G\left(0;x\right),
 \nonumber \\
 J_{2}^{(2)} 
 & = &
    2 G\left(-1,0;x\right) - G\left(0,0;x\right) + \zeta_2,
 \nonumber \\
 J_{2}^{(3)} 
 & = &
    - 4 G\left(-1,-1,0;x\right) 
    + 2 G\left(-1,0,0;x\right) 
    + 2 G\left(0,-1,0;x\right) 
    - G\left(0,0,0;x\right)
    - 2 \zeta_2  G\left(-1;x\right)
    \nonumber \\
    & &  
    + 2 \zeta_3,
 \nonumber \\
 J_{2}^{(4)} 
 & = &
      8 G\left(-1,-1,-1,0;x\right)
    - 4 G\left(-1,-1,0,0;x\right)
    - 4 G\left(-1,0,-1,0;x\right)
    - 4 G\left(0,-1,-1,0;x\right)
    \nonumber \\
    & &
    + 2 G\left(-1,0,0,0;x\right)
    + 2 G\left(0,-1,0,0;x\right)
    + 2 G\left(0,0,-1,0;x\right)
    - G\left(0,0,0,0;x\right)
    \nonumber \\
    & &
    + 4 \zeta_2 G\left(-1,-1;x \right)
    - 2 \zeta_2 G\left(0,-1;x\right)
    - 4 \zeta_3 G\left(-1;x\right)
    + \frac{8}{3} \zeta_3 G\left(0;x\right)
    + \frac{19}{4} \zeta_4,
 \nonumber \\
 J_{3}^{(0)} 
 & = &
 0,
 \nonumber \\
 J_{3}^{(1)} 
 & = &
 G\left( 0; x \right),
 \nonumber \\
 J_{3}^{(2)} 
 & = &
 4\,G\left( 0, 0; x \right)
 -6\,G\left( -1, 0; x \right)
 -2\,G\left( 1, 0; x \right)
 -\zeta_2,
 \nonumber \\
 J_{3}^{(3)} 
 & = &
 10\,G\left( 0, 0, 0; x \right)
 -8\,G\left( 0, 1, 0; x \right)
 -24\,G\left( 0, -1, 0; x \right)
 -24\,G\left( -1, 0, 0; x \right)
 -8\,G\left( 1, 0, 0; x \right)
 \nonumber \\
 & &
 +12\,G\left( -1, 1, 0; x \right)
 +36\,G\left( -1, -1, 0; x \right)
 +4\,G\left( 1, 1, 0; x \right)
 +12\,G\left( 1, -1, 0; x \right)
 \nonumber \\
 & &
 +6\,\zeta_2\,G\left( -1; x \right)
 +2\,\zeta_2\,G\left( 1; x \right)
 -3\,\zeta_2\,G\left( 0; x \right)
 -11\,\zeta_3,
 \nonumber \\
 J_{3}^{(4)} 
 & = &
 -20\,G\left( 0, 0, 1, 0; x \right)
 +48\,G\left( 0, -1, 1, 0; x \right)
 +16\,G\left( 0, 1, 1, 0; x \right)
 +48\,G\left( 0, 1, -1, 0; x \right)
 \nonumber \\
 & &
 -44\,G\left( 0, 1, 0, 0; x \right)
 -24\,G\left( 1, 1, -1, 0; x \right)
 +16\,G\left( 1, 1, 0, 0; x \right)
 -8\,G\left( 1, 1, 1, 0; x \right)
 \nonumber \\
 & &
 +16\,G\left( 1, 0, 1, 0; x \right)
 +48\,G\left( 1, -1, 0, 0; x \right)
 -24\,G\left( 1, -1, 1, 0; x \right)
 -72\,G\left( 1, -1, -1, 0; x \right)
 \nonumber \\
 & &
 -20\,G\left( 1, 0, 0, 0; x \right)
 +48\,G\left( 1, 0, -1, 0; x \right)
 -72\,G\left( -1, 1, -1, 0; x \right)
 +48\,G\left( -1, 1, 0, 0; x \right)
 \nonumber \\
 & &
 -24\,G\left( -1, 1, 1, 0; x \right)
 -72\,G\left( -1, -1, 1, 0; x \right)
 +48\,G\left( -1, 0, 1, 0; x \right)
 -60\,G\left( -1, 0, 0, 0; x \right)
 \nonumber \\
 & &
 +144\,G\left( -1, 0, -1, 0; x \right)
 +144\,G\left( -1, -1, 0, 0; x \right)
 -216\,G\left( -1, -1, -1, 0; x \right)
 \nonumber \\
 & &
 +144\,G\left( 0, -1, -1, 0; x \right)
 +22\,G\left( 0, 0, 0, 0; x \right)
 -60\,G\left( 0, 0, -1, 0; x \right)
 -96\,G\left( 0, -1, 0, 0; x \right)
 \nonumber \\
 & &
 +18\,\zeta_2\,G\left( -1, 0; x \right)
 -12\,\zeta_2\,G\left( -1, 1; x \right)
 -12\,\zeta_2\,G\left( 1, -1; x \right)
 +22\,\zeta_3\,G\left( 1; x \right)
 \nonumber \\
 & &
 -4\,\zeta_2\,G\left( 1, 1; x \right)
 +6\,\zeta_2\,G\left( 1, 0; x \right)
 +8\,\zeta_2\,G\left( 0, 1; x \right)
 -6\,\zeta_2\,G\left( 0, 0; x \right)
 +66\,\zeta_3\,G\left( -1; x \right)
 \nonumber \\
 & &
 +24\,\zeta_2\,G\left( 0, -1; x \right)
 -36\,\zeta_2\,G\left( -1, -1; x \right)
 -{\frac {80}{3}}\,\zeta_3\,G\left( 0; x \right)
 -{\frac {29}{2}}\,\zeta_4,
 \nonumber \\
 J_{4}^{(0)} 
 & = &
 0,
 \nonumber \\
 J_{4}^{(1)} 
 & = &
 0,
 \nonumber \\
 J_{4}^{(2)} 
 & = &
 2\,G\left( 0, 0; x \right),
 \nonumber \\
 J_{4}^{(3)} 
 & = &
 6\,G\left( 0, 0, 0; x \right)
 -4\,G\left( 0, 1, 0; x \right)
 -12\,G\left( 0, -1, 0; x \right)
 -2\,\zeta_2\,G\left( 0; x \right)
 +4\,G\left( 1, 0, 0; x \right)
 -6\,\zeta_3,
 \nonumber \\
 J_{4}^{(4)} 
 & = &
 8\,G\left( 1, 1, 0, 0; x \right)
 +8\,G\left( 0, 1, 1, 0; x \right)
 +24\,G\left( 0, 1, -1, 0; x \right)
 +24\,G\left( 0, -1, 1, 0; x \right)
 \nonumber \\
 & &
 -20\,G\left( 0, 1, 0, 0; x \right)
 -12\,G\left( 0, 0, 1, 0; x \right)
 +12\,G\left( 1, 0, 0, 0; x \right)
 -24\,G\left( 1, 0, -1, 0; x \right)
 \nonumber \\
 & &
 -8\,G\left( 1, 0, 1, 0; x \right)
 +72\,G\left( 0, -1, -1, 0; x \right)
 +14\,G\left( 0, 0, 0, 0; x \right)
 -36\,G\left( 0, 0, -1, 0; x \right)
 \nonumber \\
 & &
 -48\,G\left( 0, -1, 0, 0; x \right)
 -4\,\zeta_2\,G\left( 1, 0; x \right)
 +4\,\zeta_2\,G\left( 0, 1; x \right)
 -4\,\zeta_2\,G\left( 0, 0; x \right)
 \nonumber \\
 & &
 +12\,\zeta_2\,G\left( 0, -1; x \right)
 -16\,\zeta_3\,G\left( 0; x \right)
 -12\,\zeta_3\,G\left( 1; x \right)
 -\frac{13}{2}\,\zeta_4,
 \nonumber \\
 J_{5}^{(0)} 
 & = &
    1,
 \nonumber \\
 J_{5}^{(1)} 
 & = &
    -2 G\left(1;x\right) + G\left(0;x\right),
 \nonumber \\
 J_{5}^{(2)} 
 & = &
    4 G\left(1,1;x\right) 
    - 2 G\left(1;x\right) G\left(0;x\right)
    + G\left(0,0;x\right),
 \nonumber \\
 J_{5}^{(3)} 
 & = &
    - 8 G\left(1,1,1;x\right)
    + 4 G\left(1,1;x\right) G\left(0;x\right)
    - 2 G\left(1;x\right) G\left(0,0;x\right)
    + G\left(0,0,0;x\right)
    - \frac{8}{3} \zeta_3,
 \nonumber \\
 J_{5}^{(4)} 
 & = &
    16 G\left(1,1,1,1;x\right)
    - 8 G\left(1,1,1;x\right) G\left(0;x\right)
    + 4 G\left(1,1;x\right) G\left(0,0;x\right)
    - 2 G\left(1;x\right) G\left(0,0,0;x\right)
    \nonumber \\
    & &
    + G\left(0,0,0,0;x\right)
    + \frac{16}{3} \zeta_3 G\left(1,x\right)
    - \frac{8}{3} \zeta_3 G\left(0;x\right)
    - 3 \zeta_4,
 \nonumber \\
 J_{9}^{(0)} 
 & = &
 0,
 \nonumber \\
 J_{9}^{(1)} 
 & = &
 0,
 \nonumber \\
 J_{9}^{(2)} 
 & = &
 G\left( 0, 0; x \right),
 \nonumber \\
 J_{9}^{(3)} 
 & = &
 G\left( 0, 0, 0; x \right)
 -2\,G\left( 0, -1, 0; x \right)
 -\zeta_2\,G\left( 0; x \right)
 -3\,\zeta_3,
 \nonumber \\
 J_{9}^{(4)} 
 & = &
 G\left( 0, 0, 0, 0; x \right)
 -2\,G\left( 0, 0, -1, 0; x \right)
 -2\,G\left( 0, -1, 0, 0; x \right)
 +4\,G\left( 0, -1, -1, 0; x \right)
 \nonumber \\
 & &
 +2\,\zeta_2\,G\left( 0, -1; x \right)
 -2\,\zeta_3\,G\left( 0; x \right)
 -\frac{5}{4}\,\zeta_4,
 \nonumber \\
 J_{10}^{(0)} 
 & = &
 0,
 \nonumber \\
 J_{10}^{(1)} 
 & = &
 0,
 \nonumber \\
 J_{10}^{(2)} 
 & = &
 0,
 \nonumber \\
 J_{10}^{(3)} 
 & = &
 2\,G\left( 0, 0, 0; x \right)-4\,G\left( 1, 0, 0; x \right)-4\,\zeta_3,
 \nonumber \\
 J_{10}^{(4)} 
 & = &
 4\,G\left( 0, 1, 0, 0; x \right)
 -4\,G\left( 0, 0, 1, 0; x \right)
 +6\,G\left( 0, 0, 0, 0; x \right)
 -12\,G\left( 0, 0, -1, 0; x \right)
 \nonumber \\
 & &
 -12\,G\left( 1, 0, 0, 0; x \right)
 +24\,G\left( 1, 0, -1, 0; x \right)
 +8\,G\left( 1, 0, 1, 0; x \right)
 -8\,G\left( 1, 1, 0, 0; x \right)
 \nonumber \\
 & &
 -2\,\zeta_2\,G\left( 0, 0; x \right)
 +4\,\zeta_2\,G\left( 1, 0; x \right)
 +12\,\zeta_3\,G\left( 1; x \right)
 -6\,\zeta_3\,G\left( 0; x \right)
 +4\,\zeta_4,
 \nonumber \\
 J_{11}^{(0)}
 & = &
    0,
 \nonumber \\
 J_{11}^{(1)}
 & = &
  - G\left(0;x\right),
 \nonumber \\
 J_{11}^{(2)}
 & = &
    2 G\left(-1, 0;x\right)
    -3 G\left(0, 0;x\right)
    +2 G\left(1;x\right) G\left(0;x\right)
    +\zeta_2,
 \nonumber \\
 J_{11}^{(3)}
 & = &
    6 G\left(-1, 0, 0;x\right)
    -4 G\left(-1, -1, 0;x\right)
    +4 G\left(0, -1, 0;x\right)
    -7 G\left(0, 0, 0;x\right)
    \nonumber \\
    & &
    +6 G\left(1;x\right) G\left(0, 0;x\right)
    -4 G\left(0;x\right) G\left(1, 1;x\right)
    -4 G\left(-1, 0;x\right) G\left(1;x\right)
    +2 \zeta_2 G\left(0;x\right)
    \nonumber \\
    & &
    -2 \zeta_2 G\left(-1;x\right)
    -2 \zeta_2 G\left(1;x\right)
    +2 \zeta_3,
 \nonumber \\
 J_{11}^{(4)}
 & = &
    12 G\left(0, -1, 0, 0;x\right)
    -15 G\left(0, 0, 0, 0;x\right)
    +8 G\left(0, 0, -1, 0;x\right)
    -8 G\left(0, -1, -1, 0;x\right)
    \nonumber \\
    & &
    +14 G\left(-1, 0, 0, 0;x\right)
    -8 G\left(-1, 0, -1, 0;x\right)
    -12 G\left(-1, -1, 0, 0;x\right)
    \nonumber \\
    & &
    +8 G\left(-1, -1, -1, 0 ;x\right)
    +8 G\left(0;x\right) G\left(1, 1, 1;x\right)
    -12 G\left(1, 1;x\right) G\left(0, 0;x\right)
    \nonumber \\
    & &
    +14 G\left(1;x\right) G\left(0, 0, 0;x\right)
    -8 G\left(0, -1, 0;x\right) G\left(1; x\right)
    -12 G\left(-1, 0, 0;x\right) G\left(1;x\right)
    \nonumber \\
    & &
    +8 G\left(-1, -1, 0;x\right) G\left(1;x\right)
    +8 G\left(1, 1;x\right) G\left(-1, 0;x\right)
    +4 \zeta_2 G\left(1, 1;x\right)
    +4 \zeta_2 G\left(0, 0;x\right)
    \nonumber \\
    & &
    +4 \zeta_2 G\left(-1, -1;x\right)
    +4 \zeta_2 G\left(-1;x\right) G\left(1;x\right)
    -4 \zeta_2 G\left(-1; x\right) G\left(0;x\right)
    -4 \zeta_2 G\left(1;x\right) G\left(0;x\right)
    \nonumber \\
    & &
    -4 \zeta_3 G\left(-1;x\right)
    -4 \zeta_3 G\left(1;x\right)
    +\frac{20}{3} \zeta_3 G\left(0;x\right)
    + \frac{9}{4} \zeta_4,
 \nonumber \\
 J_{15}^{(0)} 
 & = &
 0,
 \nonumber \\
 J_{15}^{(1)} 
 & = &
 0,
 \nonumber \\
 J_{15}^{(2)} 
 & = &
 -G\left( 0, 0; x \right),
 \nonumber \\
 J_{15}^{(3)} 
 & = &
 4\,G\left( 1, 0, 0; x \right)
 -4\,G\left( 0, 0, 0; x \right)
 +2\,G\left( -1, 0, 0; x \right)
 +2\,G\left( 0, -1, 0; x \right)
 -2\,\zeta_2\,G\left( 0; x \right)
 \nonumber \\
 & &
 +6\,\zeta_2\,G\left( -1; x \right)
 +\frac{11}{2}\,\zeta_3
 -6\,\zeta_2\,\ln  \left( 2 \right), 
 \nonumber \\
 J_{15}^{(4)} 
 & = &
 8\,G\left( 0, 0, -1, 0; x \right)
 -10\,G\left( 0, 0, 0, 0; x \right)
 -4\,G\left( 0, -1, -1, 0; x \right)
 +8\,G\left( 0, -1, 0, 0; x \right)
 \nonumber \\
 & &
 -4\,G\left( -1, 0, -1, 0; x \right)
 +8\,G\left( -1, 0, 0, 0; x \right)
 -4\,G\left( -1, -1, 0, 0; x \right)
 -8\,G\left( -1, 1, 0, 0; x \right)
 \nonumber \\
 & &
 -16\,G\left( 1, 1, 0, 0; x \right)
 -8\,G\left( 1, -1, 0, 0; x \right)
 +16\,G\left( 1, 0, 0, 0; x \right)
 -8\,G\left( 1, 0, -1, 0; x \right)
 \nonumber \\
 & &
 +12\,G\left( 0, 1, 0, 0; x \right)
 -3\,\zeta_2\,G\left( 0, 0; x \right)
 -24\,\zeta_2\,G\left( 1, -1; x \right)
 +8\,\zeta_2\,G\left( 1, 0; x \right)
 \nonumber \\
 & &
 -12\,\zeta_2\,G\left( -1, -1; x \right)
 +16\,\zeta_2\,G\left( 0, -1; x \right)
 +4\,\zeta_2\,G\left( -1, 0; x \right)
 +8\,\zeta_3\,G\left( 0; x \right)
 \nonumber \\
 & &
 -22\,\zeta_3\,G\left( 1; x \right)
 +10\,\zeta_3\,G\left( -1; x \right)
 +24\,\zeta_2\,\ln  \left( 2 \right) \,G\left( 1; x \right)
 -24\,\zeta_2\,\ln  \left( 2 \right) \,G\left( -1; x \right)
 \nonumber \\
 & &
 -\frac{23}{2}\,\zeta_4
 +12\,\zeta_2\, \ln^2  \left( 2 \right) 
 +24\,\mathrm{Li}_4\left(\frac{1}{2}\right)
 + \ln^4  \left( 2 \right),
 \nonumber \\
 J_{16}^{(0)} 
 & = &
 0,
 \nonumber \\
 J_{16}^{(1)} 
 & = &
 0,
 \nonumber \\
 J_{16}^{(2)} 
 & = &
 0,
 \nonumber \\
 J_{16}^{(3)} 
 & = &
 -G\left( 0, 0, 0; x \right)
 -\zeta_2\,G\left( 0; x \right),
 \nonumber \\
 J_{16}^{(4)} 
 & = &
 4\,G\left( 0, 1, 0, 0; x \right)
 +2\,G\left( 0, 0, -1, 0; x \right)
 +2\,G\left( 0, -1, 0, 0; x \right)
 -3\,G\left( 0, 0, 0, 0; x \right)
 \nonumber \\
 & &
 -2\,G\left( -1, 0, 0, 0; x \right)
 -\zeta_2\,G\left( 0, 0; x \right)
 -2\,\zeta_2\,G\left( -1, 0; x \right)
 +6\,\zeta_2\,G\left( 0, -1; x \right)
 +2\,\zeta_3\,G\left( 0; x \right)
 \nonumber \\
 & &
 +\frac{1}{4}\,\zeta_4,
 \nonumber \\
 J_{18}^{(0)} 
 & = &
 0,
 \nonumber \\
 J_{18}^{(1)} 
 & = &
 -\frac{1}{2}\,G\left( 0; x \right),
 \nonumber \\
 J_{18}^{(2)} 
 & = &
 G\left( 1, 0; x \right)
 +3\,G\left( -1, 0; x \right)
 -\frac{5}{2}\,G\left( 0, 0; x \right)
 -\zeta_2,
 \nonumber \\
 J_{18}^{(3)} 
 & = &
 -\frac{17}{2}\,G\left( 0, 0, 0; x \right)
 +15\,G\left( 0, -1, 0; x \right)
 +14\,G\left( -1, 0, 0; x \right)
 -18\,G\left( -1, -1, 0; x \right)
 \nonumber \\
 & &
 +5\,G\left( 0, 1, 0; x \right)
 -6\,G\left( -1, 1, 0; x \right)
 +5\,G\left( 1, 0, 0; x \right)
 -2\,G\left( 1, 1, 0; x \right)
 -6\,G\left( 1, -1, 0; x \right)
 \nonumber \\
 & &
 -\zeta_2\,G\left( 0; x \right)
 -\zeta_2\,G\left( 1; x \right)
 +3\,\zeta_2\,G\left( -1; x \right)
 +\frac{9}{4}\,\zeta_3
 +3\,\zeta_2\,\ln  \left( 2 \right),
 \nonumber \\
 J_{18}^{(4)} 
 & = &
 17\,G\left( 1, 0, 0, 0; x \right)
 +12\,G\left( 1, 1, -1, 0; x \right)
 -30\,G\left( 1, 0, -1, 0; x \right)
 +12\,G\left( 1, -1, 1, 0; x \right)
 \nonumber \\
 & &
 +36\,G\left( 1, -1, -1, 0; x \right)
 -10\,G\left( 1, 0, 1, 0; x \right)
 -28\,G\left( -1, 1, 0, 0; x \right)
 +12\,G\left( -1, 1, 1, 0; x \right)
 \nonumber \\
 & &
 -28\,G\left( -1, 0, 1, 0; x \right)
 +36\,G\left( -1, -1, 1, 0; x \right)
 +36\,G\left( -1, 1, -1, 0; x \right)
 +29\,G\left( 0, 1, 0, 0; x \right)
 \nonumber \\
 & &
 -30\,G\left( 0, -1, 1, 0; x \right)
 -10\,G\left( 0, 1, 1, 0; x \right)
 -30\,G\left( 0, 1, -1, 0; x \right)
 +17\,G\left( 0, 0, 1, 0; x \right)
 \nonumber \\
 & &
 -80\,G\left( -1, -1, 0, 0; x \right)
 +108\,G\left( -1, -1, -1, 0; x \right)
 -84\,G\left( -1, 0, -1, 0; x \right)
 \nonumber \\
 & &
 +44\,G\left( -1, 0, 0, 0; x \right)
 -{\frac {41}{2}}\,G\left( 0, 0, 0, 0; x \right)
 +51\,G\left( 0, 0, -1, 0; x \right)
 +66\,G\left( 0, -1, 0, 0; x \right)
 \nonumber \\
 & &
 -90\,G\left( 0, -1, -1, 0; x \right)
 -10\,G\left( 1, 1, 0, 0; x \right)
 -28\,G\left( 1, -1, 0, 0; x \right)
 +4\,G\left( 1, 1, 1, 0; x \right)
 \nonumber \\
 & &
 +3\,\zeta_2\,G\left( 0, 0; x \right)
 +\zeta_2\,G\left( -1, 0; x \right)
 +3\,\zeta_2\,G\left( 0, -1; x \right)
 -6\,\zeta_2\,G\left( -1, -1; x \right)
 -5\,\zeta_2\,G\left( 0, 1; x \right)
 \nonumber \\
 & &
 +2\,\zeta_2\,G\left( 1, 0; x \right)
 -6\,\zeta_2\,G\left( 1, -1; x \right)
 +2\,\zeta_2\,G\left( 1, 1; x \right)
 +6\,\zeta_2\,G\left( -1, 1; x \right)
 -15\,\zeta_3\,G\left( 1; x \right)
 \nonumber \\
 & &
 -20\,\zeta_3\,G\left( -1; x \right)
 +{\frac {43}{3}}\,\zeta_3\,G\left( 0; x \right)
 -12\,\zeta_2\,\ln \left( 2 \right)\,G\left( -1; x \right) 
 +12\,\zeta_2\,\ln  \left( 2 \right)\,G\left( 1; x \right) 
 +{\frac {75}{8}}\,\zeta_4
 \nonumber \\
 & &
 -12\,\mathrm{Li}_4\left(\frac{1}{2}\right)
 -6\,\zeta_2\, \ln^2  \left( 2 \right) 
 -\frac{1}{2}\, \ln^4  \left( 2 \right),
 \nonumber \\
 J_{19}^{(0)} 
 & = &
 0,
 \nonumber \\
 J_{19}^{(1)} 
 & = &
 0,
 \nonumber \\
 J_{19}^{(2)} 
 & = &
 0,
 \nonumber \\
 J_{19}^{(3)} 
 & = &
 -2\,G\left( 0, 0, 0; x \right)
 -2\,\zeta_2\,G\left( 0; x \right),
 \nonumber \\
 J_{19}^{(4)} 
 & = &
 12\,G\left( 0, 0, -1, 0; x \right)
 +4\,G\left( 0, -1, 0, 0; x \right)
 +4\,G\left( 0, 0, 1, 0; x \right)
 -4\,G\left( 1, 0, 0, 0; x \right)
 \nonumber \\
 & &
 -4\,G\left( -1, 0, 0, 0; x \right)
 -6\,G\left( 0, 0, 0, 0; x \right)
 +12\,\zeta_2\,G\left( 0, -1; x \right)
 -4\,\zeta_2\,G\left( -1, 0; x \right)
 \nonumber \\
 & &
 -4\,\zeta_2\,G\left( 1, 0; x \right)
 -\frac{1}{2}\,\zeta_4,
 \nonumber \\
 J_{20}^{(0)} 
 & = &
 0,
 \nonumber \\
 J_{20}^{(1)} 
 & = &
 0,
 \nonumber \\
 J_{20}^{(2)} 
 & = &
 2\,G\left( 0, 1; x \right)
 -G\left( 0, 0; x \right)
 -4\,\zeta_2,
 \nonumber \\
 J_{20}^{(3)} 
 & = &
 2\,G\left( 0, 0, 1; x \right)
 +2\,G\left( 0, 1, 0; x \right)
 +2\,G\left( 1, 0, 0; x \right)
 -4\,G\left( 0, 1, 1; x \right)
 -G\left( 0, 0, 0; x \right)
 \nonumber \\
 & &
 -4\,G\left( 1, 0, 1; x \right)
 -2\,G\left( -1, 0, 0; x \right)
 +4\,G\left( -1, 0, 1; x \right)
 +\zeta_2\,G\left( 0; x \right)
 +8\,\zeta_2\,G\left( 1; x \right)
 \nonumber \\
 & &
 -8\,\zeta_2\,G\left( -1; x \right)
 -5\,\zeta_3,
 \nonumber \\
 J_{20}^{(4)} 
 & = &
 2\,G\left( 0, 0, 1, 0; x \right)
 -4\,G\left( 0, 0, 1, 1; x \right)
 +2\,G\left( 0, 1, 0, 0; x \right)
 -4\,G\left( 0, 1, 0, 1; x \right)
 \nonumber \\
 & &
 -4\,G\left( 0, 1, 1, 0; x \right)
 +8\,G\left( 0, 1, 1, 1; x \right)
 +2\,G\left( 0, 0, 0, 1; x \right)
 -4\,G\left( 1, 1, 0, 0; x \right)
 \nonumber \\
 & &
 -4\,G\left( 1, 0, 0, 1; x \right)
 -4\,G\left( 1, 0, 1, 0; x \right)
 +8\,G\left( 1, 0, 1, 1; x \right)
 +4\,G\left( 1, -1, 0, 0; x \right)
 \nonumber \\
 & &
 +2\,G\left( 1, 0, 0, 0; x \right)
 +8\,G\left( 1, 1, 0, 1; x \right)
 -8\,G\left( 1, -1, 0, 1; x \right)
 +4\,G\left( -1, 0, 1, 0; x \right)
 \nonumber \\
 & &
 +4\,G\left( -1, 1, 0, 0; x \right)
 -8\,G\left( -1, 1, 0, 1; x \right)
 +8\,G\left( -1, -1, 0, 1; x \right)
 +4\,G\left( -1, 0, 0, 1; x \right)
 \nonumber \\
 & &
 -8\,G\left( -1, 0, 1, 1; x \right)
 -G\left( 0, 0, 0, 0; x \right)
 -2\,G\left( -1, 0, 0, 0; x \right)
 -4\,G\left( -1, -1, 0, 0; x \right)
 \nonumber \\
 & &
 +2\,\zeta_3\,G\left( 0; x \right)
 +10\,\zeta_3\,G\left( 1; x \right)
 -10\,\zeta_3\,G\left( -1; x \right)
 -16\,\zeta_2\,G\left( -1, -1; x \right)
 +2\,\zeta_2\,G\left( -1, 0; x \right)
 \nonumber \\
 & &
 +16\,\zeta_2\,G\left( 1, -1; x \right)
 -2\,\zeta_2\,G\left( 1, 0; x \right)
 +16\,\zeta_2\,G\left( -1, 1; x \right)
 -16\,\zeta_2\,G\left( 1, 1; x \right)
 -{\frac {99}{4}}\,\zeta_4,
 \nonumber \\
 J_{21}^{(0)} 
 & = &
 0,
 \nonumber \\
 J_{21}^{(1)} 
 & = &
 0,
 \nonumber \\
 J_{21}^{(2)} 
 & = &
 G\left( 0, 0; x \right),
 \nonumber \\
 J_{21}^{(3)} 
 & = &
 4\,G\left( 0, 0, 0; x \right)
 -2\,G\left( 0, 0, 1; x \right)
 -2\,G\left( 0, 1, 0; x \right)
 -2\,G\left( 0, -1, 0; x \right)
 -2\,G\left( 1, 0, 0; x \right)
 \nonumber \\
 & &
 -\zeta_2\,G\left( 0; x \right)
 -3\,\zeta_3,
 \nonumber \\
 J_{21}^{(4)} 
 & = &
 4\,G\left( 0, 1, -1, 0; x \right)
 +4\,G\left( 0, -1, 0, 1; x \right)
 -8\,G\left( 0, 0, 1, 0; x \right)
 +4\,G\left( 0, 0, 1, 1; x \right)
 \nonumber \\
 & &
 -8\,G\left( 0, 1, 0, 0; x \right)
 +4\,G\left( 0, 1, 0, 1; x \right)
 +4\,G\left( 0, 1, 1, 0; x \right)
 +4\,G\left( 0, -1, 1, 0; x \right)
 \nonumber \\
 & &
 -8\,G\left( 0, 0, 0, 1; x \right)
 +4\,G\left( 1, 1, 0, 0; x \right)
 +4\,G\left( 1, 0, 0, 1; x \right)
 +4\,G\left( 1, 0, 1, 0; x \right)
 \nonumber \\
 & &
 -8\,G\left( 1, 0, 0, 0; x \right)
 +4\,G\left( 1, 0, -1, 0; x \right)
 +11\,G\left( 0, 0, 0, 0; x \right)
 -6\,G\left( 0, 0, -1, 0; x \right)
 \nonumber \\
 & &
 -6\,G\left( 0, -1, 0, 0; x \right)
 +4\,G\left( 0, -1, -1, 0; x \right)
 +2\,\zeta_2\,G\left( 0, -1; x \right)
 -3\,\zeta_2\,G\left( 0, 0; x \right)
 \nonumber \\
 & &
 +2\,\zeta_2\,G\left( 1, 0; x \right)
 +2\,\zeta_2\,G\left( 0, 1; x \right)
 -5\,\zeta_3\,G\left( 0; x \right)
 +6\,\zeta_3\,G\left( 1; x \right)
 -\frac{5}{4}\,\zeta_4,
 \nonumber \\
 J_{23}^{(0)} 
 & = &
 0,
 \nonumber \\
 J_{23}^{(1)} 
 & = &
 0,
 \nonumber \\
 J_{23}^{(2)} 
 & = &
 0,
 \nonumber \\
 J_{23}^{(3)} 
 & = &
 4\,G\left( 1, 0, 0; x \right)
 -2\,G\left( 0, 0, 1; x \right)
 -2\,G\left( 0, 1, 0; x \right)
 +6\,\zeta_3,
 \nonumber \\
 J_{23}^{(4)} 
 & = &
 4\,G\left( 1, 0, 0, 1; x \right)
 +4\,G\left( 0, 1, 0, 1; x \right)
 +4\,G\left( 0, 0, 1, 1; x \right)
 -8\,G\left( 0, 0, 0, 1; x \right)
 \nonumber \\
 & &
 +10\,G\left( 0, 0, -1, 0; x \right)
 -4\,G\left( 0, -1, 0, 0; x \right)
 +4\,G\left( 0, -1, 0, 1; x \right)
 +4\,G\left( 0, -1, 1, 0; x \right)
 \nonumber \\
 & &
 -6\,G\left( 0, 1, 0, 0; x \right)
 -4\,G\left( 0, 0, 1, 0; x \right)
 +4\,G\left( 0, 1, 1, 0; x \right)
 +4\,G\left( 0, 1, -1, 0; x \right)
 \nonumber \\
 & &
 +12\,G\left( 1, 0, 0, 0; x \right)
 -24\,G\left( 1, 0, -1, 0; x \right)
 -4\,G\left( 1, 0, 1, 0; x \right)
 +2\,\zeta_2\,G\left( 0, 1; x \right)
 \nonumber \\
 & &
 -4\,\zeta_2\,G\left( 1, 0; x \right)
 +12\,\zeta_3\,G\left( 0; x \right)
 -24\,\zeta_3\,G\left( 1; x \right)
 +9\,\zeta_4,
 \nonumber \\
 J_{32}^{(0)} 
 & = &
 0,
 \nonumber \\
 J_{32}^{(1)} 
 & = &
 0,
 \nonumber \\
 J_{32}^{(2)} 
 & = &
 2\,G\left( 0, 1; x \right)
 -G\left( 0, 0; x \right)
 -4\,\zeta_2,
 \nonumber \\
 J_{32}^{(3)} 
 & = &
 2\,G\left( 0, 0, 0; x \right)
 -2\,G\left( 0, 0, 1; x \right)
 -4\,G\left( 0, 1, 1; x \right)
 +2\,G\left( 1, 0, 0; x \right)
 -4\,G\left( 1, 0, 1; x \right)
 \nonumber \\
 & &
 -2\,G\left( -1, 0, 0; x \right)
 +4\,G\left( -1, 0, 1; x \right)
 +5\,\zeta_2\,G\left( 0; x \right)
 +8\,\zeta_2\,G\left( 1; x \right)
 -8\,\zeta_2\,G\left( -1; x \right)
 -5\,\zeta_3,
 \nonumber \\
 J_{32}^{(4)} 
 & = &
 4\,G\left( 0, -1, 1, 0; x \right)
 -6\,G\left( 0, 1, 0, 0; x \right)
 +4\,G\left( 0, 1, 0, 1; x \right)
 +8\,G\left( 0, 1, 1, 1; x \right)
 \nonumber \\
 & &
 -8\,G\left( 0, 0, 1, 0; x \right)
 +4\,G\left( 0, 0, 1, 1; x \right)
 -10\,G\left( 0, 0, 0, 1; x \right)
 +4\,G\left( 0, 1, -1, 0; x \right)
 \nonumber \\
 & &
 +4\,G\left( -1, 1, 0, 0; x \right)
 +4\,G\left( -1, 0, 1, 0; x \right)
 -8\,G\left( -1, 1, 0, 1; x \right)
 +9\,G\left( 0, 0, 0, 0; x \right)
 \nonumber \\
 & &
 +8\,G\left( -1, -1, 0, 1; x \right)
 -2\,G\left( 0, 0, -1, 0; x \right)
 -2\,G\left( 0, -1, 0, 0; x \right)
 -8\,G\left( -1, 0, 1, 1; x \right)
 \nonumber \\
 & &
 +4\,G\left( -1, 0, 0, 1; x \right)
 -4\,G\left( 1, 1, 0, 0; x \right)
 +4\,G\left( 1, -1, 0, 0; x \right)
 -2\,G\left( -1, 0, 0, 0; x \right)
 \nonumber \\
 & &
 -4\,G\left( -1, -1, 0, 0; x \right)
 -8\,G\left( 1, -1, 0, 1; x \right)
 -4\,G\left( 1, 0, 0, 0; x \right)
 +4\,G\left( 1, 0, 0, 1; x \right)
 \nonumber \\
 & &
 +8\,G\left( 1, 0, 1, 1; x \right)
 +8\,G\left( 1, 1, 0, 1; x \right)
 +\zeta_2\,G\left( 0, 0; x \right)
 -16\,\zeta_2\,G\left( 1, 1; x \right)
 +16\,\zeta_2\,G\left( -1, 1; x \right)
 \nonumber \\
 & &
 -6\,\zeta_2\,G\left( 0, 1; x \right)
 -10\,\zeta_2\,G\left( 1, 0; x \right)
 +16\,\zeta_2\,G\left( 1, -1; x \right)
 -16\,\zeta_2\,G\left( -1, -1; x \right)
 \nonumber \\
 & &
 +8\,\zeta_2\,G\left( 0, -1; x \right)
 +2\,\zeta_2\,G\left( -1, 0; x \right)
 -10\,\zeta_3\,G\left( -1; x \right)
 +7\,\zeta_3\,G\left( 0; x \right)
 +10\,\zeta_3\,G\left( 1; x \right)
 \nonumber \\
 & &
 -{\frac {139}{4}}\,\zeta_4,
 \nonumber \\
 J_{37}^{(0)} 
 & = &
 0,
 \nonumber \\
 J_{37}^{(1)} 
 & = &
 0,
 \nonumber \\
 J_{37}^{(2)} 
 & = &
 0,
 \nonumber \\
 J_{37}^{(3)} 
 & = &
 0,
 \nonumber \\
 J_{37}^{(4)} 
 & = &
 -6\,G\left( 0, 0, 0, 0; x \right)
 +6\,G\left( 0, 0, 0, 1; x \right)
 +4\,G\left( 0, 0, 1, 0; x \right)
 +2\,G\left( 0, 1, 0, 0; x \right)
 \nonumber \\
 & &
 -4\,\zeta_2\,G\left( 0, 0; x \right),
 \nonumber \\
 J_{40}^{(0)} 
 & = &
 0,
 \nonumber \\
 J_{40}^{(1)} 
 & = &
 0,
 \nonumber \\
 J_{40}^{(2)} 
 & = &
 0,
 \nonumber \\
 J_{40}^{(3)} 
 & = &
 0,
 \nonumber \\
 J_{40}^{(4)} 
 & = &
 -G\left( 0, 0, 0, 0; x \right)
 -8\,G\left( 0, 0, -1, 0; x \right)
 +6\,G\left( 0, 0, 0, 1; x \right)
 +4\,G\left( 0, 1, 0, 0; x \right)
 \nonumber \\
 & &
 -4\,\zeta_2\,G\left( 0, 0; x \right)
 -2\,\zeta_3\,G\left( 0; x \right)
 -3\,\zeta_4.
\eq

\subsubsection{Integrals, which are expressed in the variable $x'$}

The integrals $J_{12}-J_{13}$ 
are most naturally expressed in terms of harmonic polylogarithms in the variable $x'$.
\bq
 J_{12}^{(0)} 
 & = &
 0,
 \nonumber \\
 J_{12}^{(1)} 
 & = &
 0,
 \nonumber \\
 J_{12}^{(2)} 
 & = &
 G\left( 0, 0; x' \right)
 -2\,G\left( 0, -1; x' \right)
 +\zeta_2,
 \nonumber \\
 J_{12}^{(3)} 
 & = &
 4\,G\left( 0, 0, 0; x' \right)
 -8\,G\left( 0, 0, -1; x' \right)
 -2\,G\left( 0, -1, 0; x' \right)
 +4\,G\left( 0, -1, -1; x' \right)
 \nonumber \\
 & &
 -4\,G\left( 1, 0, 0; x' \right)
 +8\,G\left( 1, 0, -1; x' \right)
 +4\,G\left( -1, 0, -1; x' \right)
 -2\,G\left( -1, 0, 0; x' \right)
 \nonumber \\
 & &
 +2\,\zeta_2\,G\left( 0; x' \right)
 -4\,\zeta_2\,G\left( 1; x' \right)
 -2\,\zeta_2\,G\left( -1; x' \right)
 -4\,\zeta_3,
 \nonumber \\
 J_{12}^{(4)} 
 & = &
 10\,G\left( 0, 0, 0, 0; x' \right)
 -20\,G\left( 0, 0, 0, -1; x' \right)
 -8\,G\left( 0, 0, -1, 0; x' \right)
 +16\,G\left( 0, 0, -1, -1; x' \right)
 \nonumber \\
 & &
 -32\,G\left( 1, 1, 0, -1; x' \right)
 +4\,G\left( -1, -1, 0, 0; x' \right)
 -16\,G\left( 1, 0, -1, -1; x' \right)
 \nonumber \\
 & &
 -8\,G\left( -1, -1, 0, -1; x' \right)
 +16\,G\left( 0, -1, 0, -1; x' \right)
 -8\,G\left( 0, -1, -1, -1; x' \right)
 \nonumber \\
 & &
 -8\,G\left( 0, -1, 0, 0; x' \right)
 +4\,G\left( 0, -1, -1, 0; x' \right)
 -12\,G\left( 0, 1, 0, 0; x' \right)
 -16\,G\left( 1, 0, 0, 0; x' \right)
 \nonumber \\
 & &
 -8\,G\left( -1, 0, 0, 0; x' \right)
 +32\,G\left( 1, 0, 0, -1; x' \right)
 +8\,G\left( 1, 0, -1, 0; x' \right)
 +8\,G\left( 1, -1, 0, 0; x' \right)
 \nonumber \\
 & &
 -16\,G\left( 1, -1, 0, -1; x' \right)
 +24\,G\left( 0, 1, 0, -1; x' \right)
 +8\,G\left( -1, 1, 0, 0; x' \right)
 \nonumber \\
 & &
 -16\,G\left( -1, 1, 0, -1; x' \right)
 +16\,G\left( 1, 1, 0, 0; x' \right)
 +16\,G\left( -1, 0, 0, -1; x' \right)
 \nonumber \\
 & &
 +4\,G\left( -1, 0, -1, 0; x' \right)
 -8\,G\left( -1, 0, -1, -1; x' \right)
 -6\,\zeta_2\,G\left( 0, -1; x' \right)
 +3\,\zeta_2\,G\left( 0, 0; x' \right)
 \nonumber \\
 & &
 -12\,\zeta_2\,G\left( 0, 1; x' \right)
 +16\,\zeta_2\,G\left( 1, 1; x' \right)
 +8\,\zeta_2\,G\left( 1, -1; x' \right)
 -8\,\zeta_2\,G\left( 1, 0; x' \right)
 \nonumber \\
 & &
 -4\,\zeta_2\,G\left( -1, 0; x' \right)
 +8\,\zeta_2\,G\left( -1, 1; x' \right)
 +4\,\zeta_2\,G\left( -1, -1; x' \right)
 +16\,\zeta_3\,G\left( 1; x' \right)
 \nonumber \\
 & &
 +8\,\zeta_3\,G\left( -1; x' \right)
 -8\,\zeta_3\,G\left( 0; x' \right),
 \nonumber \\
 J_{13}^{(0)} 
 & = &
 0,
 \nonumber \\
 J_{13}^{(1)} 
 & = &
 0,
 \nonumber \\
 J_{13}^{(2)} 
 & = &
 0,
 \nonumber \\
 J_{13}^{(3)} 
 & = &
 G\left( 0, 0, 0; x' \right)
 -2\,G\left( 0, 0, -1; x' \right)
 +\zeta_2\,G\left( 0; x' \right)
 -2\,\zeta_3,
 \nonumber \\
 J_{13}^{(4)} 
 & = &
 3\,G\left( 0, 0, 0, 0; x' \right)
 +8\,G\left( 0, 1, 0, -1; x' \right)
 -2\,G\left( 0, 0, -1, 0; x' \right)
 +4\,G\left( 0, -1, 0, -1; x' \right)
 \nonumber \\
 & &
 -4\,G\left( 0, 1, 0, 0; x' \right)
 +4\,G\left( 0, 0, -1, -1; x' \right)
 -2\,G\left( 0, -1, 0, 0; x' \right)
 -6\,G\left( 0, 0, 0, -1; x' \right)
 \nonumber \\
 & &
 +2\,G\left( -1, 0, 0, 0; x' \right)
 -4\,G\left( -1, 0, 0, -1; x' \right)
 +\zeta_2\,G\left( 0, 0; x' \right)
 -4\,\zeta_2\,G\left( 0, 1; x' \right)
 \nonumber \\
 & &
 -2\,\zeta_2\,G\left( 0, -1; x' \right)
 +2\,\zeta_2\,G\left( -1, 0; x' \right)
 -2\,\zeta_3\,G\left( 0; x' \right)
 -4\,\zeta_3\,G\left( -1; x' \right)
 +\frac{5}{4}\,\zeta_4.
\eq

\subsubsection{Integrals, which are expressed in the variable $\tilde{x}$}

The integrals $J_{22}$ and $J_{36}$ are expressed in terms of multiple polylogarithms in the variable $\tilde{x}$.
We use the notation of eq.~(\ref{def_G_omega}).
The differential one-forms $\omega_0, \omega_{0,4}$ and $\omega_{-4,0}$ are defined in eq.~(\ref{def_omega}).
\bq
\lefteqn{
 J_{22}^{(0)} 
 = 
 0, 
 } & &
 \nonumber \\
\lefteqn{
 J_{22}^{(1)} 
 =
 0,
 } & &
 \nonumber \\
\lefteqn{
 J_{22}^{(2)} 
 =
 0,
 } & &
 \nonumber \\
\lefteqn{
 J_{22}^{(3)} 
 =
 0,
 } & &
 \nonumber \\
\lefteqn{
 J_{22}^{(4)} 
 =
G\left( \omega_{0}, \omega_{-4,0}, \omega_{-4,0}, \omega_{0}; \tilde{x} \right)
-G\left( \omega_{0}, \omega_{0}, \omega_{0,4}, \omega_{0,4}; \tilde{x} \right)
-\zeta_2\,G\left( \omega_{0}, \omega_{-4,0}; \tilde{x} \right)
+\frac{7}{4}\,\zeta_4,
} & &
 \nonumber \\
\lefteqn{
 J_{36}^{(0)} 
 =
 0,
 } & &
 \nonumber \\
\lefteqn{
 J_{36}^{(1)} 
 =
 0,
 } & &
 \nonumber \\
\lefteqn{
 J_{36}^{(2)} 
 =
 0,
 } & &
 \nonumber \\
\lefteqn{
 J_{36}^{(3)} 
 =
 0,
 } & &
 \nonumber \\
\lefteqn{
 J_{36}^{(4)} 
 =
-7\,G\left( \omega_{0,4}, \omega_{0}, \omega_{0,4}, \omega_{0,4}; \tilde{x} \right)
+2\,G\left( \omega_{0,4}, \omega_{0,4}, \omega_{0}, \omega_{0,4}; \tilde{x} \right)
+2\,G\left( \omega_{0,4}, \omega_{0,4}, \omega_{0,4}, \omega_{0}; \tilde{x} \right)
} & & \nonumber \\
 & &
+4\,G\left( \omega_{0,4}, \omega_{-4,0}, \omega_{-4,0}, \omega_{0}; \tilde{x} \right)
-4\,\zeta_2\,G\left( \omega_{0,4}, \omega_{-4,0}; \tilde{x} \right)
-10\,\zeta_3\,G\left( \omega_{0,4}; \tilde{x} \right)
-{\frac {39}{2}}\,\zeta_4.
\hspace*{10mm}
\eq

\subsection{Integrals, which only depend on $t$}
\label{sect:t_dependent}

The integrals $J_6-J_7$, $J_{14}$ and $J_{17}$ depend
only on the variable $t$ (or equivalently only on the variable $y$).
They are expressed as iterated integrals of modular forms $\{1,f_2,f_3,f_4,g_{2,1}\}$.
The modular forms have been defined in eq.~(\ref{def_modular_forms_1}).
\bq
\lefteqn{
 J_{6}^{(0)} 
 =
 0,
 } & &
 \nonumber \\
\lefteqn{
 J_{6}^{(1)} 
 = 
 0,
 } & &
 \nonumber \\
\lefteqn{
 J_{6}^{(2)} 
 = 
 \iterintmodular{1,f_3}{q_6}
 + 3 \zeta_2, 
 } & &
 \nonumber \\
\lefteqn{
 J_{6}^{(3)} 
 = 
 - \iterintmodular{f_2,1,f_3}{q_6}
 - \iterintmodular{1,f_2,f_3}{q_6}
 + 3 \zeta_2 \iterintmodular{1}{q_6}
 - 3 \zeta_2 \iterintmodular{f_2}{q_6}
 + \frac{21}{2} \zeta_3
 - 18 \zeta_2 \ln\left(2\right),
 } & &
 \nonumber \\
\lefteqn{
 J_{6}^{(4)} 
 = 
   \iterintmodular{f_2,f_2,1,f_3}{q_6}
 + \iterintmodular{f_2,1,f_2,f_3}{q_6}
 + \iterintmodular{1,f_2,f_2,f_3}{q_6}
 + \iterintmodular{1,f_4,1,f_3}{q_6}
 } & &
 \nonumber \\
 & &
 + 3 \zeta_2 \iterintmodular{f_2,f_2}{q_6}
 - 3 \zeta_2 \iterintmodular{1,f_2}{q_6}
 - 3 \zeta_2 \iterintmodular{f_2,1}{q_6}
 + 3 \zeta_2 \iterintmodular{1,f_4}{q_6}
 + \zeta_2 \iterintmodular{1,f_3}{q_6}
 \nonumber \\
 & &
 + \left( \frac{21}{2} \zeta_3 - 18 \zeta_2 \ln\left(2\right) \right) \left( \iterintmodular{1}{q_6} - \iterintmodular{f_2}{q_6} \right)
 -39 \zeta_4 
 + 72 \mathrm{Li}_4\left(\frac{1}{2}\right)
 + 36 \zeta_2 \ln^2(2)
 \nonumber \\
 & &
 + 3 \ln^4\left(2\right),
 \nonumber \\
\lefteqn{
 J_{7}^{(0)} 
 =
 0,
 } & &
 \nonumber \\
\lefteqn{
 J_{7}^{(1)} 
 =
 \iterintmodular{f_3}{q_6},
 } & &
 \nonumber \\
\lefteqn{
 J_{7}^{(2)} 
 =
 - \iterintmodular{f_2,f_3}{q_6}
 + 3 \zeta_2,
 } & &
 \nonumber \\
\lefteqn{
 J_{7}^{(3)} 
 =
 \iterintmodular{f_2,f_2,f_3}{q_6}
 + \iterintmodular{f_4,1,f_3}{q_6}
 + 3 \zeta_2 \iterintmodular{f_4}{q_6}
 - 3 \zeta_2 \iterintmodular{f_2}{q_6}
 + \zeta_2 \iterintmodular{f_3}{q_6}
 + \frac{21}{2} \zeta_3
 } & &
 \nonumber \\
 & &
 - 18 \zeta_2 \ln\left(2\right),
 \nonumber \\
\lefteqn{
 J_{7}^{(4)} 
 =
 - \iterintmodular{f_2,f_2,f_2,f_3}{q_6}
 - \iterintmodular{f_4,f_2,1,f_3}{q_6}
 - \iterintmodular{f_4,1,f_2,f_3}{q_6}
 - \iterintmodular{f_2,f_4,1,f_3}{q_6}
 } & &
 \nonumber \\
 & &
 + 3 \zeta_2 \iterintmodular{f_2,f_2}{q_6}
 - 3 \zeta_2 \iterintmodular{f_4,f_2}{q_6}
 - 3 \zeta_2 \iterintmodular{f_2,f_4}{q_6}
 + 3 \zeta_2 \iterintmodular{f_4,1}{q_6}
 - \zeta_2 \iterintmodular{f_2,f_3}{q_6}
 \nonumber \\
 & &
 + \left( \frac{21}{2} \zeta_3 - 18 \zeta_2 \ln\left(2\right) \right) \left( \iterintmodular{f_4}{q_6} - \iterintmodular{f_2}{q_6} \right)
 - \frac{2}{3} \zeta_3 \iterintmodular{f_3}{q_6}
 -39 \zeta_4 
 + 72 \mathrm{Li}_4\left(\frac{1}{2}\right)
 \nonumber \\
 & &
 + 36 \zeta_2 \ln^2(2)
 + 3 \ln^4\left(2\right),
 \nonumber \\
\lefteqn{
 J_{14}^{(0)} 
 = 0,
 } & &
 \nonumber \\
\lefteqn{
 J_{14}^{(1)} 
 = 
 0,
 } & &
 \nonumber \\
\lefteqn{
 J_{14}^{(2)} 
 = 
 0,
 } & &
 \nonumber \\
\lefteqn{
 J_{14}^{(3)} 
 = 
 - \frac{1}{9} \iterintmodular{f_3,1,f_3}{q_6}
 - \frac{1}{3} \zeta_2 \iterintmodular{f_3}{q_6},
 } & &
 \nonumber \\
\lefteqn{
 J_{14}^{(4)} 
 = 
 \frac{1}{9} \iterintmodular{f_3,1,f_2,f_3}{q_6}
 + \frac{1}{9} \iterintmodular{f_3,f_2,1,f_3}{q_6}
 + \frac{1}{3} \zeta_2 \iterintmodular{f_3,f_2}{q_6}
 - \frac{1}{3} \zeta_2 \iterintmodular{f_3,1}{q_6}
 } & &
 \nonumber \\
 & &
 - \frac{1}{9} \left( \frac{21}{2} \zeta_3 - 18 \zeta_2 \ln\left(2\right) \right) \iterintmodular{f_3}{q_6},
 \nonumber \\
\lefteqn{
 J_{17}^{(0)} 
 = 
 0,
 } & &
 \nonumber \\
\lefteqn{
 J_{17}^{(1)} 
 = 
 0,
 } & &
 \nonumber \\
\lefteqn{
 J_{17}^{(2)} 
 = 
 0,
 } & &
 \nonumber \\
\lefteqn{
 J_{17}^{(3)} 
 = 
 - \frac{2}{9} \iterintmodular{f_3,1,f_3}{q_6}
 - \frac{2}{3} \zeta_2 \iterintmodular{f_3}{q_6},
 } & &
 \nonumber \\
\lefteqn{
 J_{17}^{(4)} 
 = 
 \frac{2}{9} \iterintmodular{f_3,1,f_2,f_3}{q_6}
 + \frac{2}{9} \iterintmodular{f_3,f_2,1,f_3}{q_6}
 - \frac{4}{9} \iterintmodular{g_{2,1},f_3,1,f_3}{q_6}
 + \frac{2}{3} \zeta_2 \iterintmodular{f_3,f_2}{q_6}
 } & &
 \nonumber \\
 & &
 - \frac{2}{3} \zeta_2 \iterintmodular{f_3,1}{q_6}
 - \frac{4}{3} \zeta_2 \iterintmodular{g_{2,1},f_3}{q_6}
 - \frac{2}{9} \left( \frac{21}{2} \zeta_3 - 18 \zeta_2 \ln\left(2\right) \right) \iterintmodular{f_3}{q_6}.
\eq

\subsection{Integrals, which depend on $s$ and $t$}
\label{sect:s_t_dependent}

The integrals $J_{24}-J_{30}$, $J_{33}-J_{35}$, $J_{38}-J_{39}$ and $J_{41}-J_{45}$
depend on $s$ and $t$ and are expressed in terms of iterated integrals
with the integration kernels discussed in section~(\ref{sect:integration_kernels}).
Due to the large number of integration kernels the explicit results  
for these integrals are rather long 
(at the order of $200-300$ terms).
For this reason we list here only a few examples.
We give the results for the integrals $J_{24}$, $J_{27}$, $J_{33}$, $J_{38}$ and $J_{39}$ up to order $\eps^3$.
These are rather compact.
In addition we give the (not so short) result for $J_{41}$ up to order $\eps^4$.
This integral starts at ${\mathcal O}(\eps^4)$.
The integral $J_{41}$ is proportional to the double box integral with unit powers of the propagators, i.e. $I_{1111111}$,
and therefore central to this article.

The results for all integrals up to order $\eps^4$ are given in an electronic file attached to this article.
More information on the electronic file accompanying this article can be found in appendix~\ref{sect:supplement}.

The results for the integrals
$J_{24}$, $J_{27}$, $J_{33}$, $J_{38}$ and $J_{39}$ up to order $\eps^3$ read
\bq
 J_{24}^{(0)} 
 & = &
 0,
 \nonumber \\
 J_{24}^{(1)} 
 & = &
 0,
 \nonumber \\
 J_{24}^{(2)} 
 & = &
 0,
 \nonumber \\
 J_{24}^{(3)} 
 & = &
 \iterint{ \eta^{(b)}_0, \eta^{(\frac{b}{a})}_2, f_3 }
 -\frac{3}{2}\,\iterint{ \eta^{(b)}_0, \eta^{(b)}_{3,5}, \omega_{0,4} }
 -3\,\iterint{ \eta^{(b)}_{1,1}, \omega_{0,4}, \omega_{0,4} }
 \nonumber \\
 & &
 +\iterint{ \eta^{(\frac{a}{b})}_2, \eta^{(a)}_0, f_3 }
 +\frac{9}{2}\,\iterint{ \eta^{(b)}_0, a^{(b)}_{3,2}, \omega_{0,4}, \omega_{0,4} }
 +\iterint{ \eta^{(b)}_0, a^{(a,b)}_{4,1}, \eta^{(a)}_0, f_3 }
 \nonumber \\
 & &
 +\frac{7}{4}\,\zeta_2\,\iterint{ \eta^{(b)}_0 }
 -2\,\zeta_2\,\iterint{ \eta^{(b)}_{1,1} }
 +3\,\zeta_2\,\iterint{ \eta^{(\frac{a}{b})}_2 }
 +3\,\zeta_2\,\iterint{ \eta^{(b)}_0, a^{(b)}_{3,2} }
 \nonumber \\
 & &
 +3\,\zeta_2\,\iterint{ \eta^{(b)}_0, a^{(a,b)}_{4,1} }
 -3\,\ln  \left( 2 \right) \zeta_2
 -\frac{7}{4}\,\zeta_3,
 \nonumber \\
 J_{27}^{(0)} 
 & = &
 0,
 \nonumber \\
 J_{27}^{(1)} 
 & = &
 0,
 \nonumber \\
 J_{27}^{(2)} 
 & = &
 0,
 \nonumber \\
 J_{27}^{(3)} 
 & = &
 2\,\iterint{ \eta^{(b)}_0, \eta^{(\frac{b}{a})}_2, f_3 }
 -3\,\iterint{ \eta^{(b)}_0, \eta^{(b)}_{3,5}, \omega_{0,4} }
 -6\,\iterint{ \eta^{(b)}_{1,1}, \omega_{0,4}, \omega_{0,4} }
 \nonumber \\
 & &
 +2\,\iterint{ \eta^{(\frac{a}{b})}_2, \eta^{(a)}_0, f_3 }
 +9\,\iterint{ \eta^{(b)}_0, a^{(b)}_{3,2}, \omega_{0,4}, \omega_{0,4} }
 +2\,\iterint{ \eta^{(b)}_0, a^{(a,b)}_{4,1}, \eta^{(a)}_0, f_3 }
 \nonumber \\
 & &
 +\frac{7}{2}\,\zeta_2\,\iterint{ \eta^{(b)}_0 }
 -4\,\zeta_2\,\iterint{ \eta^{(b)}_{1,1} }
 +6\,\zeta_2\,\iterint{ \eta^{(\frac{a}{b})}_2 }
 +6\,\zeta_2\,\iterint{ \eta^{(b)}_0, a^{(b)}_{3,2} }
 \nonumber \\
 & &
 +6\,\zeta_2\,\iterint{ \eta^{(b)}_0, a^{(a,b)}_{4,1} }
 -6\,\ln  \left( 2 \right) \zeta_2
 -\frac{7}{2}\,\zeta_3,
 \nonumber \\
 J_{33}^{(0)} 
 & = &
 0,
 \nonumber \\
 J_{33}^{(1)} 
 & = &
 0,
 \nonumber \\
 J_{33}^{(2)} 
 & = &
 0,
 \nonumber \\
 J_{33}^{(3)} 
 & = &
 \iterint{ \eta^{(c)}_0, \eta^{(\frac{c}{a})}_2, f_3 }
 +\frac{3}{2}\,\iterint{ \eta^{(c)}_0, \eta^{(c)}_{3,3}, \omega_0 }
 +\iterint{ \eta^{(c)}_{1,1}, \omega_{0,4}, \omega_0 }
 \nonumber \\
 & &
 +2\,\iterint{ \eta^{(c)}_{1,3}, \omega_{-4,0}, \omega_0 }
 +\iterint{ \eta^{(\frac{a}{c})}_2, \eta^{(a)}_0, f_3 }
 -\frac{1}{3}\,\iterint{ \eta^{(c)}_0, a^{(c)}_{3,2}, \omega_{0,4}, \omega_0 }
 \nonumber \\
 & &
 -\iterint{ \eta^{(c)}_0, a^{(c)}_{3,3}, \omega_{-4,0}, \omega_0 }
 +\iterint{ \eta^{(c)}_0, a^{(a,c)}_{4,1}, \eta^{(a)}_0, f_3 }
 +\frac{3}{4}\,\zeta_2\,\iterint{ \eta^{(c)}_0 }
 \nonumber \\
 & &
 -4\,\zeta_2\,\iterint{ \eta^{(c)}_{1,1} }
 -2\,\zeta_2\,\iterint{ \eta^{(c)}_{1,3} }
 +3\,\zeta_2\,\iterint{ \eta^{(\frac{a}{c})}_2 }
 +\frac{4}{3}\,\zeta_2\,\iterint{ \eta^{(c)}_0, a^{(c)}_{3,2} }
 \nonumber \\
 & &
 +\zeta_2\,\iterint{ \eta^{(c)}_0, a^{(c)}_{3,3} }
 +3\,\zeta_2\,\iterint{ \eta^{(c)}_0, a^{(a,c)}_{4,1} },
 \nonumber \\
 J_{38}^{(0)} 
 & = &
 0,
 \nonumber \\
 J_{38}^{(1)} 
 & = &
 0,
 \nonumber \\
 J_{38}^{(2)} 
 & = &
 0,
 \nonumber \\
 J_{38}^{(3)} 
 & = &
 0,
 \nonumber \\
 J_{39}^{(0)} 
 & = &
 0,
 \nonumber \\
 J_{39}^{(1)} 
 & = &
 0,
 \nonumber \\
 J_{39}^{(2)} 
 & = &
 0,
 \nonumber \\
 J_{39}^{(3)} 
 & = &
 0.
\eq
Finally, let us give the result the result for the integral $J_{41}$.
We recall that this integral is proportional to the double box integral with unit powers of the propagators,
e.g.
\bq
 J_{41}
 & = & 
 \eps^4 \frac{\left(1-x\right)^4}{x^2} \frac{\pi}{\psi^{\curvetwo}_1} I_{1111111}.
\eq
This integral starts at ${\mathcal O}(\eps^4)$, hence
\bq
 J_{41}^{(0)} = 0,
 \;\;\;\;\;\;
 J_{41}^{(1)} = 0,
 \;\;\;\;\;\;
 J_{41}^{(2)} = 0,
 \;\;\;\;\;\;
 J_{41}^{(3)} = 0.
\eq
Due to the differential equation we may write $J_{41}^{(4)}$ as an integral over $\eps^3$-terms $J_i^{(3)}$.
Starting the integration path as usual at $(x,y)=(0,1)$ we obtain
\bq
\label{J41_int_diff_eq}
 J_{41}^{(4)}
 & = &
 -{\frac {79}{4}}\,\zeta_4
+8\,\mathrm{Li}_4\left( \frac{1}{2} \right) 
-8\, \zeta_2 \left( \ln  \left( 2 \right)  \right) ^{2}
+\frac{1}{3}\, \left( \ln  \left( 2 \right)  \right) ^{4}
+ \int\limits_\gamma \left( \frac{1}{9} \eta_{2,9} - \frac{1}{3} g_{2,1} - 2 \omega_{0,4} \right) J_{24}^{(3)}
 \nonumber \\
 & &
+ \int\limits_\gamma \left( \frac{1}{18} \eta_{2,10} + \frac{4}{3} g_{2,1} + \frac{1}{4} \omega_0 + \frac{9}{4} \omega_{0,4} \right) J_{27}^{(3)}
+ \int\limits_\gamma \eta^{(\frac{c}{b})}_{2} J_{33}^{(3)}
 \nonumber \\
 & &
+ \int\limits_\gamma \eta^{(b)}_{1,1} 
  \left[ 
         - \frac{1}{2} J_{42}^{(3)}
         + 2 J_{32}^{(3)}
         + 2 J_{28}^{(3)}
         + 2 J_{21}^{(3)}
         - 2 J_{20}^{(3)}
         + 4 J_{18}^{(3)}
         - 4 J_{15}^{(3)}
         + \frac{1}{3} J_{14}^{(3)}
         + 4 J_{13}^{(3)}
 \right. \nonumber \\
 & & \left.
         - 3 J_{10}^{(3)}
         + J_{8}^{(3)}
         - 2 J_{4}^{(3)}
         + 2 J_{3}^{(3)}
  \right]
+ \int\limits_\gamma \eta^{(b)}_{0} J_{43}^{(3)}.
\eq
The explicit result for $J_{41}^{(4)}$ reads
\bq
\label{longlonglong}
\lefteqn{
 J_{41}^{(4)} 
 = }
 & &
 \nonumber \\
 & &
\frac{4}{3}\,\iterint{ \eta^{(b)}_0, d_{2,4}, \eta^{(\frac{b}{a})}_2, f_3 }
-2\,\iterint{ \eta^{(b)}_0, d_{2,4}, \eta^{(b)}_{3,5}, \omega_{0,4} }
-{\frac {26}{3}}\,\iterint{ \eta^{(b)}_0, \eta_{2,1}, \eta^{(\frac{b}{a})}_2, f_3 }
 \nonumber \\
 & &
+13\,\iterint{ \eta^{(b)}_0, \eta_{2,1}, \eta^{(b)}_{3,5}, \omega_{0,4} }
+\frac{7}{3}\,\iterint{ \eta^{(b)}_0, g_{2,1}, \eta^{(\frac{b}{a})}_2, f_3 }
-\frac{7}{2}\,\iterint{ \eta^{(b)}_0, g_{2,1}, \eta^{(b)}_{3,5}, \omega_{0,4} }
 \nonumber \\
 & &
+3\,\iterint{ \eta^{(b)}_0, \omega_0, \eta^{(b)}_{3,5}, \omega_{0,4} }
-{\frac {31}{6}}\,\iterint{ \eta^{(b)}_0, \omega_4, \eta^{(\frac{b}{a})}_2, f_3 }
+\frac{1}{6}\,\iterint{ \eta^{(b)}_0, \eta_{2,10}, \eta^{(b)}_{3,5}, \omega_{0,4} }
 \nonumber \\
 & &
-2\,\iterint{ \eta^{(b)}_0, \omega_0, \eta^{(\frac{b}{a})}_2, f_3 }
+{\frac {31}{4}}\,\iterint{ \eta^{(b)}_0, \omega_4, \eta^{(b)}_{3,5}, \omega_{0,4} }
-\frac{1}{9}\,\iterint{ \eta^{(b)}_0, \eta_{2,10}, \eta^{(\frac{b}{a})}_2, f_3 }
 \nonumber \\
 & &
-\frac{1}{9}\,\iterint{ \eta^{(b)}_0, \eta_{2,11}, \eta^{(\frac{b}{a})}_2, f_3 }
+\frac{1}{6}\,\iterint{ \eta^{(b)}_0, \eta_{2,11}, \eta^{(b)}_{3,5}, \omega_{0,4} }
+\iterint{ \eta^{(b)}_0, \eta^{(\frac{b}{c})}_2, \eta^{(\frac{c}{a})}_2, f_3 }
 \nonumber \\
 & &
+\frac{3}{2}\,\iterint{ \eta^{(b)}_0, \eta^{(\frac{b}{c})}_2, \eta^{(c)}_{3,3}, \omega_0 }
+\frac{1}{2}\,\iterint{ \eta^{(b)}_0, \eta^{(b)}_{3,10}, \omega_{0,4}, \omega_{0,4} }
+\iterint{ \eta^{(b)}_0, \eta^{(b)}_{3,11}, \omega_{0,4}, \omega_{0,4} }
 \nonumber \\
 & &
+\iterint{ \eta^{(b)}_0, \eta^{(b)}_{3,1}, \omega_{0,4}, \omega_0 }
-\iterint{ \eta^{(b)}_0, \eta^{(b)}_{3,24}, \omega_{-4,0}, \omega_0 }
-3\,\iterint{ \eta^{(b)}_0, \eta^{(b)}_{3,3}, \omega_{0,4}, \omega_{0,4} }
 \nonumber \\
 & &
-6\,\iterint{ \eta^{(b)}_0, \eta^{(b)}_{3,4}, \omega_{0,4}, \omega_{0,4} }
+6\,\iterint{ \eta^{(b)}_0, \eta^{(b)}_{3,5}, \omega_4, \omega_{0,4} }
-3\,\iterint{ \eta^{(b)}_0, \eta^{(b)}_{3,5}, \omega_{0,4}, \omega_0 }
 \nonumber \\
 & &
-\frac{1}{3}\,\iterint{ \eta^{(b)}_0, \eta^{(b)}_{3,9}, \omega_{0,4}, \omega_0 }
-{\frac {31}{6}}\,\iterint{ \eta^{(b)}_0, \omega_{0,4}, \eta^{(\frac{b}{a})}_2, f_3 }
-\frac{1}{6}\,\iterint{ \eta_{2,9}, \eta^{(b)}_0, \eta^{(b)}_{3,5}, \omega_{0,4} }
 \nonumber \\
 & &
+\iterint{ \eta^{(b)}_0, \eta^{(a,b)}_{4,3}, \eta^{(a)}_0, f_3 }
+{\frac {31}{4}}\,\iterint{ \eta^{(b)}_0, \omega_{0,4}, \eta^{(b)}_{3,5}, \omega_{0,4} }
+\frac{1}{9}\,\iterint{ \eta_{2,9}, \eta^{(b)}_0, \eta^{(\frac{b}{a})}_2, f_3 }
 \nonumber \\
 & &
-\frac{1}{3}\,\iterint{ \eta_{2,9}, \eta^{(b)}_{1,1}, \omega_{0,4}, \omega_{0,4} }
+\frac{1}{9}\,\iterint{ \eta_{2,9}, \eta^{(\frac{a}{b})}_2, \eta^{(a)}_0, f_3 }
+\frac{7}{3}\,\iterint{ g_{2,1}, \eta^{(b)}_0, \eta^{(\frac{b}{a})}_2, f_3 }
 \nonumber \\
 & &
-\frac{7}{2}\,\iterint{ g_{2,1}, \eta^{(b)}_0, \eta^{(b)}_{3,5}, \omega_{0,4} }
-7\,\iterint{ g_{2,1}, \eta^{(b)}_{1,1}, \omega_{0,4}, \omega_{0,4} }
+\frac{7}{3}\,\iterint{ g_{2,1}, \eta^{(\frac{a}{b})}_2, \eta^{(a)}_0, f_3 }
 \nonumber \\
 & &
+\frac{1}{2}\,\iterint{ \omega_0, \eta^{(b)}_0, \eta^{(\frac{b}{a})}_2, f_3 }
-\frac{3}{4}\,\iterint{ \omega_0, \eta^{(b)}_0, \eta^{(b)}_{3,5}, \omega_{0,4} }
-\frac{3}{2}\,\iterint{ \omega_0, \eta^{(b)}_{1,1}, \omega_{0,4}, \omega_{0,4} }
 \nonumber \\
 & &
+\frac{1}{2}\,\iterint{ \omega_0, \eta^{(\frac{a}{b})}_2, \eta^{(a)}_0, f_3 }
+4\,\iterint{ \eta^{(b)}_{1,1}, d_{2,4}, \omega_{0,4}, \omega_{0,4} }
-10\,\iterint{ \eta^{(b)}_{1,1}, \eta_{2,1}, \omega_{0,4}, \omega_{0,4} }
 \nonumber \\
 & &
-\frac{9}{2}\,\iterint{ \eta^{(b)}_{1,1}, \omega_0, \omega_{0,4}, \omega_{0,4} }
+\frac{1}{2}\,\iterint{ \eta^{(b)}_{1,1}, \omega_4, \omega_{0,4}, \omega_{0,4} }
-{\frac {16}{9}}\,\iterint{ \eta^{(b)}_{1,1}, \eta^{(b)}_{1,2}, \eta^{(\frac{b}{a})}_2, f_3 }
 \nonumber \\
 & &
+\frac{8}{3}\,\iterint{ \eta^{(b)}_{1,1}, \eta^{(b)}_{1,2}, \eta^{(b)}_{3,5}, \omega_{0,4} }
+\frac{8}{3}\,\iterint{ \eta^{(b)}_{1,1}, \eta^{(b)}_{1,4}, \eta^{(\frac{b}{a})}_2, f_3 }
-4\,\iterint{ \eta^{(b)}_{1,1}, \eta^{(b)}_{1,4}, \eta^{(b)}_{3,5}, \omega_{0,4} }
 \nonumber \\
 & &
-{\frac {2}{27}}\,\iterint{ \eta^{(b)}_{1,1}, \eta^{(a)}_{3,1}, \eta^{(a)}_0, f_3 }
-{\frac {1}{54}}\,\iterint{ \eta^{(b)}_{1,1}, \eta^{(a)}_{3,2}, \eta^{(a)}_0, f_3 }
+12\,\iterint{ \eta^{(b)}_{1,1}, \omega_{0,4}, \omega_4, \omega_{0,4} }
 \nonumber \\
 & &
-8\,\iterint{ \eta^{(b)}_{1,1}, \omega_{0,4}, \omega_{0,4}, \omega_0 }
-4\,\iterint{ \eta^{(b)}_{1,1}, \omega_{0,4}, \omega_{0,4}, \omega_{0,4} }
-4\,\iterint{ \eta^{(b)}_{1,1}, \omega_{-4,0}, \omega_{-4,0}, \omega_0 }
 \nonumber \\
 & &
+\frac{1}{9}\,\iterint{ \eta_{2,10}, \eta^{(b)}_0, \eta^{(\frac{b}{a})}_2, f_3 }
-\frac{1}{6}\,\iterint{ \eta_{2,10}, \eta^{(b)}_0, \eta^{(b)}_{3,5}, \omega_{0,4} }
+\frac{1}{9}\,\iterint{ \eta_{2,10}, \eta^{(\frac{a}{b})}_2, \eta^{(a)}_0, f_3 }
 \nonumber \\
 & &
-\frac{1}{3}\,\iterint{ \eta_{2,10}, \eta^{(b)}_{1,1}, \omega_{0,4}, \omega_{0,4} }
+\iterint{ \eta^{(\frac{c}{b})}_2, \eta^{(c)}_0, \eta^{(\frac{c}{a})}_2, f_3 }
+\frac{3}{2}\,\iterint{ \eta^{(\frac{c}{b})}_2, \eta^{(c)}_0, \eta^{(c)}_{3,3}, \omega_0 }
 \nonumber \\
 & &
+\iterint{ \eta^{(\frac{c}{b})}_2, \eta^{(c)}_{1,1}, \omega_{0,4}, \omega_0 }
+2\,\iterint{ \eta^{(\frac{c}{b})}_2, \eta^{(c)}_{1,3}, \omega_{-4,0}, \omega_0 }
-\frac{15}{2}\,\iterint{ \omega_{0,4}, \eta^{(b)}_{1,1}, \omega_{0,4}, \omega_{0,4} }
 \nonumber \\
 & &
+\iterint{ \eta^{(\frac{c}{b})}_2, \eta^{(\frac{a}{c})}_2, \eta^{(a)}_0, f_3 }
+\frac{5}{2}\,\iterint{ \omega_{0,4}, \eta^{(b)}_0, \eta^{(\frac{b}{a})}_2, f_3 }
-{\frac {15}{4}}\,\iterint{ \omega_{0,4}, \eta^{(b)}_0, \eta^{(b)}_{3,5}, \omega_{0,4} }
 \nonumber \\
 & &
+\frac{5}{2}\,\iterint{ \omega_{0,4}, \eta^{(\frac{a}{b})}_2, \eta^{(a)}_0, f_3 }
-6\,\iterint{ \eta^{(b)}_0, a^{(b)}_{3,2}, d_{2,4}, \omega_{0,4}, \omega_{0,4} }
 \nonumber \\
 & &
+15\,\iterint{ \eta^{(b)}_0, a^{(b)}_{3,2}, \eta_{2,1}, \omega_{0,4}, \omega_{0,4} }
+{\frac {27}{4}}\,\iterint{ \eta^{(b)}_0, a^{(b)}_{3,2}, \omega_0, \omega_{0,4}, \omega_{0,4} }
 \nonumber \\
 & &
-\frac{3}{4}\,\iterint{ \eta^{(b)}_0, a^{(b)}_{3,2}, \omega_4, \omega_{0,4}, \omega_{0,4} }
+\frac{8}{3}\,\iterint{ \eta^{(b)}_0, a^{(b)}_{3,2}, \eta^{(b)}_{1,2}, \eta^{(\frac{b}{a})}_2, f_3 }
 \nonumber \\
 & &
-4\,\iterint{ \eta^{(b)}_0, a^{(b)}_{3,2}, \eta^{(b)}_{1,2}, \eta^{(b)}_{3,5}, \omega_{0,4} }
-4\,\iterint{ \eta^{(b)}_0, a^{(b)}_{3,2}, \eta^{(b)}_{1,4}, \eta^{(\frac{b}{a})}_2, f_3 }
 \nonumber \\
 & &
+6\,\iterint{ \eta^{(b)}_0, a^{(b)}_{3,2}, \eta^{(b)}_{1,4}, \eta^{(b)}_{3,5}, \omega_{0,4} }
+\frac{1}{9}\,\iterint{ \eta^{(b)}_0, a^{(b)}_{3,2}, \eta^{(a)}_{3,1}, \eta^{(a)}_0, f_3 }
 \nonumber \\
 & &
+\frac{1}{36}\,\iterint{ \eta^{(b)}_0, a^{(b)}_{3,2}, \eta^{(a)}_{3,2}, \eta^{(a)}_0, f_3 }
-18\,\iterint{ \eta^{(b)}_0, a^{(b)}_{3,2}, \omega_{0,4}, \omega_4, \omega_{0,4} }
 \nonumber \\
 & &
+12\,\iterint{ \eta^{(b)}_0, a^{(b)}_{3,2}, \omega_{0,4}, \omega_{0,4}, \omega_0 }
+6\,\iterint{ \eta^{(b)}_0, a^{(b)}_{3,2}, \omega_{0,4}, \omega_{0,4}, \omega_{0,4} }
 \nonumber \\
 & &
+6\,\iterint{ \eta^{(b)}_0, a^{(b)}_{3,2}, \omega_{-4,0}, \omega_{-4,0}, \omega_0 }
+6\,\iterint{ \eta^{(b)}_0, d_{2,4}, a^{(b)}_{3,2}, \omega_{0,4}, \omega_{0,4} }
 \nonumber \\
 & &
+\frac{4}{3}\,\iterint{ \eta^{(b)}_0, d_{2,4}, a^{(a,b)}_{4,1}, \eta^{(a)}_0, f_3 }
-39\,\iterint{ \eta^{(b)}_0, \eta_{2,1}, a^{(b)}_{3,2}, \omega_{0,4}, \omega_{0,4} }
 \nonumber \\
 & &
-{\frac {26}{3}}\,\iterint{ \eta^{(b)}_0, \eta_{2,1}, a^{(a,b)}_{4,1}, \eta^{(a)}_0, f_3 }
+\frac{21}{2}\,\iterint{ \eta^{(b)}_0, g_{2,1}, a^{(b)}_{3,2}, \omega_{0,4}, \omega_{0,4} }
 \nonumber \\
 & &
+\frac{7}{3}\,\iterint{ \eta^{(b)}_0, g_{2,1}, a^{(a,b)}_{4,1}, \eta^{(a)}_0, f_3 }
-9\,\iterint{ \eta^{(b)}_0, \omega_0, a^{(b)}_{3,2}, \omega_{0,4}, \omega_{0,4} }
 \nonumber \\
 & &
-2\,\iterint{ \eta^{(b)}_0, \omega_0, a^{(a,b)}_{4,1}, \eta^{(a)}_0, f_3 }
-{\frac {93}{4}}\,\iterint{ \eta^{(b)}_0, \omega_4, a^{(b)}_{3,2}, \omega_{0,4}, \omega_{0,4} }
 \nonumber \\
 & &
-{\frac {31}{6}}\,\iterint{ \eta^{(b)}_0, \omega_4, a^{(a,b)}_{4,1}, \eta^{(a)}_0, f_3 }
+\iterint{ \eta^{(b)}_0, a^{(b,c)}_{4,1}, \eta^{(c)}_0, \eta^{(\frac{c}{a})}_2, f_3 }
 \nonumber \\
 & &
+\frac{3}{2}\,\iterint{ \eta^{(b)}_0, a^{(b,c)}_{4,1}, \eta^{(c)}_0, \eta^{(c)}_{3,3}, \omega_0 }
+\iterint{ \eta^{(b)}_0, a^{(b,c)}_{4,1}, \eta^{(c)}_{1,1}, \omega_{0,4}, \omega_0 }
 \nonumber \\
 & &
+2\,\iterint{ \eta^{(b)}_0, a^{(b,c)}_{4,1}, \eta^{(c)}_{1,3}, \omega_{-4,0}, \omega_0 }
+\iterint{ \eta^{(b)}_0, a^{(b,c)}_{4,1}, \eta^{(\frac{a}{c})}_2, \eta^{(a)}_0, f_3 }
 \nonumber \\
 & &
+\iterint{ \eta^{(b)}_0, a^{(b,b)}_{4,3}, \eta^{(b)}_0, \eta^{(\frac{b}{a})}_2, f_3 }
-\frac{3}{2}\,\iterint{ \eta^{(b)}_0, a^{(b,b)}_{4,3}, \eta^{(b)}_0, \eta^{(b)}_{3,5}, \omega_{0,4} }
 \nonumber \\
 & &
-3\,\iterint{ \eta^{(b)}_0, a^{(b,b)}_{4,3}, \eta^{(b)}_{1,1}, \omega_{0,4}, \omega_{0,4} }
+\iterint{ \eta^{(b)}_0, a^{(b,b)}_{4,3}, \eta^{(\frac{a}{b})}_2, \eta^{(a)}_0, f_3 }
 \nonumber \\
 & &
+2\,\iterint{ \eta^{(b)}_0, a^{(b,b)}_{4,4}, \eta^{(b)}_0, \eta^{(\frac{b}{a})}_2, f_3 }
-3\,\iterint{ \eta^{(b)}_0, a^{(b,b)}_{4,4}, \eta^{(b)}_0, \eta^{(b)}_{3,5}, \omega_{0,4} }
 \nonumber \\
 & &
-6\,\iterint{ \eta^{(b)}_0, a^{(b,b)}_{4,4}, \eta^{(b)}_{1,1}, \omega_{0,4}, \omega_{0,4} }
+2\,\iterint{ \eta^{(b)}_0, a^{(b,b)}_{4,4}, \eta^{(\frac{a}{b})}_2, \eta^{(a)}_0, f_3 }
 \nonumber \\
 & &
-\frac{1}{2}\,\iterint{ \eta^{(b)}_0, \eta_{2,10}, a^{(b)}_{3,2}, \omega_{0,4}, \omega_{0,4} }
-\frac{1}{9}\,\iterint{ \eta^{(b)}_0, \eta_{2,10}, a^{(a,b)}_{4,1}, \eta^{(a)}_0, f_3 }
 \nonumber \\
 & &
-\frac{1}{2}\,\iterint{ \eta^{(b)}_0, \eta_{2,11}, a^{(b)}_{3,2}, \omega_{0,4}, \omega_{0,4} }
-\frac{1}{9}\,\iterint{ \eta^{(b)}_0, \eta_{2,11}, a^{(a,b)}_{4,1}, \eta^{(a)}_0, f_3 }
 \nonumber \\
 & &
-\frac{1}{3}\,\iterint{ \eta^{(b)}_0, \eta^{(\frac{b}{c})}_2, a^{(c)}_{3,2}, \omega_{0,4}, \omega_0 }
-\iterint{ \eta^{(b)}_0, \eta^{(\frac{b}{c})}_2, a^{(c)}_{3,3}, \omega_{-4,0}, \omega_0 }
 \nonumber \\
 & &
+\iterint{ \eta^{(b)}_0, \eta^{(\frac{b}{c})}_2, a^{(a,c)}_{4,1}, \eta^{(a)}_0, f_3 }
-{\frac {93}{4}}\,\iterint{ \eta^{(b)}_0, \omega_{0,4}, a^{(b)}_{3,2}, \omega_{0,4}, \omega_{0,4} }
 \nonumber \\
 & &
-{\frac {31}{6}}\,\iterint{ \eta^{(b)}_0, \omega_{0,4}, a^{(a,b)}_{4,1}, \eta^{(a)}_0, f_3 }
+\frac{1}{2}\,\iterint{ \eta_{2,9}, \eta^{(b)}_0, a^{(b)}_{3,2}, \omega_{0,4}, \omega_{0,4} }
 \nonumber \\
 & &
+\frac{1}{9}\,\iterint{ \eta_{2,9}, \eta^{(b)}_0, a^{(a,b)}_{4,1}, \eta^{(a)}_0, f_3 }
+\frac{21}{2}\,\iterint{ g_{2,1}, \eta^{(b)}_0, a^{(b)}_{3,2}, \omega_{0,4}, \omega_{0,4} }
 \nonumber \\
 & &
+\frac{7}{3}\,\iterint{ g_{2,1}, \eta^{(b)}_0, a^{(a,b)}_{4,1}, \eta^{(a)}_0, f_3 }
+\frac{9}{4}\,\iterint{ \omega_0, \eta^{(b)}_0, a^{(b)}_{3,2}, \omega_{0,4}, \omega_{0,4} }
 \nonumber \\
 & &
+\frac{1}{2}\,\iterint{ \omega_0, \eta^{(b)}_0, a^{(a,b)}_{4,1}, \eta^{(a)}_0, f_3 }
+\frac{8}{3}\,\iterint{ \eta^{(b)}_{1,1}, a^{(b)}_{3,1}, \eta^{(b)}_0, \eta^{(\frac{b}{a})}_2, f_3 }
 \nonumber \\
 & &
-4\,\iterint{ \eta^{(b)}_{1,1}, a^{(b)}_{3,1}, \eta^{(b)}_0, \eta^{(b)}_{3,5}, \omega_{0,4} }
-8\,\iterint{ \eta^{(b)}_{1,1}, a^{(b)}_{3,1}, \eta^{(b)}_{1,1}, \omega_{0,4}, \omega_{0,4} }
 \nonumber \\
 & &
+\frac{8}{3}\,\iterint{ \eta^{(b)}_{1,1}, a^{(b)}_{3,1}, \eta^{(\frac{a}{b})}_2, \eta^{(a)}_0, f_3 }
+4\,\iterint{ \eta^{(b)}_{1,1}, a^{(b)}_{3,3}, \eta^{(b)}_0, \eta^{(\frac{b}{a})}_2, f_3 }
 \nonumber \\
 & &
-6\,\iterint{ \eta^{(b)}_{1,1}, a^{(b)}_{3,3}, \eta^{(b)}_0, \eta^{(b)}_{3,5}, \omega_{0,4} }
-12\,\iterint{ \eta^{(b)}_{1,1}, a^{(b)}_{3,3}, \eta^{(b)}_{1,1}, \omega_{0,4}, \omega_{0,4} }
 \nonumber \\
 & &
+4\,\iterint{ \eta^{(b)}_{1,1}, a^{(b)}_{3,3}, \eta^{(\frac{a}{b})}_2, \eta^{(a)}_0, f_3 }
-8\,\iterint{ \eta^{(b)}_{1,1}, \eta^{(b)}_{1,2}, a^{(b)}_{3,2}, \omega_{0,4}, \omega_{0,4} }
 \nonumber \\
 & &
-{\frac {16}{9}}\,\iterint{ \eta^{(b)}_{1,1}, \eta^{(b)}_{1,2}, a^{(a,b)}_{4,1}, \eta^{(a)}_0, f_3 }
+12\,\iterint{ \eta^{(b)}_{1,1}, \eta^{(b)}_{1,4}, a^{(b)}_{3,2}, \omega_{0,4}, \omega_{0,4} }
 \nonumber \\
 & &
+\frac{8}{3}\,\iterint{ \eta^{(b)}_{1,1}, \eta^{(b)}_{1,4}, a^{(a,b)}_{4,1}, \eta^{(a)}_0, f_3 }
+\frac{1}{2}\,\iterint{ \eta_{2,10}, \eta^{(b)}_0, a^{(b)}_{3,2}, \omega_{0,4}, \omega_{0,4} }
 \nonumber \\
 & &
+\frac{1}{9}\,\iterint{ \eta_{2,10}, \eta^{(b)}_0, a^{(a,b)}_{4,1}, \eta^{(a)}_0, f_3 }
-\frac{1}{3}\,\iterint{ \eta^{(\frac{c}{b})}_2, \eta^{(c)}_0, a^{(c)}_{3,2}, \omega_{0,4}, \omega_0 }
 \nonumber \\
 & &
-\iterint{ \eta^{(\frac{c}{b})}_2, \eta^{(c)}_0, a^{(c)}_{3,3}, \omega_{-4,0}, \omega_0 }
+\iterint{ \eta^{(\frac{c}{b})}_2, \eta^{(c)}_0, a^{(a,c)}_{4,1}, \eta^{(a)}_0, f_3 }
 \nonumber \\
 & &
+{\frac {45}{4}}\,\iterint{ \omega_{0,4}, \eta^{(b)}_0, a^{(b)}_{3,2}, \omega_{0,4}, \omega_{0,4} }
+\frac{5}{2}\,\iterint{ \omega_{0,4}, \eta^{(b)}_0, a^{(a,b)}_{4,1}, \eta^{(a)}_0, f_3 }
 \nonumber \\
 & &
+12\,\iterint{ \eta^{(b)}_0, a^{(b)}_{3,2}, \eta^{(b)}_{1,2}, a^{(b)}_{3,2}, \omega_{0,4}, \omega_{0,4} }
+\frac{8}{3}\,\iterint{ \eta^{(b)}_0, a^{(b)}_{3,2}, \eta^{(b)}_{1,2}, a^{(a,b)}_{4,1}, \eta^{(a)}_0, f_3 }
 \nonumber \\
 & &
-18\,\iterint{ \eta^{(b)}_0, a^{(b)}_{3,2}, \eta^{(b)}_{1,4}, a^{(b)}_{3,2}, \omega_{0,4}, \omega_{0,4} }
-4\,\iterint{ \eta^{(b)}_0, a^{(b)}_{3,2}, \eta^{(b)}_{1,4}, a^{(a,b)}_{4,1}, \eta^{(a)}_0, f_3 }
 \nonumber \\
 & &
-\frac{1}{3}\,\iterint{ \eta^{(b)}_0, a^{(b,c)}_{4,1}, \eta^{(c)}_0, a^{(c)}_{3,2}, \omega_{0,4}, \omega_0 }
-\iterint{ \eta^{(b)}_0, a^{(b,c)}_{4,1}, \eta^{(c)}_0, a^{(c)}_{3,3}, \omega_{-4,0}, \omega_0 }
 \nonumber \\
 & &
+\iterint{ \eta^{(b)}_0, a^{(b,c)}_{4,1}, \eta^{(c)}_0, a^{(a,c)}_{4,1}, \eta^{(a)}_0, f_3 }
+\frac{9}{2}\,\iterint{ \eta^{(b)}_0, a^{(b,b)}_{4,3}, \eta^{(b)}_0, a^{(b)}_{3,2}, \omega_{0,4}, \omega_{0,4} }
 \nonumber \\
 & &
+\iterint{ \eta^{(b)}_0, a^{(b,b)}_{4,3}, \eta^{(b)}_0, a^{(a,b)}_{4,1}, \eta^{(a)}_0, f_3 }
+9\,\iterint{ \eta^{(b)}_0, a^{(b,b)}_{4,4}, \eta^{(b)}_0, a^{(b)}_{3,2}, \omega_{0,4}, \omega_{0,4} }
 \nonumber \\
 & &
+2\,\iterint{ \eta^{(b)}_0, a^{(b,b)}_{4,4}, \eta^{(b)}_0, a^{(a,b)}_{4,1}, \eta^{(a)}_0, f_3 }
+12\,\iterint{ \eta^{(b)}_{1,1}, a^{(b)}_{3,1}, \eta^{(b)}_0, a^{(b)}_{3,2}, \omega_{0,4}, \omega_{0,4} }
 \nonumber \\
 & &
+\frac{8}{3}\,\iterint{ \eta^{(b)}_{1,1}, a^{(b)}_{3,1}, \eta^{(b)}_0, a^{(a,b)}_{4,1}, \eta^{(a)}_0, f_3 }
+18\,\iterint{ \eta^{(b)}_{1,1}, a^{(b)}_{3,3}, \eta^{(b)}_0, a^{(b)}_{3,2}, \omega_{0,4}, \omega_{0,4} }
 \nonumber \\
 & &
+4\,\iterint{ \eta^{(b)}_{1,1}, a^{(b)}_{3,3}, \eta^{(b)}_0, a^{(a,b)}_{4,1}, \eta^{(a)}_0, f_3 }
-6\,\iterint{ \eta^{(b)}_0, a^{(b)}_{3,2}, a^{(b)}_{3,3}, \eta^{(b)}_0, \eta^{(\frac{b}{a})}_2, f_3 }
 \nonumber \\
 & &
+18\,\iterint{ \eta^{(b)}_0, a^{(b)}_{3,2}, a^{(b)}_{3,3}, \eta^{(b)}_{1,1}, \omega_{0,4}, \omega_{0,4} }
-6\,\iterint{ \eta^{(b)}_0, a^{(b)}_{3,2}, a^{(b)}_{3,3}, \eta^{(\frac{a}{b})}_2, \eta^{(a)}_0, f_3 }
 \nonumber \\
 & &
+9\,\iterint{ \eta^{(b)}_0, a^{(b)}_{3,2}, a^{(b)}_{3,3}, \eta^{(b)}_0, \eta^{(b)}_{3,5}, \omega_{0,4} }
-27\,\iterint{ \eta^{(b)}_0,a^{(b)}_{3,2},a^{(b)}_{3,3},\eta^{(b)}_0,a^{(b)}_{3,2},\omega_{0,4},\omega_{0,4} }
 \nonumber \\
 & &
-6\,\iterint{ \eta^{(b)}_0,a^{(b)}_{3,2},a^{(b)}_{3,3},\eta^{(b)}_0,a^{(a,b)}_{4,1},\eta^{(a)}_0,f_3 }
+6\,\iterint{ \eta^{(b)}_0, a^{(b)}_{3,2}, a^{(b)}_{3,1}, \eta^{(b)}_0, \eta^{(b)}_{3,5}, \omega_{0,4} }
 \nonumber \\
 & &
+12\,\iterint{ \eta^{(b)}_0, a^{(b)}_{3,2}, a^{(b)}_{3,1}, \eta^{(b)}_{1,1}, \omega_{0,4}, \omega_{0,4} }
-4\,\iterint{ \eta^{(b)}_0, a^{(b)}_{3,2}, a^{(b)}_{3,1}, \eta^{(\frac{a}{b})}_2, \eta^{(a)}_0, f_3 }
 \nonumber \\
 & &
-4\,\iterint{ \eta^{(b)}_0, a^{(b)}_{3,2}, a^{(b)}_{3,1}, \eta^{(b)}_0, \eta^{(\frac{b}{a})}_2, f_3 }
-18\,\iterint{ \eta^{(b)}_0,a^{(b)}_{3,2},a^{(b)}_{3,1},\eta^{(b)}_0,a^{(b)}_{3,2},\omega_{0,4},\omega_{0,4} }
 \nonumber \\
 & &
-4\,\iterint{ \eta^{(b)}_0,a^{(b)}_{3,2},a^{(b)}_{3,1},\eta^{(b)}_0,a^{(a,b)}_{4,1},\eta^{(a)}_0,f_3 }
 \nonumber \\
 & &
+ \left[
 \frac{7}{3}\,\iterint{ \eta^{(b)}_0, d_{2,4} }
-{\frac {91}{6}}\,\iterint{ \eta^{(b)}_0, \eta_{2,1} }
+{\frac {49}{12}}\,\iterint{ \eta^{(b)}_0, g_{2,1} }
-\frac{7}{2}\,\iterint{ \eta^{(b)}_0, \omega_0 }
 \right. \nonumber \\
 & & \left.
-{\frac {217}{24}}\,\iterint{ \eta^{(b)}_0, \omega_4 }
-{\frac {7}{36}}\,\iterint{ \eta^{(b)}_0, \eta_{2,10} }
-{\frac {7}{36}}\,\iterint{ \eta^{(b)}_0, \eta_{2,11} }
+\frac{3}{4}\,\iterint{ \eta^{(b)}_0, \eta^{(\frac{b}{c})}_2 }
 \right. \nonumber \\
 & & \left.
+\frac{1}{3}\,\iterint{ \eta^{(b)}_0, \eta^{(b)}_{3,10} }
+\frac{2}{3}\,\iterint{ \eta^{(b)}_0, \eta^{(b)}_{3,11} }
-4\,\iterint{ \eta^{(b)}_0, \eta^{(b)}_{3,1} }
+\iterint{ \eta^{(b)}_0, \eta^{(b)}_{3,24} }
 \right. \nonumber \\
 & & \left.
+\frac{4}{3}\,\iterint{ \eta^{(b)}_0, \eta^{(b)}_{3,9} }
+3\,\iterint{ \eta^{(b)}_0, \eta^{(a,b)}_{4,3} }
-{\frac {217}{24}}\,\iterint{ \eta^{(b)}_0, \omega_{0,4} }
+{\frac {7}{36}}\,\iterint{ \eta_{2,9}, \eta^{(b)}_0 }
 \right. \nonumber \\
 & & \left.
-\frac{2}{9}\,\iterint{ \eta_{2,9}, \eta^{(b)}_{1,1} }
+\frac{1}{3}\,\iterint{ \eta_{2,9}, \eta^{(\frac{a}{b})}_2 }
+{\frac {49}{12}}\,\iterint{ g_{2,1}, \eta^{(b)}_0 }
-\frac{14}{3}\,\iterint{ g_{2,1}, \eta^{(b)}_{1,1} }
 \right. \nonumber \\
 & & \left.
+7\,\iterint{ g_{2,1}, \eta^{(\frac{a}{b})}_2 }
+{\frac {7}{8}}\,\iterint{ \omega_0, \eta^{(b)}_0 }
-\iterint{ \omega_0, \eta^{(b)}_{1,1} }
+\frac{3}{2}\,\iterint{ \omega_0, \eta^{(\frac{a}{b})}_2 }
 \right. \nonumber \\
 & & \left.
+\frac{8}{3}\,\iterint{ \eta^{(b)}_{1,1}, d_{2,4} }
-{\frac {20}{3}}\,\iterint{ \eta^{(b)}_{1,1}, \eta_{2,1} }
-\frac{8}{3}\,\iterint{ \eta^{(b)}_{1,1}, g_{2,1} }
+\iterint{ \eta^{(b)}_{1,1}, \omega_0 }
 \right. \nonumber \\
 & & \left.
+\frac{1}{3}\,\iterint{ \eta^{(b)}_{1,1}, \omega_4 }
-{\frac {28}{9}}\,\iterint{ \eta^{(b)}_{1,1}, \eta^{(b)}_{1,2} }
+\frac{14}{3}\,\iterint{ \eta^{(b)}_{1,1}, \eta^{(b)}_{1,4} }
-\frac{2}{9}\,\iterint{ \eta^{(b)}_{1,1}, \eta^{(a)}_{3,1} }
 \right. \nonumber \\
 & & \left.
-\frac{1}{18}\,\iterint{ \eta^{(b)}_{1,1}, \eta^{(a)}_{3,2} }
+{\frac {16}{3}}\,\iterint{ \eta^{(b)}_{1,1}, \omega_{0,4} }
+4\,\iterint{ \eta^{(b)}_{1,1}, \omega_{-4,0} }
+{\frac {7}{36}}\,\iterint{ \eta_{2,10}, \eta^{(b)}_0 }
 \right. \nonumber \\
 & & \left.
-\frac{2}{9}\,\iterint{ \eta_{2,10}, \eta^{(b)}_{1,1} }
+\frac{1}{3}\,\iterint{ \eta_{2,10}, \eta^{(\frac{a}{b})}_2 }
+\frac{3}{4}\,\iterint{ \eta^{(\frac{c}{b})}_2, \eta^{(c)}_0 }
-4\,\iterint{ \eta^{(\frac{c}{b})}_2, \eta^{(c)}_{1,1} }
 \right. \nonumber \\
 & & \left.
-2\,\iterint{ \eta^{(\frac{c}{b})}_2, \eta^{(c)}_{1,3} }
+3\,\iterint{ \eta^{(\frac{c}{b})}_2, \eta^{(\frac{a}{c})}_2 }
+{\frac {35}{8}}\,\iterint{ \omega_{0,4}, \eta^{(b)}_0 }
-5\,\iterint{ \omega_{0,4}, \eta^{(b)}_{1,1} }
 \right. \nonumber \\
 & & \left.
+\frac{15}{2}\,\iterint{ \omega_{0,4}, \eta^{(\frac{a}{b})}_2 }
-4\,\iterint{ \eta^{(b)}_0, a^{(b)}_{3,2}, d_{2,4} }
+10\,\iterint{ \eta^{(b)}_0, a^{(b)}_{3,2}, \eta_{2,1} }
 \right. \nonumber \\
 & & \left.
+4\,\iterint{ \eta^{(b)}_0, a^{(b)}_{3,2}, g_{2,1} }
-\frac{3}{2}\,\iterint{ \eta^{(b)}_0, a^{(b)}_{3,2}, \omega_0 }
-\frac{1}{2}\,\iterint{ \eta^{(b)}_0, a^{(b)}_{3,2}, \omega_4 }
 \right. \nonumber \\
 & & \left.
+\frac{14}{3}\,\iterint{ \eta^{(b)}_0, a^{(b)}_{3,2}, \eta^{(b)}_{1,2} }
-7\,\iterint{ \eta^{(b)}_0, a^{(b)}_{3,2}, \eta^{(b)}_{1,4} }
+\frac{1}{3}\,\iterint{ \eta^{(b)}_0, a^{(b)}_{3,2}, \eta^{(a)}_{3,1} }
 \right. \nonumber \\
 & & \left.
+\frac{1}{12}\,\iterint{ \eta^{(b)}_0, a^{(b)}_{3,2}, \eta^{(a)}_{3,2} }
-8\,\iterint{ \eta^{(b)}_0, a^{(b)}_{3,2}, \omega_{0,4} }
-6\,\iterint{ \eta^{(b)}_0, a^{(b)}_{3,2}, \omega_{-4,0} }
 \right. \nonumber \\
 & & \left.
+4\,\iterint{ \eta^{(b)}_0, d_{2,4}, a^{(b)}_{3,2} }
+4\,\iterint{ \eta^{(b)}_0, d_{2,4}, a^{(a,b)}_{4,1} }
-26\,\iterint{ \eta^{(b)}_0, \eta_{2,1}, a^{(b)}_{3,2} }
 \right. \nonumber \\
 & & \left.
-26\,\iterint{ \eta^{(b)}_0, \eta_{2,1}, a^{(a,b)}_{4,1} }
+7\,\iterint{ \eta^{(b)}_0, g_{2,1}, a^{(b)}_{3,2} }
+7\,\iterint{ \eta^{(b)}_0, g_{2,1}, a^{(a,b)}_{4,1} }
 \right. \nonumber \\
 & & \left.
-6\,\iterint{ \eta^{(b)}_0, \omega_0, a^{(b)}_{3,2} }
-6\,\iterint{ \eta^{(b)}_0, \omega_0, a^{(a,b)}_{4,1} }
-{\frac {31}{2}}\,\iterint{ \eta^{(b)}_0, \omega_4, a^{(b)}_{3,2} }
 \right. \nonumber \\
 & & \left.
-{\frac {31}{2}}\,\iterint{ \eta^{(b)}_0, \omega_4, a^{(a,b)}_{4,1} }
+\frac{3}{4}\,\iterint{ \eta^{(b)}_0, a^{(b,c)}_{4,1}, \eta^{(c)}_0 }
-4\,\iterint{ \eta^{(b)}_0, a^{(b,c)}_{4,1}, \eta^{(c)}_{1,1} }
 \right. \nonumber \\
 & & \left.
-2\,\iterint{ \eta^{(b)}_0, a^{(b,c)}_{4,1}, \eta^{(c)}_{1,3} }
+3\,\iterint{ \eta^{(b)}_0, a^{(b,c)}_{4,1}, \eta^{(\frac{a}{c})}_2 }
+\frac{7}{4}\,\iterint{ \eta^{(b)}_0, a^{(b,b)}_{4,3}, \eta^{(b)}_0 }
 \right. \nonumber \\
 & & \left.
-2\,\iterint{ \eta^{(b)}_0, a^{(b,b)}_{4,3}, \eta^{(b)}_{1,1} }
+3\,\iterint{ \eta^{(b)}_0, a^{(b,b)}_{4,3}, \eta^{(\frac{a}{b})}_2 }
+\frac{7}{2}\,\iterint{ \eta^{(b)}_0, a^{(b,b)}_{4,4}, \eta^{(b)}_0 }
 \right. \nonumber \\
 & & \left.
-4\,\iterint{ \eta^{(b)}_0, a^{(b,b)}_{4,4}, \eta^{(b)}_{1,1} }
+6\,\iterint{ \eta^{(b)}_0, a^{(b,b)}_{4,4}, \eta^{(\frac{a}{b})}_2 }
-\frac{1}{3}\,\iterint{ \eta^{(b)}_0, \eta_{2,10}, a^{(b)}_{3,2} }
 \right. \nonumber \\
 & & \left.
-\frac{1}{3}\,\iterint{ \eta^{(b)}_0, \eta_{2,10}, a^{(a,b)}_{4,1} }
-\frac{1}{3}\,\iterint{ \eta^{(b)}_0, \eta_{2,11}, a^{(b)}_{3,2} }
-\frac{1}{3}\,\iterint{ \eta^{(b)}_0, \eta_{2,11}, a^{(a,b)}_{4,1} }
 \right. \nonumber \\
 & & \left.
+\frac{4}{3}\,\iterint{ \eta^{(b)}_0, \eta^{(\frac{b}{c})}_2, a^{(c)}_{3,2} }
+\iterint{ \eta^{(b)}_0, \eta^{(\frac{b}{c})}_2, a^{(c)}_{3,3} }
+3\,\iterint{ \eta^{(b)}_0, \eta^{(\frac{b}{c})}_2, a^{(a,c)}_{4,1} }
 \right. \nonumber \\
 & & \left.
-{\frac {31}{2}}\,\iterint{ \eta^{(b)}_0, \omega_{0,4}, a^{(b)}_{3,2} }
-{\frac {31}{2}}\,\iterint{ \eta^{(b)}_0, \omega_{0,4}, a^{(a,b)}_{4,1} }
+\frac{1}{3}\,\iterint{ \eta_{2,9}, \eta^{(b)}_0, a^{(b)}_{3,2} }
 \right. \nonumber \\
 & & \left.
+\frac{1}{3}\,\iterint{ \eta_{2,9}, \eta^{(b)}_0, a^{(a,b)}_{4,1} }
+7\,\iterint{ g_{2,1}, \eta^{(b)}_0, a^{(b)}_{3,2} }
+7\,\iterint{ g_{2,1}, \eta^{(b)}_0, a^{(a,b)}_{4,1} }
 \right. \nonumber \\
 & & \left.
+\frac{3}{2}\,\iterint{ \omega_0, \eta^{(b)}_0, a^{(b)}_{3,2} }
+\frac{3}{2}\,\iterint{ \omega_0, \eta^{(b)}_0, a^{(a,b)}_{4,1} }
+\frac{14}{3}\,\iterint{ \eta^{(b)}_{1,1}, a^{(b)}_{3,1}, \eta^{(b)}_0 }
 \right. \nonumber \\
 & & \left.
-\frac{16}{3}\,\iterint{ \eta^{(b)}_{1,1}, a^{(b)}_{3,1}, \eta^{(b)}_{1,1} }
+8\,\iterint{ \eta^{(b)}_{1,1}, a^{(b)}_{3,1}, \eta^{(\frac{a}{b})}_2 }
+7\,\iterint{ \eta^{(b)}_{1,1}, a^{(b)}_{3,3}, \eta^{(b)}_0 }
 \right. \nonumber \\
 & & \left.
-8\,\iterint{ \eta^{(b)}_{1,1}, a^{(b)}_{3,3}, \eta^{(b)}_{1,1} }
+12\,\iterint{ \eta^{(b)}_{1,1}, a^{(b)}_{3,3}, \eta^{(\frac{a}{b})}_2 }
-\frac{16}{3}\,\iterint{ \eta^{(b)}_{1,1}, \eta^{(b)}_{1,2}, a^{(b)}_{3,2} }
 \right. \nonumber \\
 & & \left.
-\frac{16}{3}\,\iterint{ \eta^{(b)}_{1,1}, \eta^{(b)}_{1,2}, a^{(a,b)}_{4,1} }
+8\,\iterint{ \eta^{(b)}_{1,1}, \eta^{(b)}_{1,4}, a^{(b)}_{3,2} }
+8\,\iterint{ \eta^{(b)}_{1,1}, \eta^{(b)}_{1,4}, a^{(a,b)}_{4,1} }
 \right. \nonumber \\
 & & \left.
+\frac{1}{3}\,\iterint{ \eta_{2,10}, \eta^{(b)}_0, a^{(b)}_{3,2} }
+\frac{1}{3}\,\iterint{ \eta_{2,10}, \eta^{(b)}_0, a^{(a,b)}_{4,1} }
+\frac{4}{3}\,\iterint{ \eta^{(\frac{c}{b})}_2, \eta^{(c)}_0, a^{(c)}_{3,2} }
 \right. \nonumber \\
 & & \left.
+\iterint{ \eta^{(\frac{c}{b})}_2, \eta^{(c)}_0, a^{(c)}_{3,3} }
+3\,\iterint{ \eta^{(\frac{c}{b})}_2, \eta^{(c)}_0, a^{(a,c)}_{4,1} }
+\frac{15}{2}\,\iterint{ \omega_{0,4}, \eta^{(b)}_0, a^{(b)}_{3,2} }
 \right. \nonumber \\
 & & \left.
+\frac{15}{2}\,\iterint{ \omega_{0,4}, \eta^{(b)}_0, a^{(a,b)}_{4,1} }
+8\,\iterint{ \eta^{(b)}_0, a^{(b)}_{3,2}, \eta^{(b)}_{1,2}, a^{(b)}_{3,2} }
+8\,\iterint{ \eta^{(b)}_0, a^{(b)}_{3,2}, \eta^{(b)}_{1,2}, a^{(a,b)}_{4,1} }
 \right. \nonumber \\
 & & \left.
-12\,\iterint{ \eta^{(b)}_0, a^{(b)}_{3,2}, \eta^{(b)}_{1,4}, a^{(b)}_{3,2} }
-12\,\iterint{ \eta^{(b)}_0, a^{(b)}_{3,2}, \eta^{(b)}_{1,4}, a^{(a,b)}_{4,1} }
 \right. \nonumber \\
 & & \left.
+\frac{4}{3}\,\iterint{ \eta^{(b)}_0, a^{(b,c)}_{4,1}, \eta^{(c)}_0, a^{(c)}_{3,2} }
+\iterint{ \eta^{(b)}_0, a^{(b,c)}_{4,1}, \eta^{(c)}_0, a^{(c)}_{3,3} }
 \right. \nonumber \\
 & & \left.
+3\,\iterint{ \eta^{(b)}_0, a^{(b,c)}_{4,1}, \eta^{(c)}_0, a^{(a,c)}_{4,1} }
+3\,\iterint{ \eta^{(b)}_0, a^{(b,b)}_{4,3}, \eta^{(b)}_0, a^{(b)}_{3,2} }
 \right. \nonumber \\
 & & \left.
+3\,\iterint{ \eta^{(b)}_0, a^{(b,b)}_{4,3}, \eta^{(b)}_0, a^{(a,b)}_{4,1} }
+6\,\iterint{ \eta^{(b)}_0, a^{(b,b)}_{4,4}, \eta^{(b)}_0, a^{(b)}_{3,2} }
 \right. \nonumber \\
 & & \left.
+6\,\iterint{ \eta^{(b)}_0, a^{(b,b)}_{4,4}, \eta^{(b)}_0, a^{(a,b)}_{4,1} }
+8\,\iterint{ \eta^{(b)}_{1,1}, a^{(b)}_{3,1}, \eta^{(b)}_0, a^{(b)}_{3,2} }
 \right. \nonumber \\
 & & \left.
+8\,\iterint{ \eta^{(b)}_{1,1}, a^{(b)}_{3,1}, \eta^{(b)}_0, a^{(a,b)}_{4,1} }
+12\,\iterint{ \eta^{(b)}_{1,1}, a^{(b)}_{3,3}, \eta^{(b)}_0, a^{(b)}_{3,2} }
 \right. \nonumber \\
 & & \left.
+12\,\iterint{ \eta^{(b)}_{1,1}, a^{(b)}_{3,3}, \eta^{(b)}_0, a^{(a,b)}_{4,1} }
-12\,\iterint{ \eta^{(b)}_0, a^{(b)}_{3,2}, a^{(b)}_{3,1}, \eta^{(\frac{a}{b})}_2 }
 \right. \nonumber \\
 & & \left.
-18\,\iterint{ \eta^{(b)}_0, a^{(b)}_{3,2}, a^{(b)}_{3,3}, \eta^{(\frac{a}{b})}_2 }
-18\,\iterint{ \eta^{(b)}_0, a^{(b)}_{3,2}, a^{(b)}_{3,3}, \eta^{(b)}_0, a^{(a,b)}_{4,1} }
 \right. \nonumber \\
 & & \left.
-18\,\iterint{ \eta^{(b)}_0, a^{(b)}_{3,2}, a^{(b)}_{3,3}, \eta^{(b)}_0, a^{(b)}_{3,2} }
-\frac{21}{2}\,\iterint{ \eta^{(b)}_0, a^{(b)}_{3,2}, a^{(b)}_{3,3}, \eta^{(b)}_0 }
 \right. \nonumber \\
 & & \left.
+12\,\iterint{ \eta^{(b)}_0, a^{(b)}_{3,2}, a^{(b)}_{3,3}, \eta^{(b)}_{1,1} }
-12\,\iterint{ \eta^{(b)}_0, a^{(b)}_{3,2}, a^{(b)}_{3,1}, \eta^{(b)}_0, a^{(a,b)}_{4,1} }
 \right. \nonumber \\
 & & \left.
-12\,\iterint{ \eta^{(b)}_0, a^{(b)}_{3,2}, a^{(b)}_{3,1}, \eta^{(b)}_0, a^{(b)}_{3,2} }
-7\,\iterint{ \eta^{(b)}_0, a^{(b)}_{3,2}, a^{(b)}_{3,1}, \eta^{(b)}_0 }
 \right. \nonumber \\
 & & \left.
+8\,\iterint{ \eta^{(b)}_0, a^{(b)}_{3,2}, a^{(b)}_{3,1}, \eta^{(b)}_{1,1} }
\right] \zeta_2
 \nonumber \\
 & &
+ \left[
  {\frac {49}{12}}\,\iterint{ \eta^{(b)}_0 }
        -{\frac {7}{36}}\,\iterint{ \eta_{2,9} }
        -{\frac {49}{12}}\,\iterint{ g_{2,1} }
        -{\frac {7}{8}}\,\iterint{ \omega_0 }
        +{\frac {22}{3}}\,\iterint{ \eta^{(b)}_{1,1} }
 \right. \nonumber \\
 & & \left.
        -{\frac {7}{36}}\,\iterint{ \eta_{2,10} }
        -{\frac {35}{8}}\,\iterint{ \omega_{0,4} }
        -11\,\iterint{ \eta^{(b)}_0, a^{(b)}_{3,2} }
        -\frac{7}{4}\,\iterint{ \eta^{(b)}_0, a^{(b,b)}_{4,3} }
 \right. \nonumber \\
 & & \left.
        -\frac{7}{2}\,\iterint{ \eta^{(b)}_0, a^{(b,b)}_{4,4} }
        -\frac{14}{3}\,\iterint{ \eta^{(b)}_{1,1}, a^{(b)}_{3,1} }
        -7\,\iterint{ \eta^{(b)}_{1,1}, a^{(b)}_{3,3} } 
        +7\,\iterint{ \eta^{(b)}_0, a^{(b)}_{3,2}, a^{(b)}_{3,1} }
 \right. \nonumber \\
 & & \left.
        +\frac{21}{2}\,\iterint{ \eta^{(b)}_0, a^{(b)}_{3,2}, a^{(b)}_{3,3} }
 \right] \zeta_3
 \nonumber \\
 & &
+ \left[
7\,\iterint{ \eta^{(b)}_0 }
-\frac{1}{3}\,\iterint{ \eta_{2,9} }
-7\,\iterint{ g_{2,1} }
-\frac{3}{2}\,\iterint{ \omega_0 }
-8\,\iterint{ \eta^{(b)}_{1,1} }
-\frac{1}{3}\,\iterint{ \eta_{2,10} }
 \right. \nonumber \\
 & & \left.
-\frac{15}{2}\,\iterint{ \omega_{0,4} }
+12\,\iterint{ \eta^{(b)}_0, a^{(b)}_{3,2} }
-3\,\iterint{ \eta^{(b)}_0, a^{(b,b)}_{4,3} }
-6\,\iterint{ \eta^{(b)}_0, a^{(b,b)}_{4,4} }
 \right. \nonumber \\
 & & \left.
-8\,\iterint{ \eta^{(b)}_{1,1}, a^{(b)}_{3,1} }
-12\,\iterint{ \eta^{(b)}_{1,1}, a^{(b)}_{3,3} }
 +18\,\iterint{ \eta^{(b)}_0, a^{(b)}_{3,2}, a^{(b)}_{3,3} }
 \right. \nonumber \\
 & & \left.
 +12\,\iterint{ \eta^{(b)}_0, a^{(b)}_{3,2}, a^{(b)}_{3,1} }
\right] \zeta_2 \ln  \left( 2 \right)
-{\frac {79}{4}}\,\zeta_4
+8\,\mathrm{Li}_4\left( \frac{1}{2} \right) 
-8\, \zeta_2 \left( \ln  \left( 2 \right)  \right) ^{2}
+\frac{1}{3}\, \left( \ln  \left( 2 \right)  \right) ^{4}.
\eq
The large number of terms for $J_{41}^{(4)}$ can be traced back to the already large number of terms of $J_{42}^{(3)}$ and $J_{43}^{(3)}$,
which are according to eq.~(\ref{J41_int_diff_eq}) integrated further with $\eta^{(b)}_{1,1}$ and $\eta^{(b)}_{0}$, respectively.
The expression in eq.~(\ref{longlonglong}) involves $47$ of the $107$ integration kernels.

\subsection{Numerical checks}
\label{sect:numerical_checks}

All results have been verified numerically with the help of the program \verb|sector_decomposition| \cite{Bogner:2007cr}.
\begin{table}[!htbp]
\begin{center}
\begin{tabular}{|l|lllll|}
 \hline 
 & $\eps^0$ & $\eps^1$ & $\eps^2$ & $\eps^3$ & $\eps^4$ \\
 \hline 
$J_{ 1}$ & $        1$ & $        0$ & $ 1.6449341$ & $-0.80137127$ & $ 1.8940657$ \\ 
$J_{ 2}$ & $        0$ & $ 2.8415816$ & $-2.8295758$ & $ 6.4116869$ & $-7.7009279$ \\ 
$J_{ 3}$ & $        0$ & $-2.8415816$ & $ 15.355894$ & $-39.817554$ & $ 97.903278$ \\ 
$J_{ 4}$ & $        0$ & $        0$ & $ 8.074586$ & $-20.479468$ & $ 55.140667$ \\ 
$J_{ 5}$ & $        1$ & $-2.7213737$ & $ 3.7029375$ & $-6.5645107$ & $ 7.7616443$ \\ 
$J_{ 6}$ & $        0$ & $        0$ & $ 4.7951687$ & $-7.7339091$ & $ 23.583241$ \\ 
$J_{ 7}$ & $        0$ & $-0.13797489$ & $ 5.0760627$ & $-8.5195954$ & $ 25.27333$ \\ 
$J_{ 8}$ & $        0$ & $        0$ & $ 9.8696044$ & $-15.803336$ & $ 48.383357$ \\ 
$J_{ 9}$ & $        0$ & $        0$ & $ 4.037293$ & $-2.1975902$ & $ 7.5750011$ \\ 
$J_{ 10}$ & $        0$ & $        0$ & $        0$ & $-10.577768$ & $ 19.861743$ \\ 
$J_{ 11}$ & $        0$ & $ 2.8415816$ & $-10.562581$ & $ 19.960005$ & $-34.628948$ \\ 
$J_{ 12}$ & $        0$ & $        0$ & $ 4.8094349$ & $-23.163298$ & $ 56.79741$ \\ 
$J_{ 13}$ & $        0$ & $        0$ & $        0$ & $-9.6340372$ & $ 18.255071$ \\ 
$J_{ 14}$ & $        0$ & $        0$ & $        0$ & $ 0.074587202$ & $-0.1198646$ \\ 
$J_{ 15}$ & $        0$ & $        0$ & $-4.037293$ & $ 23.437914$ & $-62.690651$ \\ 
$J_{ 16}$ & $        0$ & $        0$ & $        0$ & $ 8.4983135$ & $-20.922966$ \\ 
$J_{ 17}$ & $        0$ & $        0$ & $        0$ & $ 0.1491744$ & $ 0.058984085$ \\ 
$J_{ 18}$ & $        0$ & $ 1.4207908$ & $-12.163995$ & $ 44.930917$ & $-88.809767$ \\ 
$J_{ 19}$ & $        0$ & $        0$ & $        0$ & $ 16.996627$ & $-15.625817$ \\ 
$J_{ 20}$ & $        0$ & $        0$ & $-10.735443$ & $-9.8004674$ & $-37.795989$ \\ 
$J_{ 21}$ & $        0$ & $        0$ & $ 4.037293$ & $-13.184573$ & $ 21.864228$ \\ 
$J_{ 22}$ & $        0$ & $        0$ & $        0$ & $        0$ & $ 8.4599162$ \\ 
$J_{ 23}$ & $        0$ & $        0$ & $        0$ & $ 4.8796692$ & $-25.793413$ \\ 
$J_{ 24}$ & $        0$ & $        0$ & $        0$ & $ 2.6138189$ & $-0.23796592$ \\ 
$J_{ 25}$ & $        0$ & $        0$ & $ 4.037293$ & $-9.2635254$ & $ 25.950914$ \\ 
$J_{ 26}$ & $        0$ & $        0$ & $ 2.7276656$ & $-1.848663$ & $ 13.397014$ \\ 
$J_{ 27}$ & $        0$ & $        0$ & $ 0$ & $ 5.2276379$ & $ 8.7055971$ \\ 
$J_{ 28}$ & $        0$ & $        0$ & $ 0$ & $ 8.9388561$ & $ 12.795847$ \\ 
$J_{ 29}$ & $        0$ & $        0$ & $        0$ & $ 0$ & $ 18.80581$ \\ 
$J_{ 30}$ & $        0$ & $ 0$ & $ 5.4553312$ & $-3.9355497$ & $ 35.856907$ \\ 
$J_{ 32}$ & $        0$ & $        0$ & $-10.735443$ & $-40.306104$ & $-35.268067$ \\ 
$J_{ 33}$ & $        0$ & $        0$ & $        0$ & $ 7.3822471$ & $ 10.116064$ \\ 
$J_{ 34}$ & $        0$ & $        0$ & $ 10.735443$ & $-3.4643927$ & $ 53.616756$ \\ 
$J_{ 35}$ & $        0$ & $        0$ & $ 0.97741243$ & $ 5.1104476$ & $ 15.424638$ \\ 
 \hline 
\end{tabular}
\end{center}
\caption{
Numerical results for the first five terms of the $\eps$-expansion of the master integrals $J_{1}$-$J_{35}$ at the kinematic point $s=-\frac{12769}{840} m^2$, $t=\frac{10}{11} m^2$.
}
\label{table_numerical_results_I}
\end{table}
The program \verb|sector_decomposition| allows 
(as {\tt SecDec} \cite{Carter:2010hi,Borowka:2012yc,Borowka:2015mxa,Borowka:2017idc} or {\tt FIESTA} \cite{Smirnov:2008py,Smirnov:2009pb}) the numerical evaluation of multi-loop integrals.
On the one hand we evaluated all master integrals of the basis $\vec{I}$ at a few kinematic points numerically with the program
\verb|sector_decom-| \verb|position|.
On the other hand, we evaluated our results in the basis $\vec{J}$ at the same kinematic points, converted to the basis $\vec{I}$
and compared the two results.
We find good agreement.

The evaluation of the iterated integrals appearing in our results is done as follows:
We split the integration path into two pieces: 
First we integrate in $(\tilde{x},\tilde{y})$-space from $(0,0)$ to $(\tilde{x},0)$, then from 
$(\tilde{x},0)$ to $(\tilde{x},\tilde{y})$.
The integration along the first part gives only multiple polylogarithms, which can be evaluated to high precision \cite{Vollinga:2004sn}.
We use these results as new boundary constants for the integration along the second part.
Assuming that $\tilde{y}$ is small, we may expand for the integration along the second part all integration kernels in $\tilde{y}$.

As a reference we give numerical results for the master integrals in the basis $\vec{J}$ at the kinematic point
\bq
 s \; = \; -\frac{12769}{840} m^2,  & & t \; = \; \frac{10}{11} m^2.
\eq
This point corresponds to
\bq
 x \; = \; \frac{7}{120}, & & y \; = \; \frac{10}{11},
\eq
or equivalently
\bq
 \tilde{x} \; = \; \frac{1}{15}, & & \tilde{y} \; = \; \frac{1}{11}.
\eq
\begin{table}
\begin{center}
\begin{tabular}{|l|lllll|}
 \hline 
 & $\eps^0$ & $\eps^1$ & $\eps^2$ & $\eps^3$ & $\eps^4$ \\
 \hline 
$J_{ 36}$ & $        0$ & $        0$ & $        0$ & $        0$ & $-13.214347$ \\ 
$J_{ 37}$ & $        0$ & $        0$ & $        0$ & $        0$ & $-43.342128$ \\ 
$J_{ 38}$ & $        0$ & $        0$ & $        0$ & $        0$ & $ 44.194787$ \\ 
$J_{ 39}$ & $        0$ & $        0$ & $        0$ & $ 0$ & $ 0.28609557$ \\ 
$J_{ 40}$ & $        0$ & $        0$ & $        0$ & $        0$ & $-26.330837$ \\ 
$J_{ 41}$ & $        0$ & $        0$ & $ 0$ & $ 0$ & $ 11.147258$ \\ 
$J_{ 42}$ & $        0$ & $        0$ & $ 0$ & $-19.021429$ & $-320.23817$ \\ 
$J_{ 43}$ & $        0$ & $ 0$ & $ 0$ & $ 0.89070327$ & $ 9.183764$ \\ 
$J_{ 44}$ & $        0$ & $        0$ & $ 0$ & $ 0$ & $ 21.040337$ \\ 
$J_{ 45}$ & $        0$ & $        0$ & $        0$ & $ 0$ & $-1.4008206$ \\ 
 \hline 
\end{tabular}
\end{center}
\caption{
Numerical results for the first five terms of the $\eps$-expansion of the master integrals $J_{36}$-$J_{45}$ at the kinematic point $s=-\frac{12769}{840} m^2$, $t=\frac{10}{11} m^2$.
}
\label{table_numerical_results_II}
\end{table}
The numerical results for the first five terms of $\eps$-expansion are given in tables~\ref{table_numerical_results_I}
and \ref{table_numerical_results_II}.


\FloatBarrier

\section{Conclusions}
\label{sect:conclusions}

In this article we gave a detailed account on the analytic calculation of the 
master integrals for the planar double box integral relevant to top-pair production with a closed top loop.
The planar double box integral depends on two scales and involves several elliptic sub-sectors.
We showed that the associated elliptic curves can be extracted from the maximal cuts.
We demonstrated that the system involves three inequivalent elliptic curves.
To the best of our knowledge, this is the first time an integral involving more than one elliptic curve has been
calculated.
Elliptic polylogarithms are by definition iterated integrals on a single elliptic curve.
Since the system discussed in this article involves three elliptic curves, we do not expect that our results
are naturally expressible in terms of elliptic polylogarithms.
We showed that the system of differential equations can be transformed to a form linear in $\eps$,
where the $\eps^0$-term is strictly lower-triangular.
This system of differential equations is easily solved to any desired order in $\eps$.
We expressed our results in terms of iterated integrals and discussed the occurring integration kernels.
We believe that the techniques used in this paper are applicable to a wider class of Feynman integrals.

\subsection*{Acknowledgements}

We would like to thank Andreas von Manteuffel for advice on {\tt Reduze}.
L.A. and E.C. are grateful for financial support from the research training group GRK 1581.
S.W. would like to thank the Hausdorff Research Institute for Mathematics in Bonn
for hospitality, where part of this work was carried out.
The Feynman diagrams in this article have been made with the program {\tt axodraw} \cite{Vermaseren:1994je}.

\newpage

\begin{appendix}

\section{Master topologies}
\label{sect:master_topologies}

\begin{figure}[!hbp]
\begin{center}
\includegraphics[scale=1.0]{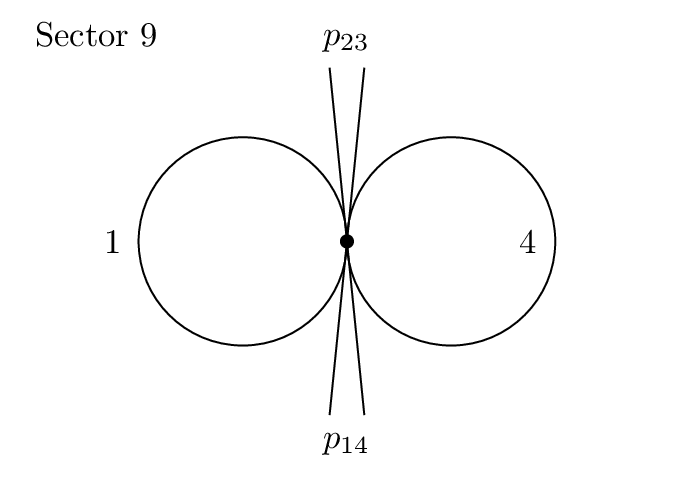}
\includegraphics[scale=1.0]{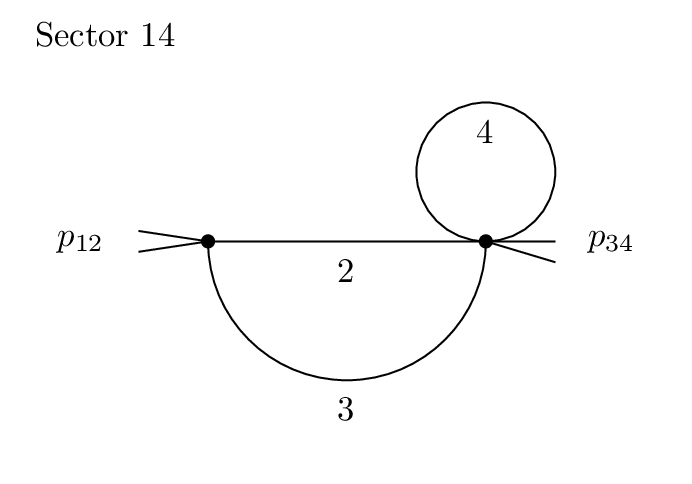}
\includegraphics[scale=1.0]{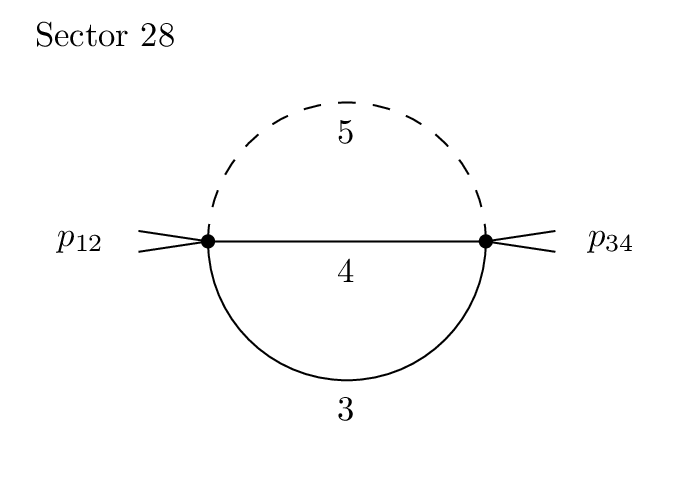}
\includegraphics[scale=1.0]{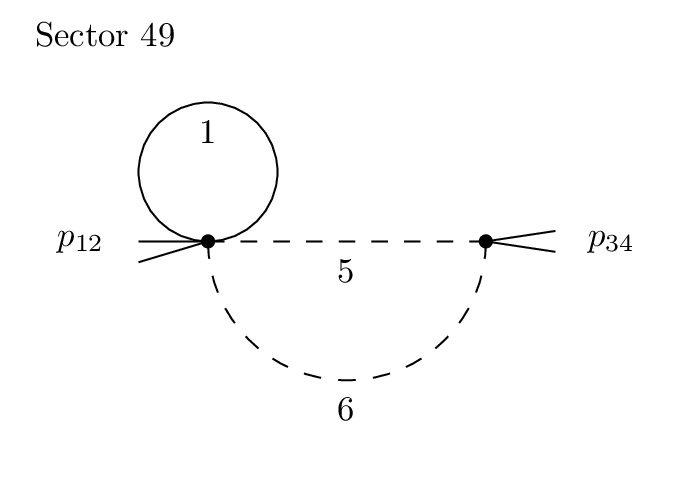}
\includegraphics[scale=1.0]{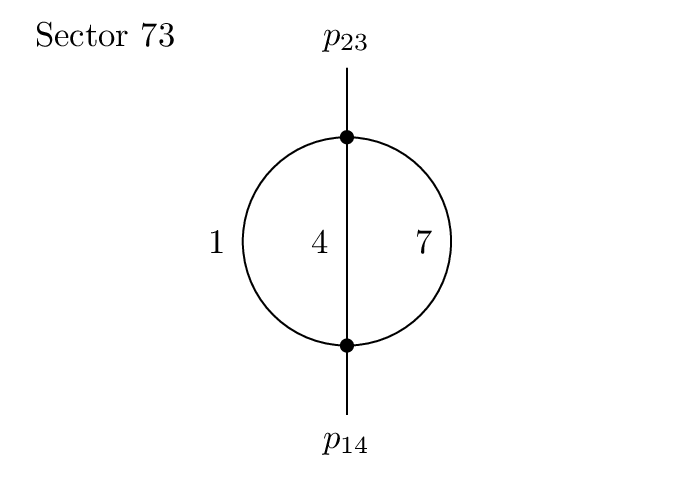}
\includegraphics[scale=1.0]{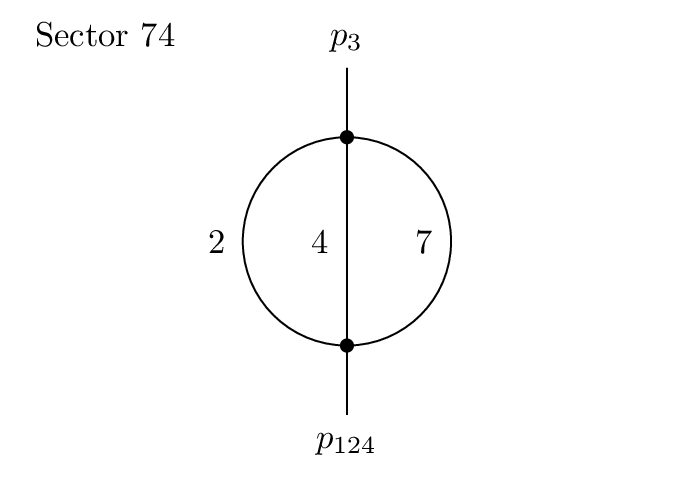}
\end{center}
\caption{
Master topologies (part 1).
}
\label{fig_master_topologies_part1}
\end{figure}

\begin{figure}
\begin{center}
\includegraphics[scale=1.0]{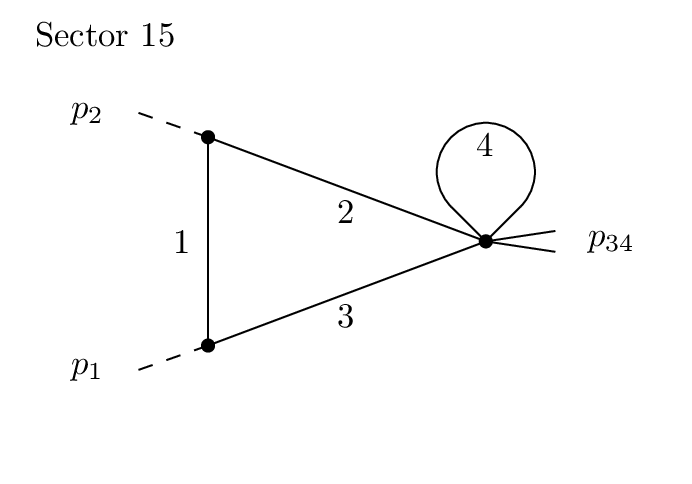}
\includegraphics[scale=1.0]{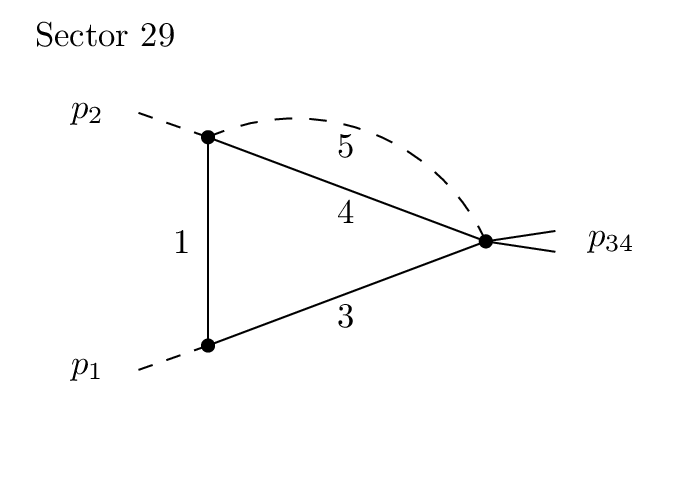}
\includegraphics[scale=1.0]{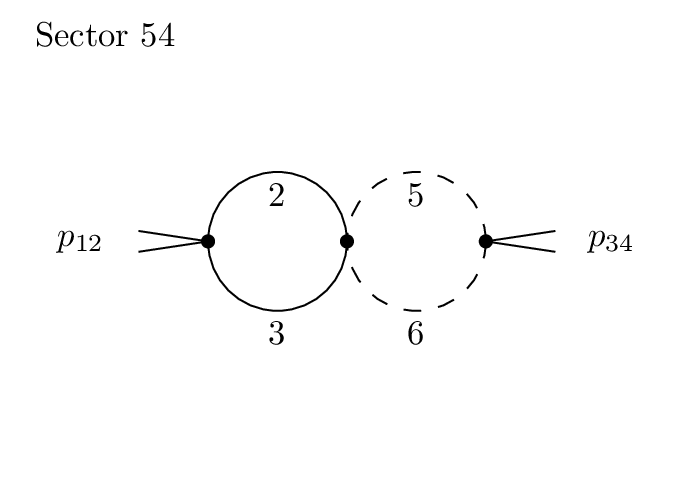}
\includegraphics[scale=1.0]{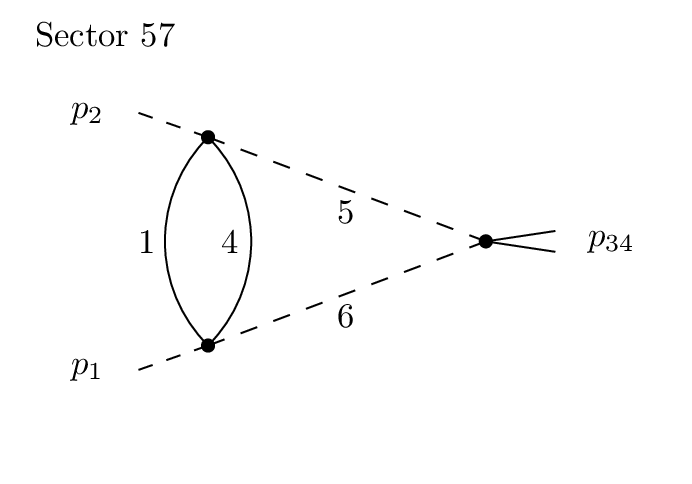}
\includegraphics[scale=1.0]{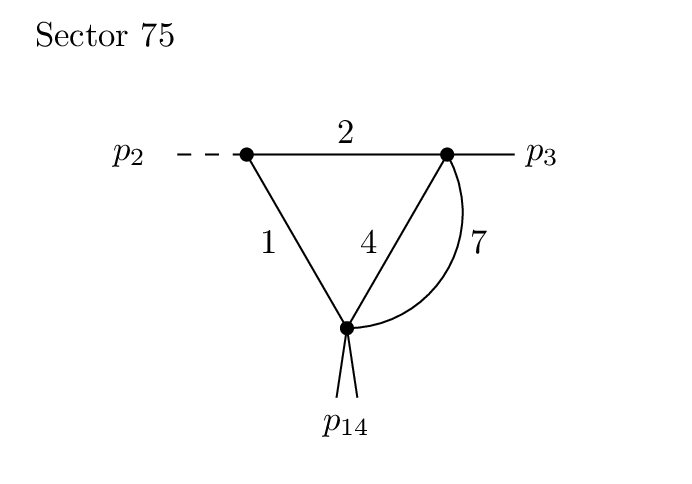}
\includegraphics[scale=1.0]{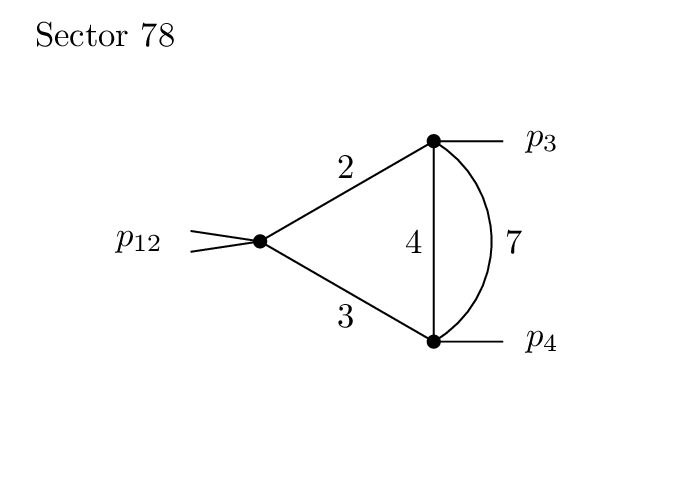}
\includegraphics[scale=1.0]{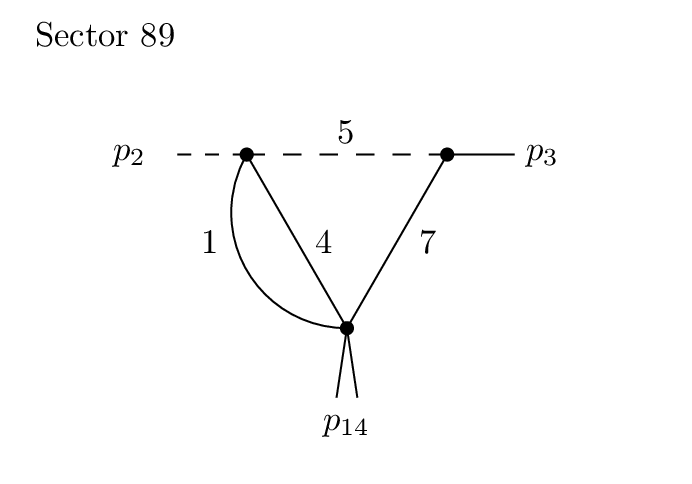}
\includegraphics[scale=1.0]{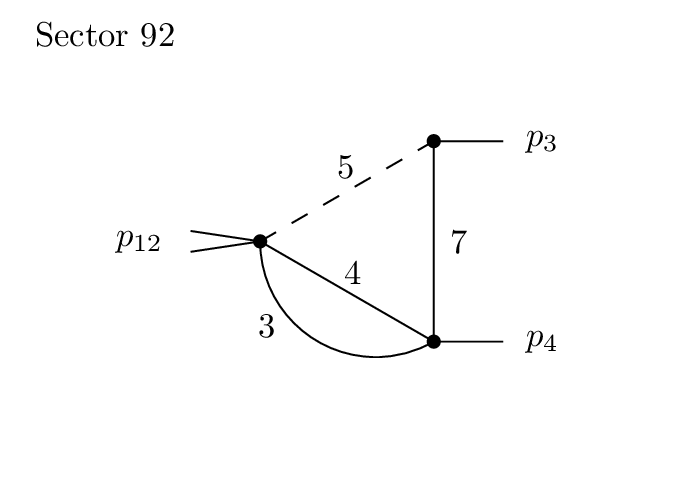}
\end{center}
\caption{
Master topologies (part 2).
}
\label{fig_master_topologies_part2}
\end{figure}

\begin{figure}
\begin{center}
\includegraphics[scale=1.0]{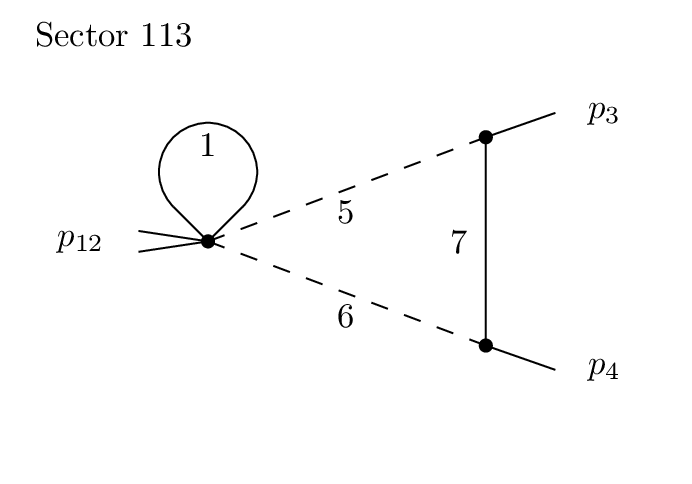}
\includegraphics[scale=1.0]{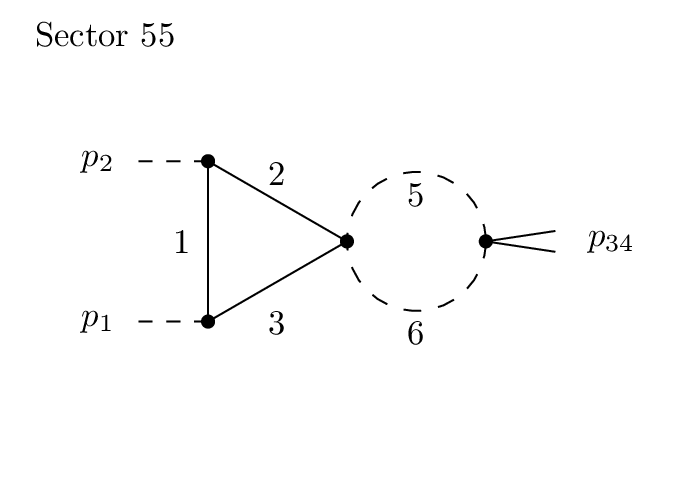}
\includegraphics[scale=1.0]{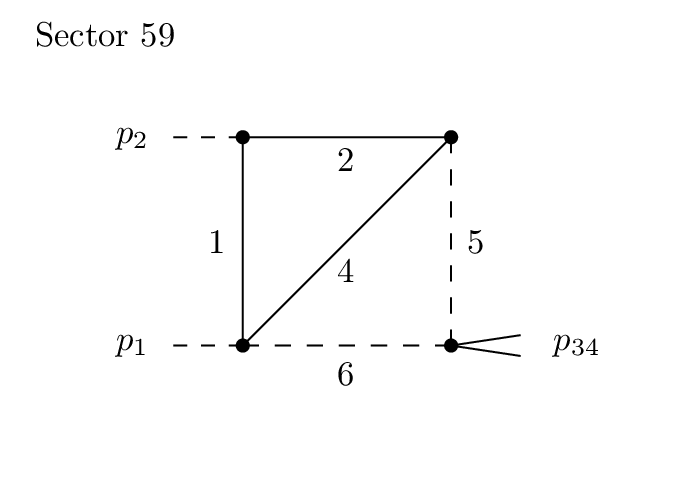}
\includegraphics[scale=1.0]{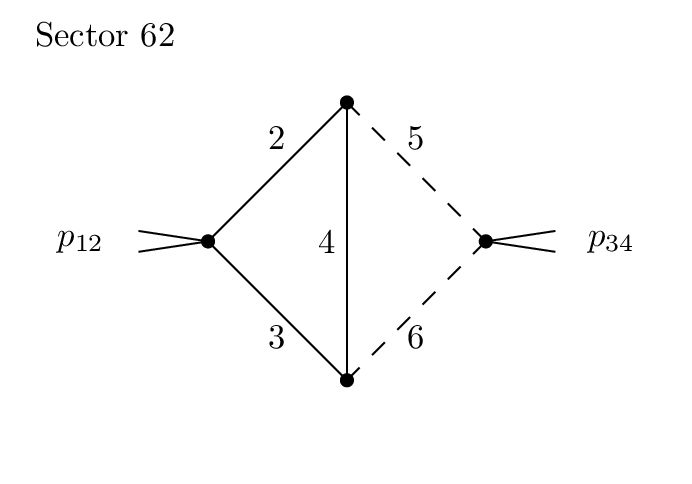}
\includegraphics[scale=1.0]{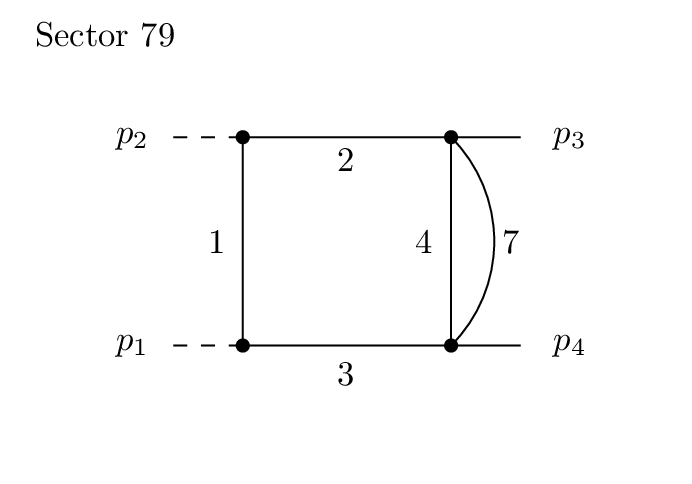}
\includegraphics[scale=1.0]{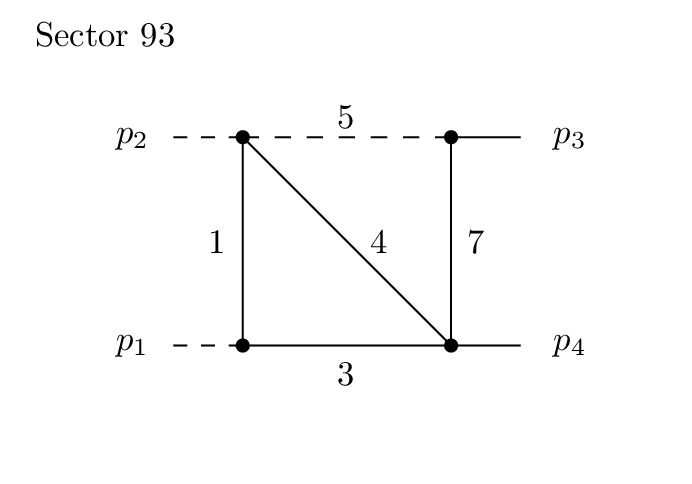}
\includegraphics[scale=1.0]{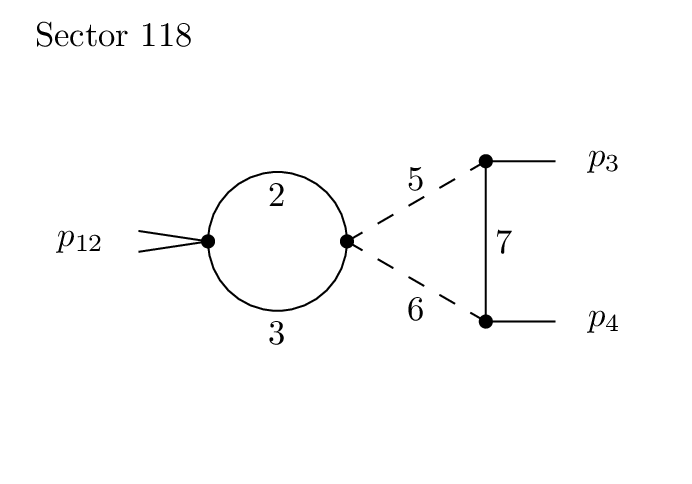}
\includegraphics[scale=1.0]{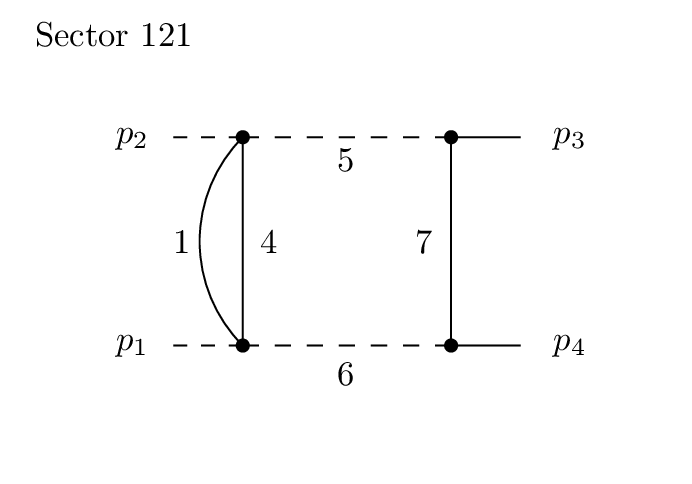}
\end{center}
\caption{
Master topologies (part 3).
}
\label{fig_master_topologies_part3}
\end{figure}

\begin{figure}
\begin{center}
\includegraphics[scale=1.0]{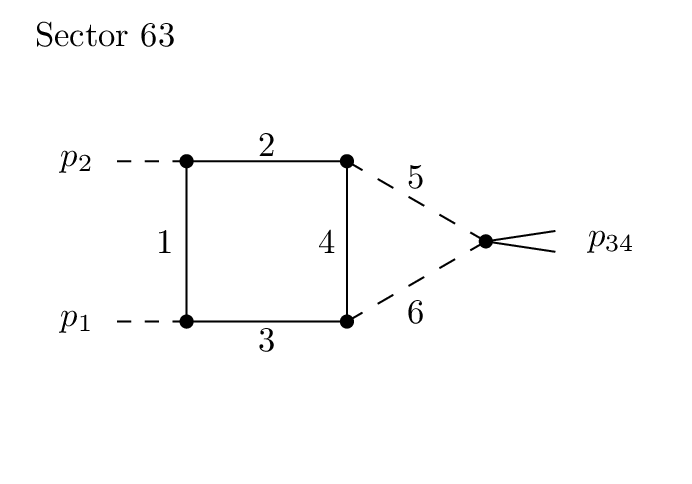}
\includegraphics[scale=1.0]{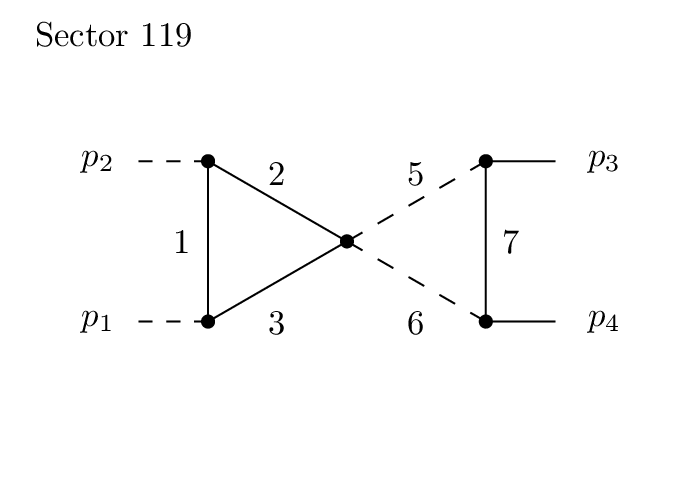}
\includegraphics[scale=1.0]{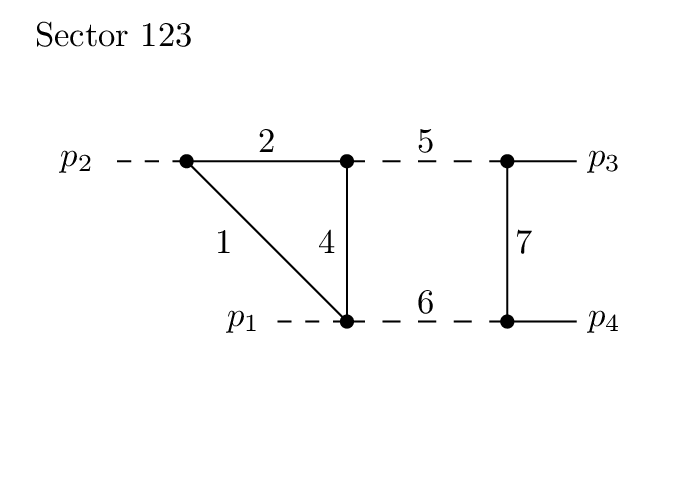}
\includegraphics[scale=1.0]{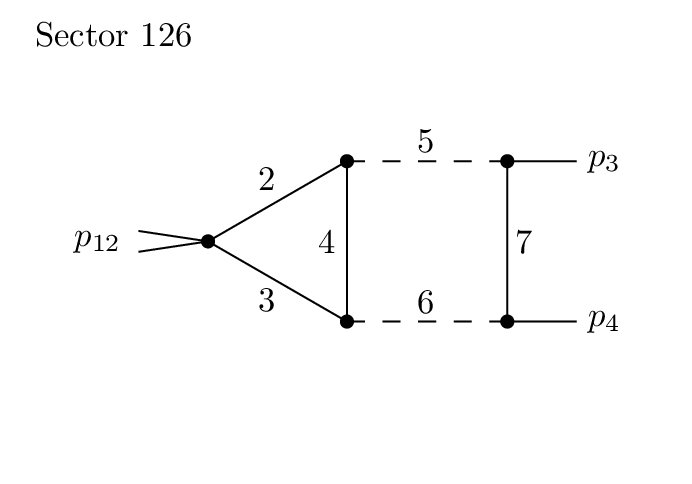}
\includegraphics[scale=1.0]{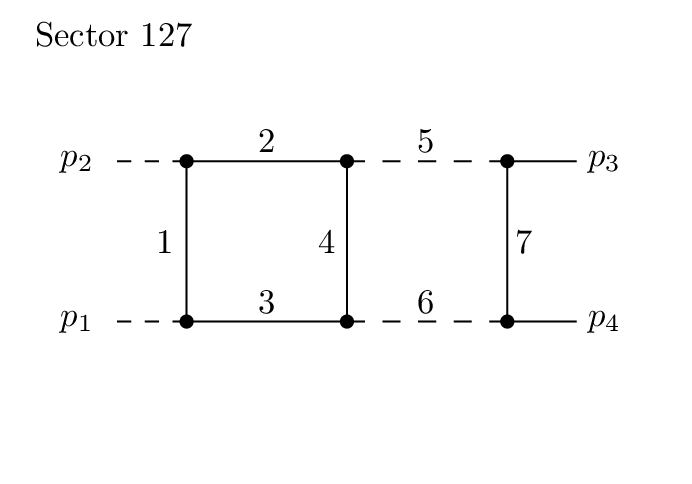}
\end{center}
\caption{
Master topologies (part 4).
}
\label{fig_master_topologies_part4}
\end{figure}

In this appendix we show diagrams of all master topologies.
In total there $27$ master topologies.
A master topology may contain several master integrals.
The total number of master integrals is $44$.
The number of master integrals within a master topology can be inferred from table~(\ref{table_master_integrals}).
The diagrams for the master topologies are shown in figs.~(\ref{fig_master_topologies_part1})-(\ref{fig_master_topologies_part4}).


\FloatBarrier

\section{The extra relation}
\label{sect:extra_relation}

In this appendix we give the relation which can be used to eliminate in sector $93$ one of the five integrals
$I_{1011101}$, $I_{2011101}$, $I_{1021101}$, $I_{1012101}$, $I_{1011201}$. The relation reads
\bq
\label{extra_relation}
 & &
 48\, 
 \left( D-3 \right)  \left( D-4 \right) ^{2} \left( D-5 \right)  \left( 2\,D-9 \right)  
 \left( 4\,{m}^{2}-s \right) ^{2}\left( {m}^{2}-t \right) ^{2} \left( {m}^{2}-s-t \right) {m}^{6} I_{1011101}
 \nonumber \\
 & &
 + 32\, 
 \left( D-3 \right)  \left( D-4 \right)\left( D-5 \right)  \left( 2\,D-9 \right)    
 \left( 4\,{m}^{2}-s \right) ^{2}\left( {m}^{2}-t \right) ^{2} \left( {m}^{2}-s-t \right) {m}^{8} I_{2011101}
 \nonumber \\
 & &
 + 32\, 
 \left( D-3 \right)  \left( D-4 \right)  \left( D-5 \right)  \left( 2\,D-9 \right)  
 \left( 4\,{m}^{2}-s \right) ^{2} \left( {m}^{2}-t \right) ^{2} \left( {m}^{2}-s-t \right) {m}^{8} I_{1021101}
 \nonumber \\
 & &
 + 16\, 
 \left( D-3 \right)  \left( D-4 \right)\left( D-5 \right)  \left( 2\,D-9 \right)    
 \left( 4\,{m}^{2}-s \right) ^{2} \left( {m}^{2}-t \right) ^{2} \left( 2\,{m}^{4}-2\,{m}^{2}t-3\,{m}^{2}s+st \right) {m}^{6} 
 \nonumber \\
 & & 
 I_{1012101}
 \nonumber \\
 & &
 + 64\, 
 \left( D-3 \right)  \left( D-4 \right) \left( D-5 \right)  \left( 2\,D-9 \right)    
 \left( 4\,{m}^{2}-s \right) ^{2} \left( {m}^{2}-t \right)  {s}^{2} {m}^{8} I_{1011201}
 \nonumber \\
 & &
 + 32\, 
  \left( D-3 \right)  \left( D-4 \right) \left( 2\,D-9 \right)    
 \left( 4\,{m}^{2}-s \right) ^{2} \left( {m}^{2}-t \right) \left[ -4\,s-6\,{m}^{2}+ \left( s+2\,{m}^{2} \right)  D \right] 
 {m}^{6}s I_{0021101}
 \nonumber \\
 & &
 + 16\, 
 \left( D-3 \right) ^{2} \left( D-4 \right) \left( 2\,D-9 \right) \left( 3\,D-10 \right)    
 \left( 4\,{m}^{2}-s \right) ^{2} \left( {m}^{2}-t \right)   
 s {m}^{6} I_{0011101}
 \nonumber \\
 & &
 -6\, 
 \left( D-3 \right)  \left( D-4 \right) ^{3} \left( 3\,D-10 \right) 
 \left( 4\,{m}^{2}-s \right) ^{2} \left( {m}^{2}-t \right) ^{3} s {m}^{2} I_{1001101}
 \nonumber \\
 & &
 + 4\, 
 \left( D-2 \right)  \left( D-3 \right)  \left( D-4 \right) \left( 2\,D-9 \right)    
 \left( 4\,{m}^{2}-s \right) ^{2} \left( {m}^{2}-t \right) ^{2} 
 \left[ 15\,t-15\,{m}^{2}+8\,s 
 \right. \nonumber \\
 & & \left.
 + \left( 3\,{m}^{2} -3\,t-2\,s \right)  D \right] {m}^{4} I_{1101001}
 \nonumber \\
 & &
 -8\, 
 \left( D-2 \right) \left( D-3 \right)  \left( D-4 \right) \left( D-5 \right)  \left( 2\,D-9 \right)    
 \left( 4\,{m}^{2}-s \right) ^{2} \left( {m}^{2}-t \right) ^{2} s {m}^{4} I_{1011100}
 \nonumber \\
 & &
 + 2\, 
 \left( D-3 \right)  \left( D-4 \right)  \left( 2\,D-9 \right)  \left( 3\,D-8 \right) 
 \left( 4\,{m}^{2}-s \right) ^{2}{m}^{4} \left( {m}^{2}-t \right)  
 \left[ 15\,t-15\,{m}^{2}+8\,s
 \right. \nonumber \\
 & & \left.
 + \left( -3\,t-2\,s+3\,{m}^{2} \right)  D  \right]  I_{0101001}
 \nonumber \\
 & &
 -4\, 
 \left( D-3 \right)  \left( D-4 \right)  
 \left( 4\,{m}^{2}-s \right) ^{2} 
 \left[
         -432\,{m}^{6}s-1890\,{m}^{6}t-270\,{m}^{2}{t}^{3}+810\,{m}^{8}+1350\,{m}^{4}{t}^{2}
 \right. \nonumber \\
 & & \left.
         +48\,s{t}^{3}+192\,{m}^{4}st-288\,{m}^{2}s{t}^{2}
         + \left( 798\,{m}^{6}t-24\,s{t}^{3}-92\,{m}^{4}st+114\,{m}^{2}{t}^{3}-570\,{m}^{4}{t}^{2}
 \right. \right. \nonumber \\
 & & \left. \left.
                  -342\,{m}^{8}
                  +156\,{m}^{2}s{t}^{2}+216\,{m}^{6}s \right)  D
         + \left( 3\,s{t}^{3}-12\,{m}^{2}{t}^{3}-84\,{m}^{6}t-27\,{m}^{6}s+60\,{m}^{4}{t}^{2}
 \right. \right. \nonumber \\
 & & \left. \left.
                  -21\,{m}^{2}s{t}^{2}+36\,{m}^{8}
                  +13\,{m}^{4}st \right) {D}^{2} \right] {m}^{2} I_{2001001}
 \nonumber \\
 & &
 -2\, 
 \left( D-3 \right)  \left( D-4 \right) \left( 3\,D-8 \right)    
 \left( 4\,{m}^{2}-s \right) ^{2} 
 \left[ 48\,s{t}^{2}-540\,{m}^{4}t+270\,{m}^{6}-144\,{m}^{4}s-24\,{m}^{2}st
 \right. \nonumber \\
 & & \left.
        +270\,{m}^{2}{t}^{2}
        + \left( 228\,{m}^{4}t+16\,{m}^{2}st-114\,{m}^{2}{t}^{2}+72\,{m}^{4}s-114\,{m}^{6}-24\,s{t}^{2} \right) D 
        + \left( 3\,s{t}^{2}
 \right. \right. \nonumber \\
 & & \left. \left.
        +12\,{m}^{2}{t}^{2}-9\,{m}^{4}s+12\,{m}^{6}-2\,{m}^{2}st-24\,{m}^{4}t \right) {D}^{2} \right] 
 {m}^{2} I_{1001001}
 \nonumber \\
 & &
 -16\, 
 \left( D-3 \right) \left( 2\,D-9 \right)  \left( {m}^{2}-t \right)   
 \left[ -140\,{m}^{2}{s}^{2}t+520\,{m}^{4}st+1120\,{m}^{8}+28\,{m}^{2}{s}^{3}-856\,{m}^{6}s
 \right. \nonumber \\
 & & \left.
        -1120\,{m}^{6}t+20\,t{s}^{3}+428\,{m}^{4}{s}^{2}
        + \left( -19\,{m}^{2}{s}^{3}-224\,{m}^{4}st+432\,{m}^{6}s+58\,{m}^{2}{s}^{2}t-238\,{m}^{4}{s}^{2}
 \right. \right. \nonumber \\
 & & \left. \left.
                 +544\,{m}^{6}t-544\,{m}^{8}-9\,t{s}^{3} \right)  D 
        + \left( 24\,{m}^{4}st-56\,{m}^{6}s-6\,{m}^{2}{s}^{2}t+3\,{m}^{2}{s}^{3}-64\,{m}^{6}t+64\,{m}^{8}
 \right. \right. \nonumber \\
 & & \left. \left. +t{s}^{3}
                 +34\,{m}^{4}{s}^{2} \right) {D}^{2}
 \right] {m}^{4} I_{0021100}
 \nonumber \\
 & &
 -8\, 
 \left( D-3 \right) \left( 2\,D-9 \right)  \left( 3\,D-8 \right)  
 \left( {m}^{2}-t \right)  
 \left[ 128\,{m}^{2}{s}^{2}+560\,{m}^{6}+220\,{m}^{2}st-560\,{m}^{4}t-20\,{s}^{2}t
 \right. \nonumber \\
 & & \left.
        -388\,{m}^{4}s
        + \left( -75\,{m}^{2}{s}^{2}-272\,{m}^{6}+9\,{s}^{2}t+208\,{m}^{4}s-104\,{m}^{2}st+272\,{m}^{4}t \right)  D 
        + \left( 12\,{m}^{2}st
 \right. \right. \nonumber \\
 & & \left. \left.
        -{s}^{2}t+11\,{m}^{2}{s}^{2}-28\,{m}^{4}s-32\,{m}^{4}t+32\,{m}^{6} \right) {D}^{2} 
 \right] {m}^{4} I_{0011100}
 \nonumber \\
 & &
 - \left( D-2 \right) ^{2}
 \left[ 
  -768\,{s}^{3}{t}^{2}+57600\,{m}^{8}t-28800\,{m}^{6}{t}^{2}+20304\,{m}^{8}s-5592\,{m}^{4}{s}^{2}t+288\,{m}^{4}{s}^{3}
 \right. \nonumber \\
 & & \left.
  +4272\,{m}^{4}s{t}^{2}-16896\,{m}^{6}st-28800\,{m}^{10}+3804\,{m}^{2}{s}^{2}{t}^{2}-2052\,{m}^{6}{s}^{2}+960\,t{m}^{2}{s}^{3}
 \right. \nonumber \\
 & & \left.
  + \left( 20080\,{m}^{10}+20080\,{m}^{6}{t}^{2}-14776\,{m}^{8}s+4970\,{m}^{4}{s}^{2}t-224\,{m}^{4}{s}^{3}+1367\,{m}^{6}{s}^{2}
 \right. \right. \nonumber \\
 & & \left. \left.
           -1688\,{m}^{4}s{t}^{2}-40160\,{m}^{8}t-3329\,{m}^{2}{s}^{2}{t}^{2}+624\,{s}^{3}{t}^{2}-776\,t{m}^{2}{s}^{3}
           +10448\,{m}^{6}st \right)  D 
 \right. \nonumber \\
 & & \left.
  + \left( 3576\,{m}^{8}s+9248\,{m}^{8}t-2064\,{m}^{6}st+955\,{m}^{2}{s}^{2}{t}^{2}-301\,{m}^{6}{s}^{2}+206\,t{m}^{2}{s}^{3}
           +58\,{m}^{4}{s}^{3}
 \right. \right. \nonumber \\
 & & \left. \left.
           +24\,{m}^{4}s{t}^{2}-4624\,{m}^{6}{t}^{2}-1422\,{m}^{4}{s}^{2}t-4624\,{m}^{10}-168\,{s}^{3}{t}^{2} \right) {D}^{2} 
  + \left( -288\,{m}^{8}s+22\,{m}^{6}{s}^{2}
 \right. \right. \nonumber \\
 & & \left. \left.
           +15\,{s}^{3}{t}^{2}+352\,{m}^{6}{t}^{2}-704\,{m}^{8}t-5\,{m}^{4}{s}^{3}+32\,{m}^{4}s{t}^{2}
           -18\,t{m}^{2}{s}^{3}-90\,{m}^{2}{s}^{2}{t}^{2}+132\,{m}^{4}{s}^{2}t
 \right. \right. \nonumber \\
 & & \left. \left.
           +128\,{m}^{6}st+352\,{m}^{10} \right) {D}^{3}
  \right]
 I_{1001000}
 \;\; = \;\; 0.
\eq


\section{The modular forms relevant to the elliptic curve $E^{\curveone}$}
\label{sect:modular_forms}

In this appendix we write for simplicity $\psi_1$ for $\psi_1^{\curveone}$ and $\tau_6$ for $\tau_6^{\curveone}$.
We further write
\bq
 r_n & = & \exp\left(\frac{2\pi i}{n}\right)
\eq
for the $n$-th root of unity.

Let us first introduce a set of modular forms for the congruence subgroup $\Gamma_1(6)$, 
which was already encountered in the calculation of the sunrise / kite system \cite{Adams:2017ejb,Adams:2018yfj}.
We consider the set
\bq
\label{mod_forms_sunrise_kite}
 \left\{ \frac{\psi_1}{\pi}, f_1, f_2, f_3, g_{2,0}, g_{2,1}, g_{2,9} \right\}.
\eq
$\psi_1/\pi$ is a modular form of modular weight $1$, $f_i$ is of modular weight $i$ and the modular weight of $g_{2,j}$ is $2$.
In this article we use as boundary point $(x,y)=(0,1)$ (or equivalently $(s,t)=(\infty,m^2)$).
Therefore we choose our periods such that $y=1$ (or $t=m^2$) for the curve $E^{\curveone}$ corresponds to $\tau_6=i \infty$ and $q_6=0$.
This corresponds to a $q_6$-expansion around the cusp $t=m^2$.
The Hauptmodul is given by
\bq
 y-1 & = &
 -8 \frac{\eta\left(\tau_6\right)^3 \eta\left(6\tau_6\right)^9}{\eta\left(2\tau_6\right)^3 \eta\left(3\tau_6\right)^9}.
\eq
For the modular forms from the set~(\ref{mod_forms_sunrise_kite}) we give
a representation in the form of an eta-quotient (where such a representation is known), 
a representation in the form of a polynomial in the two generators $e_1$ and $e_2$ of the Eisenstein subspace ${\mathcal E}_1(\Gamma_1(6))$,
a representation in terms of generalised Eisenstein series and
a representation in terms of $\overline{\mathrm{E}}_{n;m}$-functions.
We have
\bq
 \frac{\psi_1}{\pi} 
 & = &
 \frac{1}{2} \frac{\eta\left(2 \tau_6\right) \eta\left(3\tau_6\right)^6}
                  {\eta\left(\tau_6\right)^2 \eta\left(6\tau_6\right)^3}
 \nonumber \\
 & = &
 e_1 + 2 e_2 
 \nonumber \\
 & = &
 E_1\left(\tau_6;\chi_0,\chi_1\right)
 + 2 E_1\left(2\tau_6;\chi_0,\chi_1\right)
 \nonumber \\
 & = & 
 \frac{1}{2}
  + \frac{1}{\sqrt{3}} \overline{\mathrm{E}}_{0;0}\left(r_6;1;q_6\right),
 \nonumber \\
 f_1
 & = &
 \frac{1}{2} \left(y+3\right) \frac{\psi_1}{\pi}
 \nonumber \\
 & = &
 6 e_2
 \nonumber \\
 & = &
 6 E_1\left(2\tau_6;\chi_0,\chi_1\right)
 \nonumber \\
 & = &
 1
 - \sqrt{3} \overline{\mathrm{E}}_{0;0}\left(r_3;1;q_6\right)
 + \sqrt{3} \overline{\mathrm{E}}_{0;0}\left(r_6;1;q_6\right),
 \nonumber \\
 f_2 
 & = &
 - \frac{1}{4} \left( 3 y^2 - 10 y - 9 \right) \frac{\psi_1^2}{\pi^2}
 \nonumber \\
 & = &
 -12 e_1^2 + 36 e_1 e_2 + 12 e_2^2
 \nonumber \\
 & = &
 4 B_{2,2}(\tau_6) - 10 B_{2,3}(\tau_6) + 8 B_{2,6}(\tau_6)
 \nonumber \\
  & = &
 1
 - 6\overline{\mathrm{E}}_{0;-1}\left(r_2;1;q_6\right)
 + 2 \overline{\mathrm{E}}_{0;-1}\left(r_3;1;q_6\right)
 - 8 \overline{\mathrm{E}}_{0;-1}\left(r_6;1;q_6\right),
 \nonumber \\
 f_3
 & = &
 -\frac{3}{2} y \left(y-1\right) \left(y-9\right) \frac{\psi_1^3}{\pi^3}
 \nonumber \\
 & = & 
 - 12 \frac{\eta\left(\tau_6\right)^4 \eta\left(2\tau_6\right) \eta\left(6\tau_6\right)^5}{\eta\left(3\tau_6\right)^4}
 \nonumber \\
 & = &
 216 \left(e_1^3 - 2 e_1^2 e_2 - e_1 e_2^2 + 2 e_2^3\right)
 \nonumber \\
 & = &
 -12 E_3\left(\tau_6;\chi_0,\chi_1\right)
 +12 E_3\left(2\tau_6;\chi_0,\chi_1\right)
 \nonumber \\
 & = &
  - \frac{9}{2} \sqrt{3} \overline{\mathrm{E}}_{-2;0}\left(r_3;1;q_6\right)
  + \frac{1}{2} \sqrt{3} \overline{\mathrm{E}}_{-2;0}\left(r_6;1;q_6\right),
 \nonumber \\
 g_{2,0} 
 & = &
 \frac{6 \psi_1^2}{2\pi i W_y} \frac{1}{y}
 \; = \;
 - \frac{1}{2} \left(y-1\right) \left(y-9\right) \frac{\psi_1^2}{\pi^2}
 \nonumber \\
 & = &
 - 8 \frac{\eta\left(2\tau_6\right)^4 \eta\left(6\tau_6\right)^4}{\eta\left(\tau_6\right)^2 \eta\left(3\tau_6\right)^2}
 \nonumber \\
 & = &
 -24 \left( e_1^2 - e_2^2 \right)
 \nonumber \\
 & = &
 -4 B_{2,2}(\tau_6) - 8 B_{2,3}(\tau_6) + 4 B_{2,6}(\tau_6)
 \nonumber \\
 & = &
   4 \overline{\mathrm{E}}_{0;-1}\left(r_3;1;q_6\right)
 - 4 \overline{\mathrm{E}}_{0;-1}\left(r_6;1;q_6\right),
 \nonumber \\
 g_{2,1} 
 & = &
 \frac{6 \psi_1^2}{2\pi i W_y} \frac{1}{y-1}
 \; = \;
 - \frac{1}{2} y \left(y-9\right) \frac{\psi_1^2}{\pi^2}
 \nonumber \\
 & = &
 \frac{\eta\left(\tau_6\right)^3 \eta\left(2\tau_6\right)^3}{\eta\left(3\tau_6\right) \eta\left(6\tau_6\right)}
 \nonumber \\
 & = & 
 - 18 e_1^2 + 18 e_1 e_2 + 36 e_2^2 
 \nonumber \\
 & = &
 -3 B_{2,2}(\tau_6) - 9 B_{2,3}(\tau_6) + 9 B_{2,6}(\tau_6)
 \nonumber \\
 & = &
 1
 - 3 \overline{\mathrm{E}}_{0;-1}\left(r_2;1;q_6\right)
 - 9 \overline{\mathrm{E}}_{0;-1}\left(r_6;1;q_6\right),
 \nonumber \\
 g_{2,9} 
 & = &
 \frac{6 \psi_1^2}{2\pi i W_y} \frac{1}{y-9}
 \; = \;
 - \frac{1}{2} y \left(y-1\right) \frac{\psi_1^2}{\pi^2}
 \nonumber \\
 & = &
 \frac{\eta\left(\tau_6\right)^7 \eta\left(6\tau_6\right)^7}{\eta\left(2\tau_6\right)^5 \eta\left(3\tau_6\right)^5}
 \nonumber \\
 & = &
 - 6 e_1^2 + 18 e_1 e_2 - 12 e_2^2
 \nonumber \\
 & = & 
 5 B_{2,2}(\tau_6) - 5 B_{2,3}(\tau_6) + B_{2,6}(\tau_6)
 \nonumber \\
 & = &
 - 3 \overline{\mathrm{E}}_{0;-1}\left(r_2;1;q_6\right)
 + 4 \overline{\mathrm{E}}_{0;-1}\left(r_3;1;q_6\right)
 - \overline{\mathrm{E}}_{0;-1}\left(r_6;1;q_6\right),
\eq
Here we used the notation $B_{2,N}(\tau)=E_2(\tau)-NE_2(N\tau)$
and
\bq
 e_1 \; = \; E_1\left(\tau_6;\chi_0,\chi_1\right),
 & &
 e_2 \; = \; E_1\left(2\tau_6;\chi_0,\chi_1\right),
\eq
where $\chi_0$ and $\chi_1$ denote primitive Dirichlet characters with conductors $1$ and $3$, respectively.
They are given in terms of the Kronecker symbol by
\bq
 \chi_0 \; = \; \left(\frac{1}{n} \right),
 & &
 \chi_1 \; = \; \left(\frac{-3}{n} \right).
\eq
The definition of the generalised Eisenstein series is as in \cite{Stein,Adams:2017ejb}.
The two Eisenstein series $\{e_1,e_2\}$ give a basis for the modular forms of modular weight $1$ 
for the Eisenstein subspace ${\mathcal E}_1(\Gamma_1(6))$.

The $\overline{\mathrm{E}}_{n;m}$-functions are defined by \cite{Adams:2015ydq,Adams:2016xah}
\begin{align}
\label{def_Ebar_weight_1}
 \overline{\mathrm{E}}_{n;m}\left(x;y;q\right) 
 & =
 \left\{ \begin{array}{ll}
 \frac{1}{i}
 \left[
 \mathrm{ELi}_{n;m}\left(x;y;q\right) - \mathrm{ELi}_{n;m}\left(x^{-1};y^{-1};q\right)
 \right],
 & \mbox{$n+m$ even,} \\
 & \\
 \mathrm{ELi}_{n;m}\left(x;y;q\right) + \mathrm{ELi}_{n;m}\left(x^{-1};y^{-1};q\right),
 & \mbox{$n+m$ odd.} \\
 \end{array}
 \right.
\end{align}
and
\begin{align}
 \mathrm{ELi}_{n;m}\left(x;y;q\right) 
 & =  
 \sum\limits_{j=1}^\infty \sum\limits_{k=1}^\infty \; \frac{x^j}{j^n} \frac{y^k}{k^m} q^{j k}.
\end{align}
Products of modular forms are again modular forms.
In particular
\bq
 f_4 & = & f_1^4
\eq
and
\bq
 g_{3,1} & = & g_{2,1} \frac{\psi_1}{\pi},
 \nonumber \\
 g_{4,j} & = & g_{2,j} \left( \frac{\psi_1}{\pi} \right)^2,
 \;\;\;\;\;\;\;\;\;\;\;\; 
 j \in \{0,1,9\},
 \nonumber \\
 h_{3,0} & = & \frac{1}{3} f_3,
 \nonumber \\
 h_{4,0} & = & \frac{1}{3} f_3 \frac{\psi_1}{\pi},
 \nonumber \\
 h_{4,1} & = & \frac{2}{3} f_1 f_3 - \frac{4}{3} f_3 \frac{\psi_1}{\pi}.
\eq
This shows that all elements of the set in eq.~(\ref{modular_forms}) are modular forms.


\section{Boundary constants}
\label{sect:boundary_constants}

In this appendix we list the boundary constants 
at the point $(\tilde{x},\tilde{y})=(0,0)$ (or equivalently $(s,t)=(\infty,m^2)$ or $(x,y)=(0,1)$).
The boundary constants are given by
\bq
 \left. J_{1} \right|_{(\tilde{x},\tilde{y})=(0,0)}
 & = &
 1 + \zeta_2 \eps^2 - \frac{2}{3} \zeta_3 \eps^3 + \frac{7}{4} \zeta_4 \eps^4 
 + {\mathcal O}\left(\eps^5\right),
 \nonumber \\ 
 \left. J_{2} \right|_{(\tilde{x},\tilde{y})=(0,0)}
 & = &
 \zeta_2{\eps}^{2}
 +2\,\zeta_3{\eps}^{3}
 +{\frac {19}{4}}\,\zeta_4{\eps}^{4}
 + {\mathcal O}\left(\eps^5\right),
 \nonumber \\ 
 \left. J_{3} \right|_{(\tilde{x},\tilde{y})=(0,0)}
 & = &
 -\zeta_2{\eps}^{2}
 -11\,\zeta_3{\eps}^{3}
 -{\frac {29}{2}}\,\zeta_4{\eps}^{4}
 + {\mathcal O}\left(\eps^5\right),
 \nonumber \\ 
 \left. J_{4} \right|_{(\tilde{x},\tilde{y})=(0,0)}
 & = &
 -6\,\zeta_3{\eps}^{3}
 -\frac{13}{2}\,\zeta_4{\eps}^{4}
 + {\mathcal O}\left(\eps^5\right),
 \nonumber \\ 
 \left. J_{5} \right|_{(\tilde{x},\tilde{y})=(0,0)}
 & = &
 1
 -\frac{8}{3}\,\zeta_3{\eps}^{3}
 -3\,\zeta_4{\eps}^{4}
 + {\mathcal O}\left(\eps^5\right),
 \nonumber \\ 
 \left. J_{6} \right|_{(\tilde{x},\tilde{y})=(0,0)}
 & = &
 3\,\zeta_2{\eps}^{2}
 +\left[ \frac{21}{2}\,\zeta_3-18\,\zeta_2\,\ln  \left( 2 \right)  \right] {\eps}^{3} 
 +\left[ 72\,\mathrm{Li}_4\left(\frac{1}{2}\right) -39\,\zeta_4+36\,\zeta_2\, \ln^2  \left( 2 \right)  
 \right. \nonumber \\
 & & \left.
 +3\, \ln^4  \left( 2 \right) \right] {\eps}^{4} 
 + {\mathcal O}\left(\eps^5\right),
 \nonumber \\ 
 \left. J_{7} \right|_{(\tilde{x},\tilde{y})=(0,0)}
 & = &
 3\,\zeta_2{\eps}^{2}
 +\left[ \frac{21}{2}\,\zeta_3-18\,\zeta_2\,\ln  \left( 2 \right)  \right] {\eps}^{3} 
 +\left[ 72\,\mathrm{Li}_4\left(\frac{1}{2}\right) -39\,\zeta_4+36\,\zeta_2\, \ln^2  \left( 2 \right)  
 \right. \nonumber \\
 & & \left.
 +3\, \ln^4  \left( 2 \right) \right] {\eps}^{4} 
 + {\mathcal O}\left(\eps^5\right),
 \nonumber \\ 
 \left. J_{8} \right|_{(\tilde{x},\tilde{y})=(0,0)}
 & = &
 6 \zeta_2 \eps^2 + \left[ 21 \zeta_3 - 36 \zeta_2 \ln 2 \right] \eps^3 
 + \left[ 144 \, \mathrm{Li}_4\left(\frac{1}{2}\right) - 78 \zeta_4 + 72 \zeta_2 \ln^2\left(2\right) 
 \right.
 \nonumber \\
 & &
 \left. + 6 \ln^4\left(2\right)
\right] \eps^4 
 + {\mathcal O}\left(\eps^5\right),
 \nonumber \\ 
 \left. J_{9} \right|_{(\tilde{x},\tilde{y})=(0,0)}
 & = &
 -3\,\zeta_3{\eps}^{3}
 -\frac{5}{4}\,\zeta_4{\eps}^{4}
 + {\mathcal O}\left(\eps^5\right),
 \nonumber \\ 
 \left. J_{10} \right|_{(\tilde{x},\tilde{y})=(0,0)}
 & = &
 -4\,\zeta_3{\eps}^{3}
 +4\,\zeta_4{\eps}^{4}
 + {\mathcal O}\left(\eps^5\right),
 \nonumber \\ 
 \left. J_{11} \right|_{(\tilde{x},\tilde{y})=(0,0)}
 & = &
 \zeta_2{\eps}^{2}
 +2\,\zeta_3{\eps}^{3}
 +\frac{9}{4}\,\zeta_4{\eps}^{4}
 + {\mathcal O}\left(\eps^5\right),
 \nonumber \\ 
 \left. J_{12} \right|_{(\tilde{x},\tilde{y})=(0,0)}
 & = &
 \zeta_2{\eps}^{2}
 -4\,\zeta_3{\eps}^{3}
 + {\mathcal O}\left(\eps^5\right),
 \nonumber \\ 
 \left. J_{13} \right|_{(\tilde{x},\tilde{y})=(0,0)}
 & = &
 -2\,\zeta_3{\eps}^{3}
 +\frac{5}{4}\,\zeta_4{\eps}^{4}
 + {\mathcal O}\left(\eps^5\right),
 \nonumber \\ 
 \left. J_{14} \right|_{(\tilde{x},\tilde{y})=(0,0)}
 & = &
 {\mathcal O}\left(\eps^5\right),
 \nonumber \\ 
 \left. J_{15} \right|_{(\tilde{x},\tilde{y})=(0,0)}
 & = &
 \left[ \frac{11}{2}\,\zeta_3-6\,\zeta_2\,\ln  \left( 2 \right)  \right] {\eps}^{3} 
 +\left[ 24\,\mathrm{Li}_4\left(\frac{1}{2}\right) -\frac{23}{2}\,\zeta_4+12\,\zeta_2\, \ln^2  \left( 2 \right)  + \ln^4  \left( 2 \right) \right] {\eps}^{4} 
 \nonumber \\
 & &
 + {\mathcal O}\left(\eps^5\right),
 \nonumber \\ 
 \left. J_{16} \right|_{(\tilde{x},\tilde{y})=(0,0)}
 & = &
 \frac{1}{4}\,\zeta_4{\eps}^{4}
 + {\mathcal O}\left(\eps^5\right),
 \nonumber \\ 
 \left. J_{17} \right|_{(\tilde{x},\tilde{y})=(0,0)}
 & = &
 {\mathcal O}\left(\eps^5\right),
 \nonumber \\ 
 \left. J_{18} \right|_{(\tilde{x},\tilde{y})=(0,0)}
 & = &
 -\zeta_2{\eps}^{2}
 +\left[ \frac{9}{4}\,\zeta_3+3\,\zeta_2\,\ln  \left( 2 \right)  \right] {\eps}^{3} 
 +\left[ -12\,\mathrm{Li}_4\left(\frac{1}{2}\right) +{\frac {75}{8}}\,\zeta_4-6\,\zeta_2\, \ln^2  \left( 2 \right)  
 \right. \nonumber \\
 & & \left.
         -\frac{1}{2}\, \ln^4  \left( 2 \right) 
 \right] {\eps}^{4} 
 + {\mathcal O}\left(\eps^5\right),
 \nonumber \\ 
 \left. J_{19} \right|_{(\tilde{x},\tilde{y})=(0,0)}
 & = &
 -\frac{1}{2}\,\zeta_4{\eps}^{4}
 + {\mathcal O}\left(\eps^5\right),
 \nonumber \\ 
 \left. J_{20} \right|_{(\tilde{x},\tilde{y})=(0,0)}
 & = &
 -4\,\zeta_2{\eps}^{2}
 -5\,\zeta_3{\eps}^{3}
 -{\frac {99}{4}}\,\zeta_4{\eps}^{4}
 + {\mathcal O}\left(\eps^5\right),
 \nonumber \\ 
 \left. J_{21} \right|_{(\tilde{x},\tilde{y})=(0,0)}
 & = &
 -3\,\zeta_3{\eps}^{3}
 -\frac{5}{4}\,\zeta_4{\eps}^{4}
 + {\mathcal O}\left(\eps^5\right),
 \nonumber \\ 
 \left. J_{22} \right|_{(\tilde{x},\tilde{y})=(0,0)}
 & = &
 \frac{7}{4}\,\zeta_4{\eps}^{4}
 + {\mathcal O}\left(\eps^5\right),
 \nonumber \\ 
 \left. J_{23} \right|_{(\tilde{x},\tilde{y})=(0,0)}
 & = &
 6\,\zeta_3{\eps}^{3}
 +9\,\zeta_4{\eps}^{4}
 + {\mathcal O}\left(\eps^5\right),
 \nonumber \\ 
 \left. J_{24} \right|_{(\tilde{x},\tilde{y})=(0,0)}
 & = &
 \left[ -\frac{7}{4}\,\zeta_3-3\,\zeta_2\,\ln  \left( 2 \right)  \right] {\eps}^{3} 
 +\left[ {\frac {15}{4}}\,\zeta_4+18\,\zeta_2\, \ln^2  \left( 2 \right)  \right] {\eps}^{4} 
 + {\mathcal O}\left(\eps^5\right),
 \nonumber \\ 
 \left. J_{25} \right|_{(\tilde{x},\tilde{y})=(0,0)}
 & = &
 \left[ -{\frac {23}{6}}\,\zeta_3+2\,\zeta_2\,\ln  \left( 2 \right)  \right] {\eps}^{3}
 + \left[ -8\,\mathrm{Li}_4\left(\frac{1}{2}\right) +3\,\zeta_4-4\,\zeta_2\, \ln^2  \left( 2 \right)  -\frac{1}{3}\, \ln^4  \left( 2 \right) \right] {\eps}^{4}
 \nonumber \\
 & &
 + {\mathcal O}\left(\eps^5\right),
 \nonumber \\ 
 \left. J_{26} \right|_{(\tilde{x},\tilde{y})=(0,0)}
 & = &
 \frac{7}{4}\,\zeta_2{\eps}^{2}
 +\left[ {\frac {35}{8}}\,\zeta_3-{\frac {27}{2}}\,\zeta_2\,\ln  \left( 2 \right)  \right] {\eps}^{3} 
 +\left[ 42\,\mathrm{Li}_4\left(\frac{1}{2}\right) -19\,\zeta_4+39\,\zeta_2\, \ln^2  \left( 2 \right)  
 \right. \nonumber \\
 & & \left.
 +\frac{7}{4}\, \ln^4  \left( 2 \right) \right]){\eps}^{4} 
 + {\mathcal O}\left(\eps^5\right),
 \nonumber \\ 
 \left. J_{27} \right|_{(\tilde{x},\tilde{y})=(0,0)}
 & = &
 \left[ -\frac{7}{2}\,\zeta_3-6\,\zeta_2\,\ln  \left( 2 \right)  \right] {\eps}^{3} 
 +\left[ {\frac {105}{4}}\,\zeta_4+36\,\zeta_2\, \ln^2  \left( 2 \right) \right] {\eps}^{4} 
 + {\mathcal O}\left(\eps^5\right),
 \nonumber \\ 
 \left. J_{28} \right|_{(\tilde{x},\tilde{y})=(0,0)}
 & = &
 \left[ -\frac{1}{3}\,\zeta_3-4\,\zeta_2\,\ln  \left( 2 \right)  \right] {\eps}^{3} 
 +\left[ 16\,\mathrm{Li}_4\left(\frac{1}{2}\right) -12\,\zeta_4+8\,\zeta_2\, \ln^2  \left( 2 \right)  +\frac{2}{3}\, \ln^4  \left( 2 \right)  \right] {\eps}^{4} 
 \nonumber \\
 & &
 + {\mathcal O}\left(\eps^5\right),
 \nonumber \\ 
 \left. J_{29} \right|_{(\tilde{x},\tilde{y})=(0,0)}
 & = &
 {\mathcal O}\left(\eps^5\right),
 \nonumber \\ 
 \left. J_{30} \right|_{(\tilde{x},\tilde{y})=(0,0)}
 & = &
 \frac{7}{2}\,\zeta_2{\eps}^{2}
 + \left[ {\frac {35}{4}}\,\zeta_3-27\,\zeta_2\,\ln  \left( 2 \right)  \right] {\eps}^{3} 
 + \left[ 84\,\mathrm{Li}_4\left(\frac{1}{2}\right) -{\frac {77}{4}}\,\zeta_4 +78\,\zeta_2\, \ln^2  \left( 2 \right)  
 \right. \nonumber \\
 & & \left.
 +\frac{7}{2}\, \ln^4  \left( 2 \right) \right] {\eps}^{4}
 + {\mathcal O}\left(\eps^5\right),
 \nonumber \\ 
 \left. J_{32} \right|_{(\tilde{x},\tilde{y})=(0,0)}
 & = &
 -4\,\zeta_2{\eps}^{2}
 -5\,\zeta_3{\eps}^{3}
 -{\frac {139}{4}}\,\zeta_4{\eps}^{4}
 + {\mathcal O}\left(\eps^5\right),
 \nonumber \\ 
 \left. J_{33} \right|_{(\tilde{x},\tilde{y})=(0,0)}
 & = &
 {\frac {105}{4}}\,\zeta_4{\eps}^{4}
 + {\mathcal O}\left(\eps^5\right),
 \nonumber \\ 
 \left. J_{34} \right|_{(\tilde{x},\tilde{y})=(0,0)}
 & = &
 4\,\zeta_2{\eps}^{2}
 +\left[ \frac{11}{2}\,\zeta_3-6\,\zeta_2\,\ln  \left( 2 \right)  \right] {\eps}^{3} 
 +\left[ 24\,\mathrm{Li}_4\left(\frac{1}{2}\right) +\frac{7}{2}\,\zeta_4+12\,\zeta_2\, \ln^2  \left( 2 \right)  
 \right. \nonumber \\
 & & \left.
 + \ln^4  \left( 2 \right) \right] {\eps}^{4} 
 + {\mathcal O}\left(\eps^5\right),
 \nonumber \\ 
 \left. J_{35} \right|_{(\tilde{x},\tilde{y})=(0,0)}
 & = &
 \frac{3}{4}\,\zeta_2{\eps}^{2}
 + \left[ {\frac {21}{8}}\,\zeta_3-\frac{9}{2}\,\zeta_2\,\ln  \left( 2 \right)  \right] {\eps}^{3}
 + \left[ 18\,\mathrm{Li}_4\left(\frac{1}{2}\right) +{\frac {33}{2}}\,\zeta_4+9\,\zeta_2\, \ln^2  \left( 2 \right)  
 \right. \nonumber \\
 & & \left.
          +\frac{3}{4}\, \ln^4  \left( 2 \right)  \right] {\eps}^{4}
 + {\mathcal O}\left(\eps^5\right),
 \nonumber \\ 
 \left. J_{36} \right|_{(\tilde{x},\tilde{y})=(0,0)}
 & = &
 -{\frac {39}{2}}\,\zeta_4{\eps}^{4}
 + {\mathcal O}\left(\eps^5\right),
 \nonumber \\ 
 \left. J_{37} \right|_{(\tilde{x},\tilde{y})=(0,0)}
 & = &
 {\mathcal O}\left(\eps^5\right),
 \nonumber \\ 
 \left. J_{38} \right|_{(\tilde{x},\tilde{y})=(0,0)}
 & = &
 \frac{17}{2}\,{\eps}^{4}\zeta_4
 + {\mathcal O}\left(\eps^5\right),
 \nonumber \\ 
 \left. J_{39} \right|_{(\tilde{x},\tilde{y})=(0,0)}
 & = &
 {\mathcal O}\left(\eps^5\right),
 \nonumber \\ 
 \left. J_{40} \right|_{(\tilde{x},\tilde{y})=(0,0)}
 & = &
 -3\,\zeta_4{\eps}^{4}
 + {\mathcal O}\left(\eps^5\right),
 \nonumber \\ 
 \left. J_{41} \right|_{(\tilde{x},\tilde{y})=(0,0)}
 & = &
 \left[ 8\,\mathrm{Li}_4\left(\frac{1}{2}\right) -{\frac {79}{4}}\,\zeta_4-8\,\zeta_2\, \ln^2  \left( 2 \right)  +\frac{1}{3}\,
 \ln^4  \left( 2 \right) \right] {\eps}^{4} 
 + {\mathcal O}\left(\eps^5\right),
 \nonumber \\ 
 \left. J_{42} \right|_{(\tilde{x},\tilde{y})=(0,0)}
 & = &
 -24\,\zeta_3{\eps}^{3}
 -102\,\zeta_4{\eps}^{4}
 + {\mathcal O}\left(\eps^5\right),
 \nonumber \\ 
 \left. J_{43} \right|_{(\tilde{x},\tilde{y})=(0,0)}
 & = &
 \left[ {\frac {49}{12}}\,\zeta_3+7\,\zeta_2\,\ln  \left( 2 \right)  \right] {\eps}^{3} 
 + \left[ 8\,\mathrm{Li}_4\left(\frac{1}{2}\right) -{\frac {109}{4}}\,\zeta_4-50\,\zeta_2\, \ln^2  \left( 2 \right)  
 \right. \nonumber \\
 & & \left.
        +\frac{1}{3}\, \ln^4  \left( 2 \right) \right] {\eps}^{4}
 + {\mathcal O}\left(\eps^5\right),
 \nonumber \\ 
 \left. J_{44} \right|_{(\tilde{x},\tilde{y})=(0,0)}
 & = &
 \left[ \frac{16}{3}\,\mathrm{Li}_4\left(\frac{1}{2}\right) +{\frac {31}{12}}\,\zeta_4-\frac{16}{3}\,\zeta_2\, \ln^2  \left( 2 \right) 
  +\frac{2}{9} \, \ln^4  \left( 2 \right) \right] {\eps}^{4} 
 + {\mathcal O}\left(\eps^5\right),
 \nonumber \\ 
 \left. J_{45} \right|_{(\tilde{x},\tilde{y})=(0,0)}
 & = & 
 \left[ -{\frac {56}{3}}\,\mathrm{Li}_4\left(\frac{1}{2}\right) -{\frac {11}{12}}\,\zeta_4
  +{\frac {56}{3}}\,\zeta_2\, \ln^2  \left( 2\right)  -{\frac {7}{9}}\, \ln^4  \left( 2 \right) \right] {\eps}^{4} 
 + {\mathcal O}\left(\eps^5\right).
\eq


\section{Supplementary material}
\label{sect:supplement}

Attached to this article is an electronic file in ASCII format with {\tt Maple} syntax, defining the quantities
\begin{center}
 \verb|I_basis|, \; \verb|U|, \; \verb|Uinv|, \; \verb|A|, \; \verb|integration_kernels|, \; \verb|J|.
\end{center}
\verb|I_basis| is a vector defining the basis $\vec{I}$ as listed in the fourth column of table~\ref{table_master_integrals}, 
\verb|U| gives the transformation matrix to the basis $\vec{J}$, i.e.
\bq
 \vec{J} & = & U \vec{I},
\eq
\verb|Uinv| denotes the matrix inverse of $U$.
The quantity \verb|A| defines the matrix
\bq
 A & = & A^{(0)} + \eps A^{(1)}
\eq
appearing in the differential equation
\bq
 d \vec{J} & = & A \vec{J}.
\eq
The entries of $A$ are linear combinations of integration kernels (with symbolic names), whose explicit expressions are defined
in \verb|integration_kernels|.
Finally, \verb|J| contains the results for the master integrals up to order $\eps^4$ in terms of iterated integrals, including
the boundary constants for the boundary point $(x,y)=(0,1)$.

\end{appendix}

{\footnotesize
\bibliography{/home/stefanw/notes/biblio}

\begin{thebibliography}{100}

\bibitem{Broadhurst:1993mw}
D.~J. Broadhurst, J.~Fleischer, and O.~Tarasov,
\newblock Z.Phys. {\bf C60}, 287 (1993), arXiv:hep-ph/9304303.

\bibitem{Berends:1993ee}
F.~A. Berends, M.~Buza, M.~B{\"o}hm, and R.~Scharf,
\newblock Z.Phys. {\bf C63}, 227 (1994).

\bibitem{Bauberger:1994nk}
S.~Bauberger, M.~B{\"o}hm, G.~Weiglein, F.~A. Berends, and M.~Buza,
\newblock Nucl.Phys.Proc.Suppl. {\bf 37B}, 95 (1994), arXiv:hep-ph/9406404.

\bibitem{Bauberger:1994by}
S.~Bauberger, F.~A. Berends, M.~B{\"o}hm, and M.~Buza,
\newblock Nucl.Phys. {\bf B434}, 383 (1995), arXiv:hep-ph/9409388.

\bibitem{Bauberger:1994hx}
S.~Bauberger and M.~B{\"o}hm,
\newblock Nucl.Phys. {\bf B445}, 25 (1995), arXiv:hep-ph/9501201.

\bibitem{Caffo:1998du}
M.~Caffo, H.~Czyz, S.~Laporta, and E.~Remiddi,
\newblock Nuovo Cim. {\bf A111}, 365 (1998), arXiv:hep-th/9805118.

\bibitem{Laporta:2004rb}
S.~Laporta and E.~Remiddi,
\newblock Nucl. Phys. {\bf B704}, 349 (2005), hep-ph/0406160.

\bibitem{Kniehl:2005bc}
B.~A. Kniehl, A.~V. Kotikov, A.~Onishchenko, and O.~Veretin,
\newblock Nucl. Phys. {\bf B738}, 306 (2006), arXiv:hep-ph/0510235.

\bibitem{Groote:2005ay}
S.~Groote, J.~G. K{\"o}rner, and A.~A. Pivovarov,
\newblock Annals Phys. {\bf 322}, 2374 (2007), arXiv:hep-ph/0506286.

\bibitem{Groote:2012pa}
S.~Groote, J.~K{\"o}rner, and A.~Pivovarov,
\newblock Eur.Phys.J. {\bf C72}, 2085 (2012), arXiv:1204.0694.

\bibitem{Bailey:2008ib}
D.~H. Bailey, J.~M. Borwein, D.~Broadhurst, and M.~L. Glasser,
\newblock J. Phys. {\bf A41}, 205203 (2008), arXiv:0801.0891.

\bibitem{MullerStach:2011ru}
S.~M{\"u}ller-Stach, S.~Weinzierl, and R.~Zayadeh,
\newblock Commun. Num. Theor. Phys. {\bf 6}, 203 (2012), arXiv:1112.4360.

\bibitem{Adams:2013nia}
L.~Adams, C.~Bogner, and S.~Weinzierl,
\newblock J. Math. Phys. {\bf 54}, 052303 (2013), arXiv:1302.7004.

\bibitem{Bloch:2013tra}
S.~Bloch and P.~Vanhove,
\newblock J. Numb. Theor. {\bf 148}, 328 (2015), arXiv:1309.5865.

\bibitem{Adams:2014vja}
L.~Adams, C.~Bogner, and S.~Weinzierl,
\newblock J. Math. Phys. {\bf 55}, 102301 (2014), arXiv:1405.5640.

\bibitem{Adams:2015gva}
L.~Adams, C.~Bogner, and S.~Weinzierl,
\newblock J. Math. Phys. {\bf 56}, 072303 (2015), arXiv:1504.03255.

\bibitem{Adams:2015ydq}
L.~Adams, C.~Bogner, and S.~Weinzierl,
\newblock J. Math. Phys. {\bf 57}, 032304 (2016), arXiv:1512.05630.

\bibitem{Remiddi:2013joa}
E.~Remiddi and L.~Tancredi,
\newblock Nucl.Phys. {\bf B880}, 343 (2014), arXiv:1311.3342.

\bibitem{Bloch:2016izu}
S.~Bloch, M.~Kerr, and P.~Vanhove,
\newblock Adv. Theor. Math. Phys. {\bf 21}, 1373 (2017), arXiv:1601.08181.

\bibitem{Groote:2018rpb}
S.~Groote and J.~G. K{\"o}rner,
\newblock (2018), arXiv:1804.10570.

\bibitem{Adams:2017ejb}
L.~Adams and S.~Weinzierl,
\newblock Commun. Num. Theor. Phys. {\bf 12}, 193 (2018), arXiv:1704.08895.

\bibitem{Bloch:2014qca}
S.~Bloch, M.~Kerr, and P.~Vanhove,
\newblock Compos. Math. {\bf 151}, 2329 (2015), arXiv:1406.2664.

\bibitem{Remiddi:2016gno}
E.~Remiddi and L.~Tancredi,
\newblock Nucl. Phys. {\bf B907}, 400 (2016), arXiv:1602.01481.

\bibitem{Adams:2016xah}
L.~Adams, C.~Bogner, A.~Schweitzer, and S.~Weinzierl,
\newblock J. Math. Phys. {\bf 57}, 122302 (2016), arXiv:1607.01571.

\bibitem{Bogner:2017vim}
C.~Bogner, A.~Schweitzer, and S.~Weinzierl,
\newblock Nucl. Phys. {\bf B922}, 528 (2017), arXiv:1705.08952.

\bibitem{Adams:2018yfj}
L.~Adams and S.~Weinzierl,
\newblock Phys. Lett. {\bf B781}, 270 (2018), arXiv:1802.05020.

\bibitem{Sogaard:2014jla}
M.~Søgaard and Y.~Zhang,
\newblock Phys. Rev. {\bf D91}, 081701 (2015), arXiv:1412.5577.

\bibitem{Bonciani:2016qxi}
R.~Bonciani {\em et~al.},
\newblock JHEP {\bf 12}, 096 (2016), arXiv:1609.06685.

\bibitem{vonManteuffel:2017hms}
A.~von Manteuffel and L.~Tancredi,
\newblock JHEP {\bf 06}, 127 (2017), arXiv:1701.05905.

\bibitem{Primo:2017ipr}
A.~Primo and L.~Tancredi,
\newblock Nucl. Phys. {\bf B921}, 316 (2017), arXiv:1704.05465.

\bibitem{Ablinger:2017bjx}
J.~Ablinger {\em et~al.},
\newblock J. Math. Phys. {\bf 59}, 062305 (2018), arXiv:1706.01299.

\bibitem{Bourjaily:2017bsb}
J.~L. Bourjaily, A.~J. McLeod, M.~Spradlin, M.~von Hippel, and M.~Wilhelm,
\newblock Phys. Rev. Lett. {\bf 120}, 121603 (2018), arXiv:1712.02785.

\bibitem{Hidding:2017jkk}
M.~Hidding and F.~Moriello,
\newblock (2017), arXiv:1712.04441.

\bibitem{Passarino:2017EPJC}
G.~{Passarino},
\newblock European Physical Journal C {\bf 77}, 77 (2017), arXiv:1610.06207.

\bibitem{Remiddi:2017har}
E.~Remiddi and L.~Tancredi,
\newblock Nucl. Phys. {\bf B925}, 212 (2017), arXiv:1709.03622.

\bibitem{Broedel:2017kkb}
J.~Broedel, C.~Duhr, F.~Dulat, and L.~Tancredi,
\newblock JHEP {\bf 05}, 093 (2018), arXiv:1712.07089.

\bibitem{Broedel:2017siw}
J.~Broedel, C.~Duhr, F.~Dulat, and L.~Tancredi,
\newblock Phys. Rev. {\bf D97}, 116009 (2018), arXiv:1712.07095.

\bibitem{Broedel:2018iwv}
J.~Broedel, C.~Duhr, F.~Dulat, B.~Penante, and L.~Tancredi,
\newblock (2018), arXiv:1803.10256.

\bibitem{Lee:2017qql}
R.~N. Lee, A.~V. Smirnov, and V.~A. Smirnov,
\newblock JHEP {\bf 03}, 008 (2018), arXiv:1709.07525.

\bibitem{Lee:2018ojn}
R.~N. Lee, A.~V. Smirnov, and V.~A. Smirnov,
\newblock (2018), arXiv:1805.00227.

\bibitem{Broedel:2014vla}
J.~Broedel, C.~R. Mafra, N.~Matthes, and O.~Schlotterer,
\newblock JHEP {\bf 07}, 112 (2015), arXiv:1412.5535.

\bibitem{Broedel:2015hia}
J.~Broedel, N.~Matthes, and O.~Schlotterer,
\newblock J. Phys. {\bf A49}, 155203 (2016), arXiv:1507.02254.

\bibitem{Broedel:2017jdo}
J.~Broedel, N.~Matthes, G.~Richter, and O.~Schlotterer,
\newblock J. Phys. {\bf A51}, 285401 (2018), arXiv:1704.03449.

\bibitem{DHoker:2015wxz}
E.~D'Hoker, M.~B. Green, {\"O}.~G{\"u}rdogan, and P.~Vanhove,
\newblock Commun. Num. Theor. Phys. {\bf 11}, 165 (2017), arXiv:1512.06779.

\bibitem{Hohenegger:2017kqy}
S.~Hohenegger and S.~Stieberger,
\newblock Nucl. Phys. {\bf B925}, 63 (2017), arXiv:1702.04963.

\bibitem{Broedel:2018izr}
J.~Broedel, O.~Schlotterer, and F.~Zerbini,
\newblock (2018), arXiv:1803.00527.

\bibitem{Czakon:2013goa}
M.~Czakon, P.~Fiedler, and A.~Mitov,
\newblock Phys.Rev.Lett. {\bf 110}, 252004 (2013), arXiv:1303.6254.

\bibitem{Baernreuther:2013caa}
P.~B{\"a}rnreuther, M.~Czakon, and P.~Fiedler,
\newblock JHEP {\bf 02}, 078 (2014), arXiv:1312.6279.

\bibitem{Czakon:2008zk}
M.~Czakon,
\newblock Phys. Lett. {\bf B664}, 307 (2008), arXiv:0803.1400.

\bibitem{Czakon:2007wk}
M.~Czakon, A.~Mitov, and S.~Moch,
\newblock Nucl. Phys. {\bf B798}, 210 (2008), arXiv:0707.4139.

\bibitem{Czakon:2005rk}
M.~Czakon,
\newblock Comput. Phys. Commun. {\bf 175}, 559 (2006), hep-ph/0511200.

\bibitem{Adams:2018bsn}
L.~Adams, E.~Chaubey, and S.~Weinzierl,
\newblock Phys. Rev. Lett. {\bf 121}, 142001 (2018), arXiv:1804.11144.

\bibitem{Kotikov:1990kg}
A.~V. Kotikov,
\newblock Phys. Lett. {\bf B254}, 158 (1991).

\bibitem{Kotikov:1991pm}
A.~V. Kotikov,
\newblock Phys. Lett. {\bf B267}, 123 (1991).

\bibitem{Remiddi:1997ny}
E.~Remiddi,
\newblock Nuovo Cim. {\bf A110}, 1435 (1997), hep-th/9711188.

\bibitem{Gehrmann:1999as}
T.~Gehrmann and E.~Remiddi,
\newblock Nucl. Phys. {\bf B580}, 485 (2000), hep-ph/9912329.

\bibitem{Argeri:2007up}
M.~Argeri and P.~Mastrolia,
\newblock Int. J. Mod. Phys. {\bf A22}, 4375 (2007), arXiv:0707.4037.

\bibitem{MullerStach:2012mp}
S.~M{\"u}ller-Stach, S.~Weinzierl, and R.~Zayadeh,
\newblock Commun.Math.Phys. {\bf 326}, 237 (2014), arXiv:1212.4389.

\bibitem{Henn:2013pwa}
J.~M. Henn,
\newblock Phys. Rev. Lett. {\bf 110}, 251601 (2013), arXiv:1304.1806.

\bibitem{Henn:2014qga}
J.~M. Henn,
\newblock J. Phys. {\bf A48}, 153001 (2015), arXiv:1412.2296.

\bibitem{Ablinger:2015tua}
J.~Ablinger {\em et~al.},
\newblock Comput. Phys. Commun. {\bf 202}, 33 (2016), arXiv:1509.08324.

\bibitem{Adams:2017tga}
L.~Adams, E.~Chaubey, and S.~Weinzierl,
\newblock Phys. Rev. Lett. {\bf 118}, 141602 (2017), arXiv:1702.04279.

\bibitem{Bosma:2017hrk}
J.~Bosma, K.~J. Larsen, and Y.~Zhang,
\newblock Phys. Rev. {\bf D97}, 105014 (2018), arXiv:1712.03760.

\bibitem{Gehrmann:2014bfa}
T.~Gehrmann, A.~von Manteuffel, L.~Tancredi, and E.~Weihs,
\newblock JHEP {\bf 06}, 032 (2014), arXiv:1404.4853.

\bibitem{Argeri:2014qva}
M.~Argeri {\em et~al.},
\newblock JHEP {\bf 03}, 082 (2014), arXiv:1401.2979.

\bibitem{Lee:2014ioa}
R.~N. Lee,
\newblock JHEP {\bf 04}, 108 (2015), arXiv:1411.0911.

\bibitem{Prausa:2017ltv}
M.~Prausa,
\newblock Comput. Phys. Commun. {\bf 219}, 361 (2017), arXiv:1701.00725.

\bibitem{Gituliar:2017vzm}
O.~Gituliar and V.~Magerya,
\newblock Comput. Phys. Commun. {\bf 219}, 329 (2017), arXiv:1701.04269.

\bibitem{Meyer:2016slj}
C.~Meyer,
\newblock JHEP {\bf 04}, 006 (2017), arXiv:1611.01087.

\bibitem{Meyer:2017joq}
C.~Meyer,
\newblock Comput. Phys. Commun. {\bf 222}, 295 (2018), arXiv:1705.06252.

\bibitem{Lee:2017oca}
R.~N. Lee and A.~A. Pomeransky,
\newblock (2017), arXiv:1707.07856.

\bibitem{Becchetti:2017abb}
M.~Becchetti and R.~Bonciani,
\newblock JHEP {\bf 01}, 048 (2018), arXiv:1712.02537.

\bibitem{Beilinson:1994}
A.~Beilinson and A.~Levin,
\newblock in {\it Motives}, ed. U. Jannsen, S. Kleiman, J.-P. Serre, Proc. of
  Symp. in Pure Mathematics {\bf 55}, Part 2, AMS, 1994, 123-190.

\bibitem{Levin:1997}
A.~Levin,
\newblock Comp. Math. {\bf 106}, 267 (1997).

\bibitem{Levin:2007}
A.~Levin and G.~Racinet,
\newblock (2007), arXiv:math/0703237.

\bibitem{Enriquez:2010}
B.~{Enriquez},
\newblock Selecta Math. {\bf 20}, 491 (2014), arXiv:1003.1012.

\bibitem{Brown:2011}
F.~Brown and A.~Levin,
\newblock (2011), arXiv:1110.6917.

\bibitem{Wildeshaus}
J.~Wildeshaus,
\newblock Lect. Notes Math. {\bf 1650}, Springer, (1997).

\bibitem{Kinoshita:1962ur}
T.~Kinoshita,
\newblock J. Math. Phys. {\bf 3}, 650 (1962).

\bibitem{Tarasov:1996br}
O.~V. Tarasov,
\newblock Phys. Rev. {\bf D54}, 6479 (1996), hep-th/9606018.

\bibitem{Tarasov:1997kx}
O.~V. Tarasov,
\newblock Nucl. Phys. {\bf B502}, 455 (1997), hep-ph/9703319.

\bibitem{vonManteuffel:2012np}
A.~von Manteuffel and C.~Studerus,
\newblock (2012), arXiv:1201.4330.

\bibitem{Maierhoefer:2017hyi}
P.~Maierhöfer, J.~Usovitsch, and P.~Uwer,
\newblock Comput. Phys. Commun. {\bf 230}, 99 (2018), arXiv:1705.05610.

\bibitem{Smirnov:2014hma}
A.~V. Smirnov,
\newblock Comput. Phys. Commun. {\bf 189}, 182 (2015), arXiv:1408.2372.

\bibitem{Lee:2012cn}
R.~N. Lee,
\newblock (2012), arXiv:1212.2685.

\bibitem{Lee:2013mka}
R.~N. Lee,
\newblock J. Phys. Conf. Ser. {\bf 523}, 012059 (2014), arXiv:1310.1145.

\bibitem{Tkachov:1981wb}
F.~V. Tkachov,
\newblock Phys. Lett. {\bf B100}, 65 (1981).

\bibitem{Chetyrkin:1981qh}
K.~G. Chetyrkin and F.~V. Tkachov,
\newblock Nucl. Phys. {\bf B192}, 159 (1981).

\bibitem{Laporta:2001dd}
S.~Laporta,
\newblock Int. J. Mod. Phys. {\bf A15}, 5087 (2000), hep-ph/0102033.

\bibitem{Lee:2013hzt}
R.~N. Lee and A.~A. Pomeransky,
\newblock JHEP {\bf 11}, 165 (2013), arXiv:1308.6676.

\bibitem{Georgoudis:2016wff}
A.~Georgoudis, K.~J. Larsen, and Y.~Zhang,
\newblock Comput. Phys. Commun. {\bf 221}, 203 (2017), arXiv:1612.04252.

\bibitem{Chen}
K.-T. Chen,
\newblock Bull. Amer. Math. Soc. {\bf 83}, 831 (1977).

\bibitem{Goncharov_no_note}
A.~B. Goncharov,
\newblock Math. Res. Lett. {\bf 5}, 497 (1998).

\bibitem{Goncharov:2001}
A.~B. Goncharov,
\newblock (2001), math.AG/0103059.

\bibitem{Borwein}
J.~M. Borwein, D.~M. Bradley, D.~J. Broadhurst, and P.~Lisonek,
\newblock Trans. Amer. Math. Soc. {\bf 353:3}, 907 (2001), math.CA/9910045.

\bibitem{Moch:2001zr}
S.~Moch, P.~Uwer, and S.~Weinzierl,
\newblock J. Math. Phys. {\bf 43}, 3363 (2002), hep-ph/0110083.

\bibitem{Vollinga:2004sn}
J.~Vollinga and S.~Weinzierl,
\newblock Comput. Phys. Commun. {\bf 167}, 177 (2005), hep-ph/0410259.

\bibitem{Brown:2014aa}
F.~{Brown},
\newblock (2014), arXiv:1407.5167.

\bibitem{Fleischer:1998nb}
J.~Fleischer, A.~V. Kotikov, and O.~L. Veretin,
\newblock Nucl. Phys. {\bf B547}, 343 (1999), hep-ph/9808242.

\bibitem{Kotikov:2007vr}
A.~Kotikov, J.~H. Kuhn, and O.~Veretin,
\newblock Nucl. Phys. {\bf B788}, 47 (2008), arXiv:hep-ph/0703013.

\bibitem{Bonciani:2010ms}
R.~Bonciani, G.~Degrassi, and A.~Vicini,
\newblock Comput. Phys. Commun. {\bf 182}, 1253 (2011), arXiv:1007.1891.

\bibitem{Henn:2013woa}
J.~M. Henn and V.~A. Smirnov,
\newblock JHEP {\bf 11}, 041 (2013), arXiv:1307.4083.

\bibitem{Vermaseren:1998uu}
J.~A.~M. Vermaseren,
\newblock Int. J. Mod. Phys. {\bf A14}, 2037 (1999), hep-ph/9806280.

\bibitem{Remiddi:1999ew}
E.~Remiddi and J.~A.~M. Vermaseren,
\newblock Int. J. Mod. Phys. {\bf A15}, 725 (2000), hep-ph/9905237.

\bibitem{Baikov:1996iu}
P.~A. Baikov,
\newblock Nucl. Instrum. Meth. {\bf A389}, 347 (1997), arXiv:hep-ph/9611449.

\bibitem{Lee:2009dh}
R.~N. Lee,
\newblock Nucl. Phys. {\bf B830}, 474 (2010), arXiv:0911.0252.

\bibitem{Kosower:2011ty}
D.~A. Kosower and K.~J. Larsen,
\newblock Phys. Rev. {\bf D85}, 045017 (2012), arXiv:1108.1180.

\bibitem{CaronHuot:2012ab}
S.~Caron-Huot and K.~J. Larsen,
\newblock JHEP {\bf 1210}, 026 (2012), arXiv:1205.0801.

\bibitem{Frellesvig:2017aai}
H.~Frellesvig and C.~G. Papadopoulos,
\newblock JHEP {\bf 04}, 083 (2017), arXiv:1701.07356.

\bibitem{Bosma:2017ens}
J.~Bosma, M.~Sogaard, and Y.~Zhang,
\newblock JHEP {\bf 08}, 051 (2017), arXiv:1704.04255.

\bibitem{Harley:2017qut}
M.~Harley, F.~Moriello, and R.~M. Schabinger,
\newblock JHEP {\bf 06}, 049 (2017), arXiv:1705.03478.

\bibitem{Primo:2016ebd}
A.~Primo and L.~Tancredi,
\newblock Nucl. Phys. {\bf B916}, 94 (2017), arXiv:1610.08397.

\bibitem{Blumlein:2009}
J.~Bl{\"u}mlein, D.~J. Broadhurst, and J.~A.~M. Vermaseren,
\newblock Comput. Phys. Commun. {\bf 181}, 582 (2010), arXiv:0907.2557.

\bibitem{Broedel:2013tta}
J.~Broedel, O.~Schlotterer, and S.~Stieberger,
\newblock Fortsch. Phys. {\bf 61}, 812 (2013), arXiv:1304.7267.

\bibitem{Ferguson:1992}
H.~R.~P. Ferguson and D.~H. Bailey,
\newblock (1992),
\newblock RNR Technical Report RNR-91-032.

\bibitem{Davydychev:2000na}
A.~I. Davydychev and M.~Y. Kalmykov,
\newblock Nucl. Phys. {\bf B605}, 266 (2001), hep-th/0012189.

\bibitem{Davydychev:2003mv}
A.~I. Davydychev and M.~Y. Kalmykov,
\newblock Nucl. Phys. {\bf B699}, 3 (2004), hep-th/0303162.

\bibitem{Bytev:2009kb}
V.~V. Bytev, M.~{\relax Yu}. Kalmykov, and B.~A. Kniehl,
\newblock Nucl. Phys. {\bf B836}, 129 (2010), arXiv:0904.0214.

\bibitem{Bogner:2007cr}
C.~Bogner and S.~Weinzierl,
\newblock Comput. Phys. Commun. {\bf 178}, 596 (2008), arXiv:0709.4092.

\bibitem{Carter:2010hi}
J.~Carter and G.~Heinrich,
\newblock Comput. Phys. Commun. {\bf 182}, 1566 (2011), arXiv:1011.5493.

\bibitem{Borowka:2012yc}
S.~Borowka, J.~Carter, and G.~Heinrich,
\newblock Comput. Phys. Commun. {\bf 184}, 396 (2013), arXiv:1204.4152.

\bibitem{Borowka:2015mxa}
S.~Borowka {\em et~al.},
\newblock Comput. Phys. Commun. {\bf 196}, 470 (2015), arXiv:1502.06595.

\bibitem{Borowka:2017idc}
S.~Borowka {\em et~al.},
\newblock Comput. Phys. Commun. {\bf 222}, 313 (2018), arXiv:1703.09692.

\bibitem{Smirnov:2008py}
A.~V. Smirnov and M.~N. Tentyukov,
\newblock Comput. Phys. Commun. {\bf 180}, 735 (2009), arXiv:0807.4129.

\bibitem{Smirnov:2009pb}
A.~V. Smirnov, V.~A. Smirnov, and M.~Tentyukov,
\newblock Comput. Phys. Commun. {\bf 182}, 790 (2011), arXiv:0912.0158.

\bibitem{Vermaseren:1994je}
J.~A.~M. Vermaseren,
\newblock Comput. Phys. Commun. {\bf 83}, 45 (1994).

\bibitem{Stein}
W.~A. Stein,
\newblock {\em Modular Forms, a Computational Approach} (American Mathematical
  Society, 2007).

\end{thebibliography}
\bibliographystyle{/home/stefanw/latex-style/h-physrev5}
}

\end{document}